\documentclass[ALICE,manyauthors]{cernphprep}
\usepackage[comma,square,numbers,sort&compress]{natbib}
\usepackage{hyperref}
\usepackage{graphicx}

\usepackage{epstopdf}
\usepackage{xspace}
\usepackage{upgreek}
\usepackage{color}
\usepackage{doi}
\usepackage{afterpage}

\usepackage{float}
\usepackage{flexisym}

\usepackage{xargs,lipsum,caption,changepage,ifthen}
\usepackage{calc}

\usepackage{sidecap}
\usepackage[section]{placeins}

\usepackage{rotating}
\usepackage{pdflscape}

\usepackage{tabularx}
\usepackage{booktabs}
\usepackage{footnote}
\usepackage{threeparttable}
\usepackage{multirow}
\usepackage{bm}
\usepackage{relsize}
\usepackage{gensymb}

\usepackage{chngcntr}

\usepackage{amsmath}
\usepackage{amssymb}

\usepackage{array}
\newcolumntype{L}[1]{>{\raggedleft\arraybackslash}p{#1}}

\newcolumntype{C}[1]{>{\centering\arraybackslash}p{#1}}

\newcolumntype{R}[1]{>{\raggedleftlet\newline\\arraybackslashhspace{0pt}}m{#1}}
\newcolumntype{Y}{>{\raggedleft\arraybackslash}X}
\newcolumntype{K}[1]{>{\raggedright\arraybackslash}p{#1}}

\newlength\mylena %
\usepackage[T1]{fontenc}

  \usepackage{pp13macros}

\begin{document}%

\begin{titlepage}

\PHyear{2020}
\PHnumber{059}      
\PHdate{15 April}  

\title{Production of light-flavor hadrons in \pp collisions at $\sqrt{s}~=~7\text{ and }\sppt{13}$}
\ShortTitle{Production of light-flavor hadrons in \pp collisions at $\sqrt{s}~=~7\text{ and }\sppt{13}$} 

\Collaboration{ALICE Collaboration\thanks{See Appendix~\ref{app:collab} for the list of collaboration members}}
\ShortAuthor{ALICE Collaboration} 

\begin{abstract} \label{abstr}

 The production of $\pi^{\pm}$, $\rm{K}^{\pm}$, $\rm{K}^{0}_{S}$, $\rm{K}^{*}(892)^{0}$, $\rm{p}$,
 $\phi(1020)$, $\Lambda$, $\Xi^{-}$, $\Omega^{-}$, and their antiparticles was measured in
 inelastic proton-proton (pp) collisions at a center-of-mass energy of $\sqrt{s}$ = 13 TeV at
 midrapidity ($|y|<0.5$) as a function of transverse momentum ($p_{\rm{T}}$) using the ALICE
 detector at the CERN LHC. Furthermore, the single-particle $p_{\rm{T}}$ distributions of
 $\rm{K}^{0}_{S}$, $\Lambda$, and $\overline{\Lambda}$ in inelastic pp collisions at \mbox{$\sqrt{s}$ =
 7 TeV} are reported here for the first time. The $p_{\rm{T}}$ distributions are studied at
 midrapidity within the transverse momentum range $0\leq p_{\rm{T}}\leq20$ GeV/$c$, depending on
 the particle species. The $p_{\rm{T}}$ spectra, integrated yields, and particle yield ratios are
 discussed as a function of collision energy and compared with measurements at lower $\sqrt{s}$
 and with results from various general-purpose QCD-inspired Monte Carlo models. A hardening of the
 spectra at high $p_{\rm{T}}$ with increasing collision energy is observed, which is similar for
 all particle species under study. The transverse mass and $x_{\rm{T}}\equiv2p_{\rm{T}}/\sqrt{s}$
 scaling properties of hadron production are also studied. As the collision energy increases from
 $\sqrt{s}$ = 7 to 13 TeV, the yields of non- and single-strange hadrons normalized to the pion
 yields remain approximately constant as a function of $\sqrt{s}$, while ratios for multi-strange
 hadrons indicate enhancements. The $p_{\rm{T}}$-differential cross sections of $\pi^{\pm}$,
 $\rm{K}^{\pm}$ and $\rm{p}$ ($\overline{\rm{p}}$) are compared with next-to-leading order
 perturbative QCD calculations, which are found to overestimate the cross sections for $\pi^{\pm}$
 and $\rm{p}$ ($\overline{\rm{p}}$) at high $p_{\rm{T}}$.
  
\end{abstract}
\end{titlepage}
\setcounter{page}{2}


\section{Introduction} \label{sec:intro}

  Identified particle spectra and yields, which are among the most fundamental physical observables in high-energy hadronic collisions, have been 
  intensively studied in hadron-collider and cosmic-ray physics for many decades~\cite{dEnterria:2011twh}.
  Hadron production at collider energies originates from soft and hard scattering processes at the partonic level. Hard scatterings, where two partons interact 
  with a large momentum transfer, are responsible for the production of particles with high transverse momenta. This process is theoretically 
  described by perturbative Quantum Chromodynamics (pQCD) calculations based on the factorization theorem~\cite{Collins:1989gx}. In this approach, the cross section 
  is a convolution of the parton distribution function (PDF), the partonic QCD matrix elements, and the fragmentation function (FF). The PDFs describe the 
  probability densities of finding a parton with a specific flavor carrying fraction $x$ of the proton momentum, whereas the FFs encode the probability densities 
  that the parton with a specific flavor fragments into a hadron carrying a fraction of the parton's longitudinal momentum; both considered at a given energy scale.
  At the LHC, with increasing center-of-mass collision energy (\cme), the lower $x$ regime is probed and contributions from hard-scattering processes increase. 
  In the kinematic region probed by these measurements, high-\pt particles dominantly originate from the fragmentation of gluons~\cite{deFlorian:2007ekg,dEnterria:2013sgr}. 
  Parameterizations of both the PDFs and FFs are derived from global analyses~\cite{deFlorian:2014xna,deFlorian:2017lwf} based on fits to experimental data at various 
  \cme with next-to-leading order (NLO) accuracy. These include single-inclusive hadron production in semi-inclusive electron-positron annihilation data, 
  semi-inclusive deep-inelastic scattering, and single inclusive hadron spectra at high \pt, notably including results at LHC energies. Results presented in this paper 
  can be used as further input for these studies. In particular, identified particle spectra provide new constraints on the gluon-to-pion and, especially, gluon-to-kaon 
  fragmentation functions~\cite{Abelev:2014laa,Acharya:2017hyu,Koch:2011fw,Acharya:2017tlv}. While particle production at high \pt is expected to be calculable 
  with pQCD, the LHC results are in general not well reproduced by pQCD calculations, see Ref.~\cite{deFlorian:2017lwf} and references therein. Charged particle production 
  at high \pt is known to scale with $\xt\equiv2\pt/\sqrt{s}$, as observed in a wide energy range up to \sppt{7}. This has been observed by the CDF Collaboration in 
  p$\bar{\text{p}}$ collisions at the Tevatron~\cite{Aaltonen:2009ne,Abe:1988yu}, by the UA1 Collaboration at the CERN SPS~\cite{Albajar:1989an}, by the STAR Collaboration 
  in \pp collisions at RHIC~\cite{Adams:2006nd}, and by the CMS Collaboration~\cite{Chatrchyan:2011av} at the CERN LHC. Above $\xt\simeq10^{-2}$, significant deviations 
  from the leading-twist NLO pQCD predictions have been reported in Ref.~\cite{Arleo:2009ch} and are investigated in this paper.
  
  The bulk of particles produced at low transverse momenta ($\pt<\gevc{2}$) originate from soft scattering processes involving small momentum transfers. In this regime, 
  particle production cannot be calculated from first principles. Instead, calculations rely on QCD inspired phenomenological models, which are tuned to reproduce previous 
  measurements. Hence, measurements at low \pt provide further important constraints on such models. The universal transverse mass, $\mT\equiv\sqrt{m^{2}+\pt^{2}}$, scaling 
  originally proposed by R. Hagedorn~\cite{Hagedorn:346206} was first seen to hold approximately at ISR energies~\cite{Gatoff:1992cv}. It was then observed by the 
  PHENIX~\cite{Adare:2010fe, Adare:2011vy} and STAR~\cite{Abelev:2006cs} collaborations to hold only separately for mesons and baryons at RHIC energies, by applying the 
  approximate \mT scaling relation respectively for pions and protons. At $\sppg{900}$ a disagreement was observed for charged kaons and \phm mesons, which indicated a 
  breaking of the generalized scaling behavior~\cite{Khandai:2012xx}. Moreover, recent studies, based on identified particle spectra measured in \pp collisions at \sppt{7} 
  by ALICE, indicate that \mT scaling also breaks in the low-\pt region~\cite{Altenkamper:2017qot}. These observations motivate studies of the applicability of \mT scaling 
  of particle production through the precise measurement of identified hadrons at \sppt{13}.
  
  The results reported in this paper are the measurements of the production of \pix, \kx, \kzs, \ksm, \ksbm, \prp, \prm, \phm, \lap, \lam, \xim, \xip, \omm and \omp at the 
  highest collision energies, and therefore extend the studies of the energy dependence of the production of light-flavor hadrons into new territory. The study of the 
  production of the \ksm and \phm resonances, containing respectively one and two strange valence quarks, contributes to a better understanding of strange particle 
  production mechanisms. Because of their short lifetimes (${\sim}\fmc{4}$ for \ksm and ${\sim}\fmc{46}$ for \phm), their decay daughters may undergo re-scattering and/or 
  regeneration processes that affect their yields and the shapes of their \pt distributions. In addition, multi-strange baryons, \omm (\omp) and \xim (\xip), are of crucial 
  importance due to their dominant strange (s) quark content. Furthermore, the production of \kzs, \lap, and \lam in \pp collisions at \sppt{7} is reported here for the first 
  time, completing the set of reference measurements at that energy~\cite{Adam:2015qaa,Adam:2016dau,Abelev:2012hy,Abelev:2012jp}. 
  
  The present measurements serve as important baselines for studies of particle production as a function of the charged-particle multiplicity~\cite{Acharya:2019kyh} or event 
  shape (e.g. spherocity)~\cite{Acharya:2019mzb} and also provide input to tune the modeling of several contributions in Monte Carlo (MC) event generators such as 
  \py~\cite{Skands_Perugia_Tunes,Skands_Monash_Tune} and \eplhc~\cite{Pierog:2013ria}. In addition, measurements in minimum bias \pp collisions reported in this paper serve 
  as reference data to study nuclear effects in proton-lead (\ppb) and lead-lead (\pb) collisions. 
  
  The paper is organized as follows. In Section 2 the ALICE experimental apparatus and the analyzed data samples are described, focusing on the detectors which are relevant 
  for the presented measurements. In Section 3 the details of the event and track selection criteria and of the Particle IDentification (PID) techniques are discussed. The 
  results are given in Section 4, in which the \pt spectra and the extraction procedures for the \pt-integrated yield and average \pt are presented. Section 5 discusses the 
  results, followed by a summary in Section 6. For the remainder of this paper, the masses will be omitted from the symbols of the strongly decaying particles, which will be 
  denoted as \ks, \ksb, and \ph.


\section{Experimental setup} \label{sec:alice}

  A detailed description of the ALICE detector and its performance can be found in Refs.~\cite{Aamodt:2008zz, Abelev:2014ffa}. The main subsystems of the ALICE detector used 
  in this analysis are the V0 detector, the Inner Tracking System (ITS), the Time Projection Chamber (TPC), the Time of Flight (TOF) detector, and the High-Momentum Particle 
  Identification Detector (HMPID).

  The V0 detector~\cite{ALICE_VZERO} is used for triggering and beam background suppression. It is made up of two scintillator arrays placed along the beam axis on each side 
  of the interaction point (IP) at $z =340$~cm and $z=-90$ cm, covering the pseudorapidity regions $2.8<\eta<5.1$ (V0A) and $-3.7<\eta<-1.7$ (V0C), respectively.

  In the measurements of light-flavor hadrons, primary charged particles are considered. Primary particles are defined as particles with a mean proper lifetime $\tau$ that is 
  larger than 1\,cm/$c$, which are either produced directly in the interaction or from decays of particles produced at the interaction vertex with $\tau$ shorter than 1\,cm/$c$. 
  This excludes particles produced in interactions with the detector material~\cite{ALICE-PUBLIC-2017-005}. Primary charged-hadron tracks are reconstructed by the ITS and TPC 
  detectors, which have full azimuthal acceptance within $|\eta|<0.8$ for full-length tracks. They are located inside a solenoidal magnet providing a magnetic field of $B=0.5$\,T.

  The ITS~\cite{Aamodt:2008zz,ALICE_ITS} is a silicon tracking detector made up of six concentric cylindrically-shaped layers, measuring high-resolution space points near the 
  collision vertex. The two innermost layers consist of Silicon Pixel Detectors (SPD) used to reconstruct the primary vertex of the collision and short track segments called 
  ``tracklets". The four outer layers are equipped with silicon drift (SDD) and strip (SSD) detectors and allow measurement of the specific energy loss (\dedx) with a relative 
  resolution of about 10\%. The ITS is also used as a stand-alone tracking detector to reconstruct charged particles with momenta below \mevc{200} that are deflected or decay 
  before reaching the TPC.

  The TPC~\cite{ALICE_TPC} is the main tracking detector of ALICE. It is a large volume cylindrical drift detector spanning the approximate radial and longitudinal ranges 
  $85 < r < 250$~cm and $-250 < z < 250$~cm, respectively. The endcaps of the TPC are equipped with multiwire proportional chambers (MWPCs) segmented radially into pad rows. 
  Together with the measurement of the drift time, the TPC provides three dimensional space point information, with up to 159 tracking points. Charged tracks originating from 
  the primary vertex can be reconstructed down to $p{\sim}\mevc{100}$~\cite{Abelev:2014ffa}, albeit with a lower tracking efficiency for identified charged hadrons with 
  $\pt<\mevc{200}$. Combining information from the ITS and TPC allows the momenta of charged particles to be measured for momenta from 0.05 to \gevc{100} with a resolution 
  of 1--10\%, depending on \pt. The TPC provides charged-hadron identification via measurement of the specific energy loss \dedx in the fill gas, with a resolution of 
  ${\sim}5\%$~\cite{ALICE_TPC}. 

  The Time of Flight detector (TOF)~\cite{ALICE_TOF1,ALICE_TOF2,ALICE_TOF3} is a cylindrical array of multi-gap resistive plate chambers which sits outside the TPC. It covers 
  the pseudorapidity range $|\eta|<0.9$ with (almost) full azimuthal acceptance. The total time-of-flight resolution, including the resolution on the collision time, is about 
  90~ps in \pp collisions.

  The HMPID consists of seven proximity focusing Ring Imaging Cherenkov (RICH) counters. Primary charged particles penetrate the radiator volume, filled with liquid 
  $\text{C}_{6}\text{F}_{14}$, and generate Cherenkov photons that are converted into photoelectrons in thin CsI-coated photocathodes. Photo-electron clusters, together with 
  pad clusters (also called ``MIP'' clusters) associated with the primary ionization of a particle, form Cherenkov rings. The amplified signal is read out by MWPCs, filled with 
  $\text{CH}_{4}$. The detector covers $|\eta|<0.5$ and $1.2^{\circ}<\varphi<58.5^{\circ}$, which corresponds to ${\sim}5\%$ of the TPC geometrical acceptance.


\section{Event and track selection} \label{subsec:analyses:eventsel}

    \subsection{Event selection\label{subsubsec:eventsel}}

    The measurements at \sppt{13} are obtained from a minimum bias data sample of \pp collisions collected in June 2015 during a period of low pileup in LHC Run 2. The minimum 
    bias trigger required at least one hit in both of the V0 scintillator arrays in coincidence with the arrival of proton bunches from both directions along the beam axis. The 
    mean number of inelastic proton-proton interactions per single bunch crossing ranges between 2\% and 14\%. A requirement of a coincidence of signals in both V0A and V0C detectors removes 
    contamination from single-diffractive and electromagnetic events.  Contamination arising from beam-induced background events, produced outside the interaction region, is 
    removed offline by using timing information from the V0 detector, which has a time resolution better than 1~ns. Background events are further rejected by exploiting the 
    correlation between the number of clusters and the multiplicity of tracklets in the SPD. From the triggered events, only events with a reconstructed primary vertex are 
    considered for the analyses. Additionally, the position of the primary vertex along the beam axis is required to be within $\pm10$~cm with respect to the nominal interaction 
    point (center of the ALICE barrel). This requirement ensures that the vast majority of reconstructed tracks are within the central barrel acceptance ($|\eta|<0.8$) and it 
    reduces background events by removing unwanted collisions from satellite bunches. Contamination from pileup events, which have more than one \pp collision per bunch crossing,
    were rejected offline by excluding events with multiple primary vertices reconstructed in the SPD~\cite{Abelev:2014ffa}. The pileup-rejected events are less than 1\% of the total sample 
    of minimum bias events. The size of the analyzed sample after selections ranges between 40 and 60 million events (corresponding to an integrated luminosity $0.74-1.1\,\rm{nb}^{-1}$), 
    depending on the requirements of the analyses of the different particle species. 
    
    The measurements of \kzs, \lap, and \lam at \sppt{7} are obtained by analyzing a sample of about 150 million events (corresponding to an integrated luminosity $2.41\,\rm{nb}^{-1}$) 
    collected in 2010 during the LHC Run 1 data taking period. The corresponding trigger and event selection criteria applied were very similar to those used for the measurements at 
    \sppt{13}; see Refs.~\cite{Adam:2015qaa,ALICE_charged_pp13} for details on the triggering and event selection for these periods.

    All corrections are calculated using Monte Carlo events from \py~6 and \py~8. \py~6.425 (Perugia 2011 tune) and \py~8.210 (Monash 2013 tune) event generators were used for 
    \sppt{13}. The \py~6.421 (Perugia 0 tune) was used for \kzs and \lap at \sppt{7} because that production was used for correcting the other \sppt{7} analyses. The particles 
    produced using these event generators were propagated through a simulation of the ALICE detector using GEANT3~\cite{GEANT3}.

    \subsection{Track selection} \label{subsubsec:tracksel}

    Tracks from charged particles are reconstructed in the TPC and ITS detectors and then propagated to the outer detectors and matched with reconstructed points in the TOF and HMPID. 
    Additionally, in the analysis of \pix, \kx and \px, a dedicated tracking algorithm based only on ITS information (ITS stand-alone, ITS-sa) was used to reconstruct low momentum tracks. 
    In the measurements, global tracks, which are reconstructed using the combined ITS and TPC information, are distinguished from ITS-sa tracks.

    For analyses using global tracks, track selection criteria are applied to limit the contamination due to secondary particles, to maximize tracking efficiency and improve the \dedx 
    and momentum resolution for primary charged particles, and to guarantee an optimal PID quality. The number of crossed pad rows in the TPC is required to be at least 70 (out of a 
    maximum possible of 159); the ratio of the number of crossed pad rows to the number of findable clusters (that is the number of geometrically possible clusters which can be assigned 
    to a track) is restricted to be greater than 0.8, see Ref.~\cite{Abelev:2014ffa} for the details. The goodness-of-fit values $\chi^{2}$ per cluster ($\chi^{2}/N_{\rm clusters}$) of 
    the track fit in the TPC must be less than 4. Tracks must be associated with at least one cluster in the SPD and the $\chi^{2}$ values per cluster in the ITS are restricted in order 
    to select high-quality tracks.
    The distance of closest approach (DCA) to the primary vertex in the plane perpendicular to the beam axis (DCA$_{xy}$) is required to be less than 7 times the resolution of this quantity; 
    this selection is \pt dependent, i.e. $\text{DCA}_{xy} < 7 \times (0.0015+0.05\times(\pt/(\gevc{}))^{-1.01})$ cm. A loose selection criterion is also applied on the the DCA in the beam 
    direction (DCA$_{z}$), by rejecting tracks with DCA$_{z}$ larger than 2~cm, to remove tracks from possible residual pileup events. The transverse momentum of each track must be greater 
    than 150~\mvc and the pseudorapidity is restricted to the range $|\eta|<0.8$ to avoid edge effects in the TPC acceptance. Additionally, tracks produced by the reconstructed weak decays 
    of pions and kaons (the ``kink" decay topology) are rejected.
    
    For the topological reconstruction of weakly decaying particles, the selected global tracks are combined using specific algorithms, as described in Sec.~\ref{subsubsec:analyses:pidreco:weakdecaystrange}.
    Track selection criteria are the same applied for global tracks with a few exceptions: for tracks used in the reconstruction of 
    \kzs, \lap, \lam, \xim, \xip, \omm, and \omp, no ITS information is required and special selection criteria are applied on the DCA to the collision vertex, as shown in Table~\ref{tab:V0AndCascadeselection}. The kink topology tracks that are used to reconstruct the weak decays of \kx  do not have ITS information. For the latter, 
    removal of contributions from pileup collisions outside the trigger proton bunch (``out-of-bunch pileup'') is achieved by requiring that at least one charged decay track matches a hit 
    in a ``fast'' detector (either the ITS or the TOF detector).
    
    ITS stand-alone tracking uses similar selection criteria to those mentioned above. Tracks are required to have at least four ITS clusters, with at least one in the SPD, three in the 
    SSD and SDD and $\chi^{2}/N_{\rm clusters}<2.5$. This further reduces contamination from secondary tracks and provides high resolution for the track impact parameter and optimal resolution 
    for \dedx. Similar to global tracks, a \pt-dependent parameterization of the DCA$_{xy}$ selection is used, but with different parameters to account for the different resolution. For the 
    \pt ranges used in this analysis, the selected ITS-sa tracks have the same \pt resolution as those measured in \pp collisions at \sppt{7}: 6\% for pions, 8\% for kaons, and 10\% for protons~\cite{Adam:2015qaa}.


\section{Data analysis techniques} \label{subsec:analyses:pidreco}

  Table~\ref{tab:analyses:reco} lists the basic characteristics of the particles studied in this paper. This section describes the techniques used to measure the yields of the various 
  hadron species. In Sec.~\ref{common}, aspects common to all analyses are described, including the correction and normalization procedure and the common sources of systematic uncertainties. 
  Next, the analysis of each hadron species is described in detail. The measurements of charged pions, charged kaons, and (anti)protons, which are performed using several different PID techniques, 
  are described in Sec.~\ref{subsubsec:analyses:pidreco:pikp}. It is worth noting that charged kaons are also identified using the kink topology of their two-body decays. The measurements of 
  weakly decaying strange hadrons (\kzs, \lap, \xim, \omm and their antiparticles) are reported in Sec.~\ref{subsubsec:analyses:pidreco:weakdecaystrange}, followed by the strongly decaying 
  resonances (\ks, \ksb, and \ph) in Sec.~\ref{resonances}.

    \begin{table}[t]
      \let\center\empty
      \let\endcenter\relax
      \centering
      \caption{Main characteristics of the reconstructed particles: valence quark content, mass, proper decay length ($c\tau$),
      the decay channel studied in this paper, and the corresponding branching ratio (B.R.)~\cite{PDG}.}
      \resizebox{0.99\textwidth}{!}{

  \begin{threeparttable}
  \begin{tabular}{c c c c c c c c c c c c c}
  \toprule
  & \multirow{2}{*}{Particle} 		& & Valence 							& & Mass		& & \multirow{2}{*}{$c\tau$}	& & \multirow{2}{*}{Decay}	& & 	\multirow{2}{*}{B.R. (\%)}	& \\
  &  					& & Quark Content						& & (MeV/$c^{2}$) 	& &  				& & 				& & 					& \\
  \midrule
  & \pip					& & $\mathrm{u\overline{d}}$						& &  139.57		& & 7.8 m	& & ---					& & 	---		& \\ [2pt]
  & \kp						& & $\mathrm{u\overline{s}}$						& &  493.68		& & 3.7 m	& & $\kp\rightarrow\mu^{+}\nu_{\mu}$ & & 	$63.56\pm0.11$		& \\ [2pt]
  & \kzs 					& & $\frac{1}{\sqrt{2}}(\mathrm{d\overline{s}}-\mathrm{\overline{d}s})$	& &  497.61		& & 2.68 cm	& & $\kzs\rightarrow \pip+\pim$ 	& & 	$69.20\pm0.05$	& \\ [2pt]
\rot{\rlap{Mesons}} & \ks 			& & $\mathrm{d\overline{s}}$						& &  895.55		& & 4.16 fm	& & $\ks\rightarrow\pim+\kp$		& & 	$\sim66.6$	& \\ [2pt]
  & \ph 					& & $\mathrm{s\overline{s}}$						& & 1019.46		& & 46.2 fm	& & $\ph\rightarrow \kp+\km$ 		& & 	$49.2\pm0.5$	& \\ [2pt]
  \midrule
  & \prp	  				& & uud								& & 938.27 		& & ---		& & ---					& &	---	 	& \\ [2pt]
  & \lap	  				& & uds								& & 1115.68 		& & 7.89 cm	& & $\lap\rightarrow\prp+\pim$		& & 	$63.9\pm0.5$	& \\ [2pt]
  & \xim	 				& & dss								& & 1321.71		& & 4.91 cm	& & $\xim\rightarrow \lap+\pim$ 	& & 	$99.887\pm0.035$& \\ [2pt]
  \rot{\rlap{Baryons}} & \omm 	& & sss										& & 1672.45		& & 2.46 cm	& & $\omm\rightarrow\lap+\km$		& & 	$67.8\pm0.7$	& \\ [2pt]
  \bottomrule
  \end{tabular}
  \end{threeparttable}

}
      \label{tab:analyses:reco}
    \end{table}

    \subsection{Common aspects of all analyses} \label{common}
  
    In several of the analyses presented below, the measured PID signal is compared to the expected value based on various particle mass hypotheses. The difference between the measured 
    and expected values is expressed in terms of $\sigma$, the standard deviation of the corresponding measured signal distribution. The size of this difference, in multiples of $\sigma$, 
    is denoted $n_{\sigma}$. In the following, the $\sigma$ values accounting for the resolution of the PID signals measured in the TPC and TOF detectors are denoted as \stpc and \stof, respectively.

	The corrected yield of each hadron species as a function of \pt is 
	\begin{equation}\label{eq:correction}
	Y_{\mathrm{corr}} = \frac{Y_{\mathrm{raw}}}{\Delta\pt\,\Delta{y}}\times\frac{\fsl}{A\times\varepsilon}\times(1-\fcont)\times\ffluka~.
	\end{equation}
	$Y_{\mathrm{corr}}$ is obtained by following the procedure described in previous publications.
	Here, $Y_{\mathrm{raw}}$ is the number of particles measured in each \pt bin and $A\times\varepsilon$ is the product of the acceptance and the efficiency (including PID efficiency, 
	matching efficiency, detector acceptance, reconstruction, and selection efficiencies). Monte Carlo simulations are used to evaluate $A\times\varepsilon$, which takes on similar values 
	to those found in our previous analyses. The factor \fsl, also known as the ``signal-loss'' correction, accounts for reductions in the measured particle yields due to event triggering 
	and primary vertex reconstruction. Such losses are more important at low \pt, since events that fail the trigger conditions or fail to have a reconstructible primary vertex tend to have 
	softer particle \pt spectra than the average inelastic collision. For \sppt{13}, \fsl deviates from unity by a few percent at low \pt to less than one percent for $\pt\gtrsim\gevc{2}$. 
	The trigger configuration used for \sppt{7} resulted in negligible signal loss, thus \fsl is set to unity for this energy. The factor $(1-\fcont)$ is used to correct for contamination 
	from secondary and misidentified particles; \fcont is non-zero only for the measurements of \pix, \kx, \px, \lap, and \lam, and it is more important at low \pt. The computation of \fcont 
	for those species is described further in the relevant sections below. The factor \ffluka corrects for inaccuracies in the hadronic production cross sections in GEANT3, which is used in 
	the calculation of $A\times\varepsilon$ to describe the interactions of hadrons with the detector material of ALICE. GEANT4 and FLUKA~\cite{Battistoni:2007zzb}, which have more accurate 
	descriptions of the hadronic cross sections, are used to calculate the correction factor, which can be different from unity by up to a few percent. The correction \ffluka is applied only 
	for the analyses of \km, \prm, \lam, \xip, and \omp.

	After correction, the yields are normalized to the number of inelastic \pp collisions using the ratio of the ALICE visible cross section to the total inelastic cross section. This ratio 
	is $0.852^{+0.062}_{-0.030}$ for \sppt{7}~\cite{Abelev:2012sea} and $0.7448 \pm 0.0190$ for \sppt{13}~\cite{ALICE_Luminosity_pp13,Loizides:2017ack}.

	The procedures for the estimation of systematic uncertainties strictly follow those applied in our measurements from LHC Run 1. All described uncertainties are assumed to be strongly 
	correlated among adjacent \pt bins. For the evaluation of the total systematic uncertainty in every analysis, all contributions originating from different sources are considered to be 
	uncorrelated and summed in quadrature. Components of uncertainties related to the ITS-TPC matching efficiency correction and to the event selection are considered correlated among different 
	measurements. The systematic uncertainty due to the normalization to the number of inelastic collisions is $\pm~2.6\%$ for \sppt{13} and $^{+7.3\%}_{-3.5\%}$ for \sppt{7} independent of \pt. 
	This uncertainty is common to all measured \pt spectra and \dndyinline values (see Sec.~\ref{subsec:ptspectra}) at a given energy. The systematic uncertainty associated to possible residual 
	contamination from pileup events was estimated varying pileup rejection criteria and was found to be of 1\%. The singal loss correction has a small dependence on the Monte Carlo event 
	generator used to calculate it. These variations result in \pt-dependent uncertainties that are largest at low \pt, where they have values of $0.2\%$ for $\Omega$, ${\sim}1\%$ for 
	\pix, \kx, \px, and $\Xi$, and ${\sim}2\%$ for \kzs, \lam, \ks, and \ph.	
	
	The systematic uncertainty accounting for the limited knowledge of the material budget is estimated by varying the amount of detector material in the MC simulations within its expected 
	uncertainties~\cite{Abelev:2014ffa}. For the analysis of \pix, \kx, \px, \ks, and \ph, the values are taken from the studies reported in Refs.~\cite{Abelev:2013vea} and~\cite{ALICE_Kstar_phi_pPb}. 
	This uncertainty is estimated to be around $3.3\%$ for \kx, $1.1\%$ for \pix, $1.8\%$ for \px, $3\%$ for \ks, and $2\%$ for \ph; it is largest at low momenta and tends to be negligible 
	towards higher momenta. For the measurement of \kzs and \lap at \sppt{7}, the material budget uncertainty is estimated to be $4\%$, independent of \pt. For the measurements of \kzs, \lap, 
	$\Xi$ and $\Omega$ at \sppt{13}, the material budget uncertainty is \pt dependent for low \pt ($\lesssim\gevc{2}$) and constant at higher \pt. For low \pt, the uncertainty reaches maximum 
	values of about 4.7\% for \kzs, 6.7\% for \lap, 6\% for $\Xi$, and 3.5\% for $\Omega$; at high \pt, the uncertainty is less than 1\% for \kzs, \lap, and $\Xi$, and about 1.5\% for $\Omega$.
	
	The systematic uncertainty due to the limited description of the hadronic interaction cross sections in the transport code is evaluated using GEANT4 and FLUKA. This leads to uncertainties 
	of up to $2.8\%$ for \pix, $2.5\%$ for \kx, $0.8\%$ for \prp, and $5\%$ for \prm~\cite{Abelev:2013vea}. It is at most $3\%$ for \ks, $2\%$ for \ph and $1-2\%$ for the strange baryons. It is 
	negligible for \kzs at both reported collision energies. In the following sections, details are given on the contributions (specific to each analysis) related to track or topological selections 
	and signal extraction methods, as well as those related to feed-down.

    \subsection{Identification of primary charged pions, charged kaons, and (anti)protons} \label{subsubsec:analyses:pidreco:pikp}

    To measure the production of primary charged pions, kaons, and (anti)protons over a wide range of \pt, five analyses using distinct PID techniques were carried out. The individual analyses 
    follow the techniques adopted in previous measurements based on data collected at lower center-of-mass energies and for different collision systems during LHC Run 1~\cite{Aamodt:2011zj,Adam:2015kca,Adam:2015qaa,Adam:2016dau,Acharya:2018orn}. 
    The \pt spectra have been measured from $\pt=\gevc{0.1}$ for pions, $\pt=\gevc{0.2}$ for kaons, and $\pt=\gevc{0.3}$ for protons, up to \gevc{20} for all three species. The individual analyses 
    with their respective \pt reaches are summarized in Tab.~\ref{tab:pikp_combined}. All the analysis techniques are extensively described in Refs.~\cite{Aamodt:2011zj,Abelev:2013vea,Adam:2015qaa,Abelev:2014laa}. 
    Each procedure is discussed separately in Secs.~\ref{subsubsec:pidreco:pikp:itssa}--\ref{subsubsec:pidreco:pikp:kinks}, with special emphasis on those aspects that are relevant for the current 
    measurements. The results for the different analyses are then combined as described in Sec.~\ref{subsec:ptCombined}.

    The calculation of \fcont in Eq.~\ref{eq:correction}~at low \pt is performed by subtracting the secondary \pix, \kx, and \px from the primary particle sample. This method is data-driven and it 
    is based on the measured distance of closest approach to the primary vertex in the plane transverse to the beam direction ($\text{DCA}_{xy}$), following the same procedure adopted in Ref.~\cite{Adam:2015qaa}. 
    The $\text{DCA}_{xy}$ distribution of the selected tracks was fitted in every \pt bin with Monte Carlo templates composed of three ingredients: primary particles, secondaries from material and 
    secondaries from weak decays, each accounting for the expected shapes of the distribution. Because of the different track and PID selection criteria, the contributions are different for each 
    analysis. The resulting corrections are significant at low \pt and decrease towards higher \pt due to decay kinematics. Up to $\pt=\gevc{2}$, the contamination is $2-10\%$ for pions, up to $20\%$ 
    for kaons (in the narrow momentum range where the \dedx response for kaons and secondary electrons overlap), and $15-20\%$ for protons.
    
	\begin{table}[t]
	  \begin{center}
	  \let\center\empty
	  \let\endcenter\relax
	  \centering
	  \caption{Summary of the kinematic ranges ($p_{\rm T}$ (GeV/$c$) and $\eta$ or $y$) covered by the individual analyses for the measurement of $\pi^{\pm}$, $\text{K}^{\pm}$ and
	  $(\bar{\text{p}})\text{p}$ in \pp collisions at \sppt{13}.
	  }
	  \resizebox{0.75\textwidth}{!}{

  \begin{threeparttable}
  \begin{tabular}{l c c c c c c c c}
  \toprule
  \multicolumn{1}{c}{\multirow{2}{*}{\textbf{Analysis technique}}}  	& & \multicolumn{5}{c}{\pt range (GeV/$c$)} & \multicolumn{2}{c}{\multirow{2}{*}{$\eta$ or $y$ range}} \\ 
  \cmidrule{2-7} \\ [-12pt]
  & & $\pi^{+}+\pi^{-}$ 	& & ${\rm K}^{+}+{\rm K}^{-}$ 	& & ${\rm p}+\bar{\rm p}$ 	& &	    			\\
  \midrule [1pt]
  ITS-sa								& & $0.1-0.7$		& & $0.2-0.6$			& & $0.3-0.65$ 			& & $|y|<0.5$ 			\\ [4pt]

  \multicolumn{1}{l}{\multirow{2}{*}{TPC-TOF fits}}			& & \multicolumn{1}{c}{\multirow{2}{*}{$0.3-3.0$}}	& & \multicolumn{1}{c}{\multirow{2}{*}{$0.3-3.0$}}		& & \multicolumn{1}{c}{\multirow{2}{*}{$0.4-3.0$}} 		& & \multicolumn{1}{c}{\multirow{1}{*}{$|y|<0.5$}}	 		\\ [0pt]  
									& & 		 	& & 				& & 	 			& & $|\eta|<0.4$	 		\\ [4pt]  
  HMPID 								& & $1.5-4$		& & $1.5-4$			& & $1.5-6$ 			& & $|y|<0.5$ 			\\ [4pt]
  Kinks 								& & $-$			& & $0.2-7$			& & $-$ 			& & $|y|<0.5$ 			\\ [4pt]
  TPC rel. rise 							& & $2-20$		& & $3-20$			& & $3-20$ 			& & $|\eta|<0.8$ 		\\ [4pt]
  \bottomrule
  \end{tabular}
  \end{threeparttable}
  \label{table:PID}

}

	  \label{tab:pikp_combined}
	  \end{center}
	\end{table}

    The main sources of systematic uncertainties for each analysis are summarized in Tab.~\ref{tab:syst:pikp}, including contributions common to all analyses. The systematic uncertainty due to 
    the subtraction of secondary particles is estimated by changing the fit range of the $\text{DCA}_{xy}$ distribution, resulting in uncertainties of up to $4\%$ for protons and $1\%$ for pions, 
    with negligible uncertainties for kaons. The uncertainty due to the matching of TPC tracks with ITS hits is estimated to be in the range ${\sim}1-5\%$ for $\pt\lesssim\gevc{3}$ depending on \pt, 
    while it takes values around $6\%$ at higher \pt. This uncertainty together with that resulting from the variation of the track quality selection criteria lead to the systematic uncertainty 
    of the global tracking efficiency that varies from 2.2 to 7.3\% from low to high \pt, independent of particle species.
	
    \afterpage{
	\begin{table}[t]
	  \centering
	  \caption{Summary of the main sources and values of the relative systematic uncertainties (expressed in \%) for the $\pip+\pim$, $\kp+\km$, and $\prp+\prm$ \pt-differential yields.
	  A single value between two or three columns indicates that no \pt dependence is observed. Values are reported for low, intermediate (wherever they are available) and high \pt. The abbreviation ``negl.” indicates a negligible value.
	  }
	  
  \begin{threeparttable}
  \begin{tabular}{K{0.28\textwidth} C{0.005\textwidth} C{0.005\textwidth} C{0.005\textwidth} C{0.005\textwidth} C{0.005\textwidth} C{0.001\textwidth} C{0.001\textwidth} C{0.005\textwidth} C{0.005\textwidth} C{0.005\textwidth} C{0.005\textwidth} C{0.005\textwidth} C{0.001\textwidth} C{0.001\textwidth} C{0.005\textwidth} C{0.005\textwidth} C{0.005\textwidth} C{0.005\textwidth} C{0.005\textwidth} C{0.001\textwidth} C{0.001\textwidth} C{0.001\textwidth}}
  \toprule
  {Hadron species}               	 	&  \multicolumn{6}{c}{\pix} 	  & & \multicolumn{6}{c}{\kx} 		& & \multicolumn{6}{c}{\prp(\prm)}    	&  	\\ 
  \midrule
    \multicolumn{21}{c}{\textbf{Source of uncertainty common to all analyses (\%)}} \\
  \cmidrule{1-21} \\[-25pt] \\
  {\pt (GeV/$c$)}                               & 0.3   & & 2.5 & &20	&   	  & &   0.25    	& & 2.5 & &20 	 & 	& &   0.45     & & 2.5 & & 20 	 	&  	\\ [1pt]
  \cmidrule{2-7}        \cmidrule{9-14} 	\cmidrule{16-21}     												   	\\ [-10pt]
  {Feed-down correction}\tnote{1} 		& 1.0	&     & 1.0 &       & 0.3	& &     & 	    &     & negl.&      &      & &        &  4.0        &     &     & 1.3  &  	&  	\\ [1pt]
  {Hadronic interaction} 		        & 2.8   &     &     & 2.4   & 	    & &     & 	2.5 &     &      & 1.8  &      & &     0.8~(5.0)  &    &     &     & 4.6     &   	&  	\\ [1pt]
  {Material budget} 			        & 0.5 	&     & 1.1 &       & 0.2   & &     & 	3.3 &     & 1.0  &      & 0.3  & &        &  1.7        &     & 1.8    &  & 0.1  	&  	\\ [1pt]
  {Signal-loss correction} 	            & 1.0 	&     & 0.3 &       & 0.2	& &     & 0.9   &     & 0.3  &      & 0.2  & &        &  1.0        &     & 0.2 &      & 0.2  &  	\\ [1pt]
  {Global tracking efficiency}\tnote{2} 	& 2.2 	& & 5.3 & & 7.3	& &     & 	2.0 &     & 5.3  &      & 7.3  & &        &  2.0        &     & 5.3 &      & 7.3  &  	\\ [1pt]
  \midrule
  \midrule
    \multicolumn{21}{c}{\textbf{Source of uncertainty specific to an analysis (\%)}} \\
  \cmidrule{1-21} \\[-25pt] \\
  {\pt (GeV/$c$)}                               & 0.3   & & 1.6 & & 2.8 &   	  & &  0.3    	& & 1.6 & & 2.8 & 	& &   0.4    & & 1.6 & & 2.8  	&  	\\ [1pt]
  \cmidrule{2-7}        \cmidrule{9-14} 	\cmidrule{16-21}     												   	\\ [-10pt]
  {TPC-TOF fits PID}                		& 0.8   & & 4.0 & & 8.0 &   & &   2.0     & & 6.5 & & 14.0&	& &   1.0     & & 2.2 & & 7.0  	&       \\ [1pt]
  {TOF matching efficiency}          	        &       & & 3.0       & &     &   & &           & & 6.0 & &     & 	& &           & & 4.0 & &   		&       \\ [1pt]
  \midrule 
  {\pt (GeV/$c$)}                               & 1.5   & & 2.5 & & 4.0 &   	  & &   1.5    	& & 2.5 & & 4.0  & 	& &   1.5     & & 4.0 & & 6.0  		&  	\\ [1pt]
  \cmidrule{2-7}        \cmidrule{9-14} 	\cmidrule{16-21}     												   	\\ [-10pt]
  {HMPID PID} 					& 3.0 	& & 3.5 & & 12.0&	  & & 	3.0 	& & 3.5 & & 12 &   	& & 	3.0   & & 9.0 & & 11 		&  	\\ [1pt]
  {Distance cut correction} 			& 3.0 	& & 3.0 & & 2.0	&	  & & 	3.0 	& & 2.0 & & 2.0  &   	& & 	4.0   & & 4.0 & & 2.0 		&  	\\ [1pt]
  \midrule 
  {\pt (GeV/$c$)}                               & 2.0   & & 10 & &20	&   	  & &   3.0    	& & 10 & &20 	 & 	& &   3.0     & & 10 & & 20 	 	&  	\\ [1pt]
  \cmidrule{2-7}        \cmidrule{9-14} 	\cmidrule{16-21}     												   	\\ [-10pt]
  {TPC relativistic rise PID} 		& 1.4 	& & 2.2 & & 2.3	&	  & & 	15.0 	& & 7.2 & & 6.5  &   	& &  17     & & 13 & & 13  	&  	\\ [1pt]
  \midrule 
  {\pt (GeV/$c$)}                               &       & & --- & &	&   	  & &   0.25    	& & 2.0 & & 7.0  & 	& &           & & --- & &  	 	&  	\\ [-13pt]
  \cmidrule{2-7}        \cmidrule{9-14} 	\cmidrule{16-21}     												   	\\ [-13pt]
  {Kink PID} 					    &  	& & --- & & 	&	  & & 	2.5 	& & 2.2 & & 2.2  &   	& &  	      & & --- & & 	 	&  	\\ [-13pt]
  {Reconstruction efficiency} 		&  	& & --- & & 	&	  & & 	 	& & 3.0 & &      &   	& &   	      & & --- & &   		&  	\\ [-13pt]
  {Contamination} 			       	&  	& & --- & & 	&	  & & 	negl. 	& & 5.0 & & 3.0 &   	& &           & & --- & & 		&  	\\ [-13pt]
  \midrule 
  {\pt (GeV/$c$)}                               & 0.1   & &  	      & & 0.7 &   & &   0.2    	& & 	      & & 0.6 & & &   0.3     & &  	    & & 0.65  	&  	\\ [1pt]
  \cmidrule{2-7}        \cmidrule{9-14} 	\cmidrule{16-21}     												   	\\ [-10pt]
  {ITS-sa PID}                		        & 3.1   & & 	      & & 5.2 &   & &   1.3     & & 	      & & 4.6 & & &   4.5     & & 	    & & 0.6  	&       \\ [1pt]
  {$\mathbf{E}\times\mathbf{B}$ effect}         &       & & 2.1 & &           &   & &           & & 2.1       & &     & & &           & & 2.1 & &   		&       \\ [1pt]
  {ITS-sa tracking efficiency}                  & 4.1   & & 	      & & 4.3 &   & &   5.3     & & 	      & & 4.3 & & &   9.8     & & 	    & & 3.7  	&       \\ [1pt]
  \cmidrule{1-21} \\[-10pt]
  {\pt (GeV/$c$)}                               & 0.1   & & 2.5 & &20	&   	  & &   0.2    	& & 2.5 & &20 	 & 	& &   0.3     & & 2.5 & & 20 	 	&  	\\ [1pt]
  \cmidrule{2-7}        \cmidrule{9-14} 	\cmidrule{16-21}     												   	\\ [-10pt]
  {\textbf{Total}} 	        	& 8.0 	& & 6.5 & & 8.1	&	  & & 	7.0 	& & 6.3 & & 10.8  &   	& &  8.4     & & 7.3 & & 15.8  	&  	\\ [1pt]
  \bottomrule[1.5pt] \\ [1pt]
  \end{tabular}
      \begin{tablenotes}
        \item[1] Note that in TPC relativistic rise analysis the systematic uncertainty is estimated as half of the applied correction, which is not based on DCA template fits.
        \item[2] This source of uncertainty includes components related to the matching of TPC tracks with ITS hits and the track quality selection criteria. These two components were taken from the analysis of inclusive charged hadrons~\cite{Adam:2015pza}, and are excluded in ITS-sa analysis.
      \end{tablenotes}  
  \end{threeparttable}

	  \label{tab:syst:pikp}
	\end{table}
    \clearpage
    }
	
    \afterpage{
    \begin{figure}[t]
      \centering
	\includegraphics[keepaspectratio,width=0.49\columnwidth]{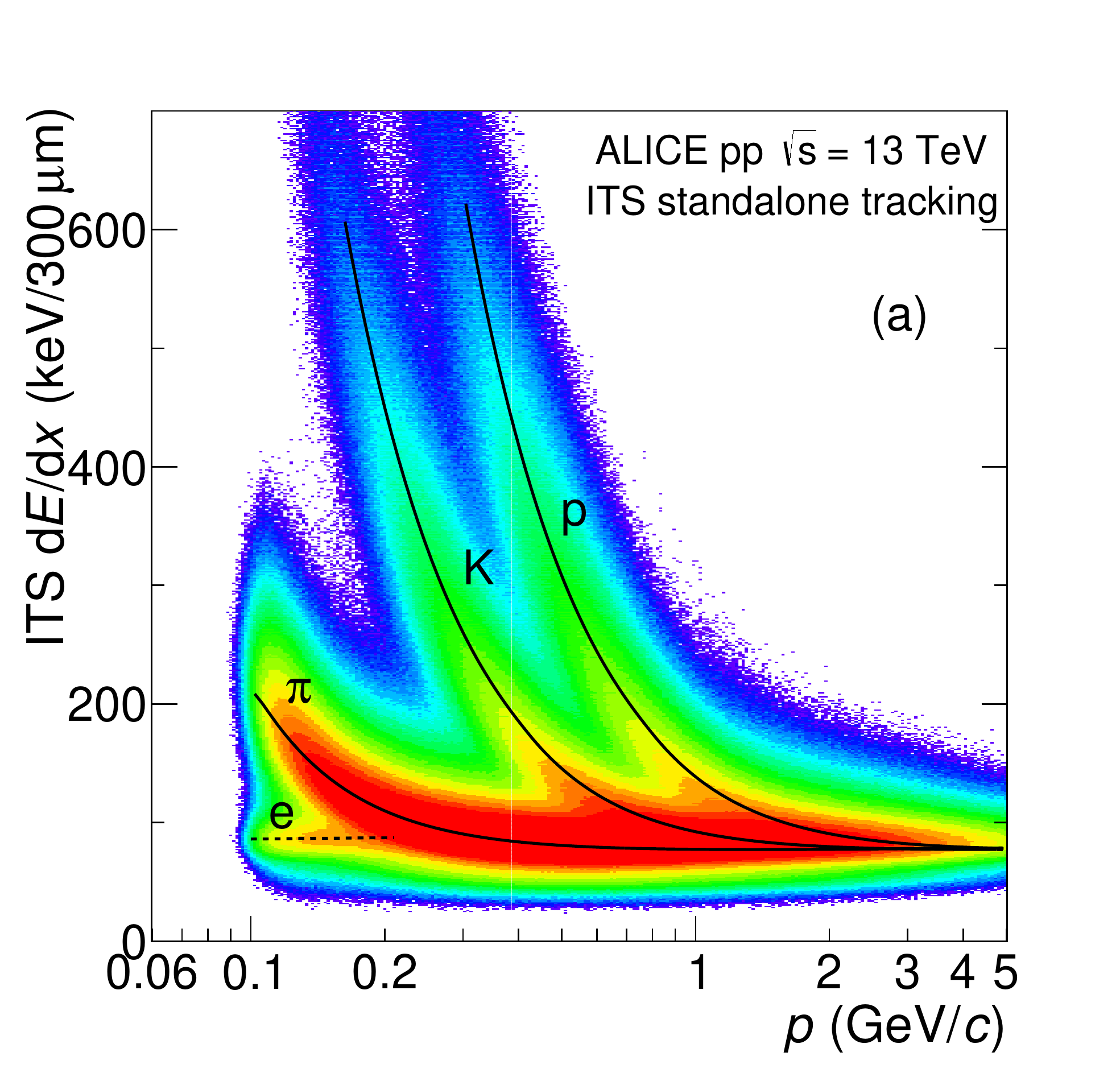}
	\includegraphics[keepaspectratio,width=0.49\columnwidth]{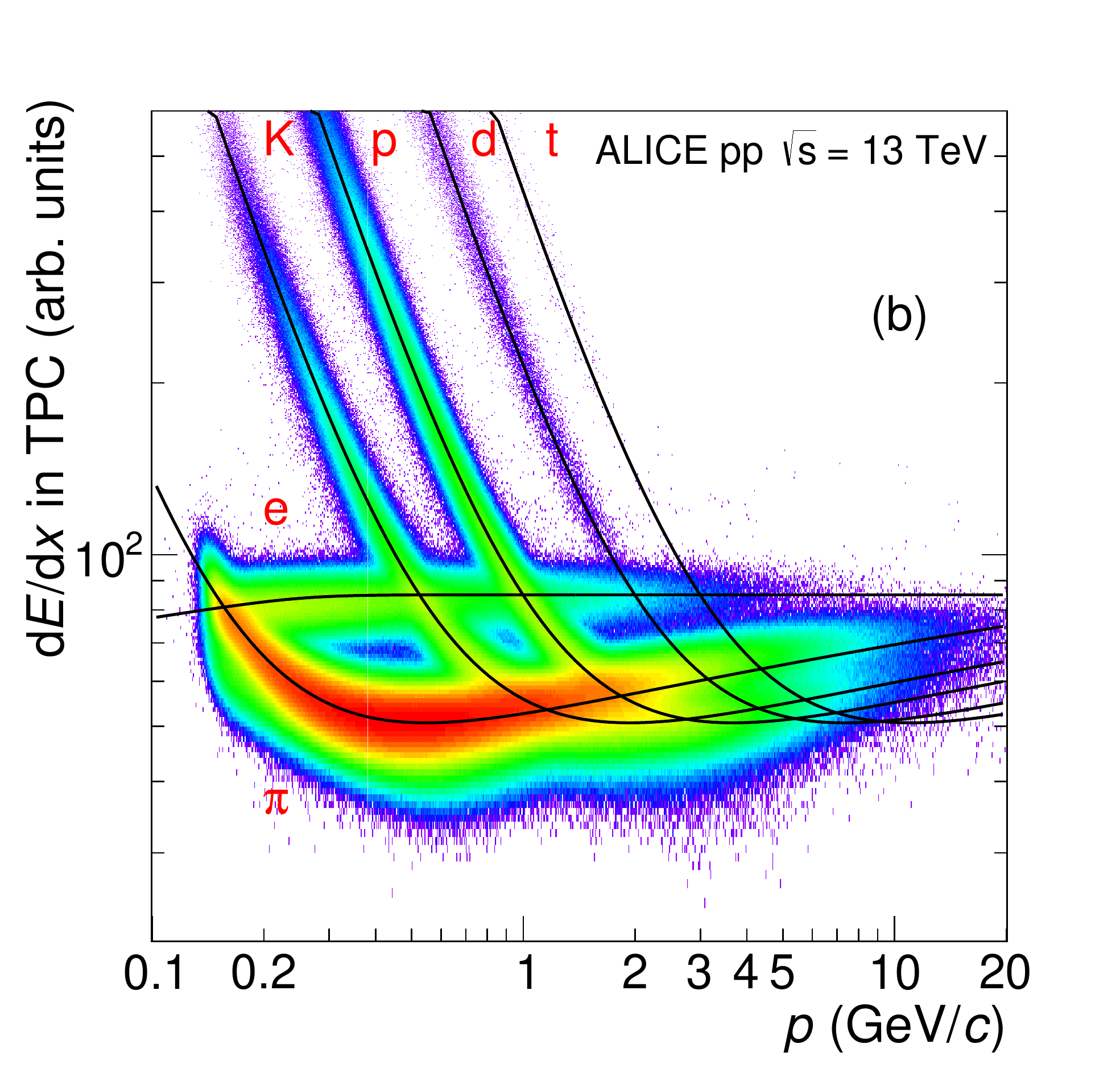}
	\includegraphics[keepaspectratio,width=0.49\columnwidth]{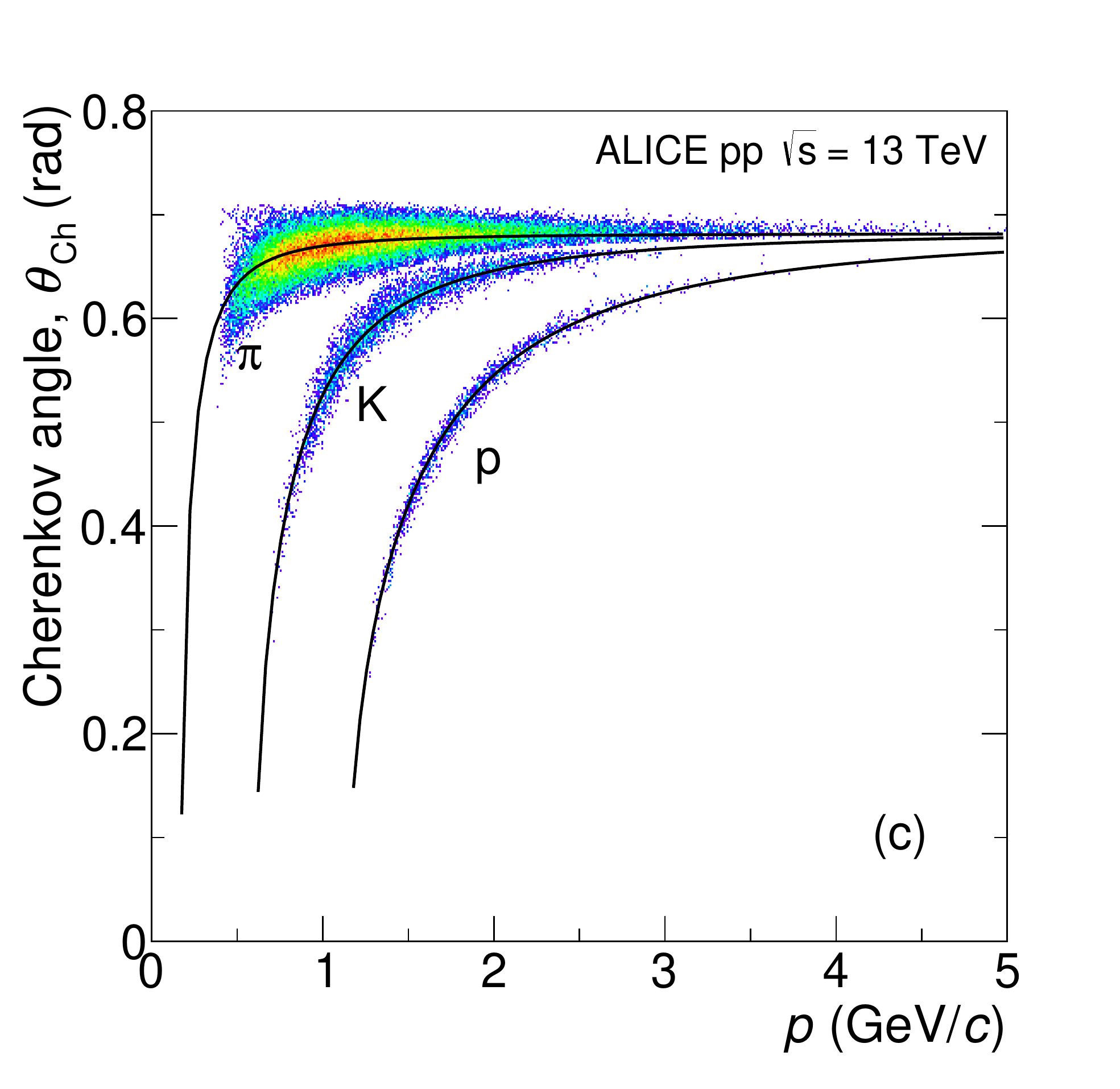}
	\includegraphics[keepaspectratio,width=0.49\columnwidth]{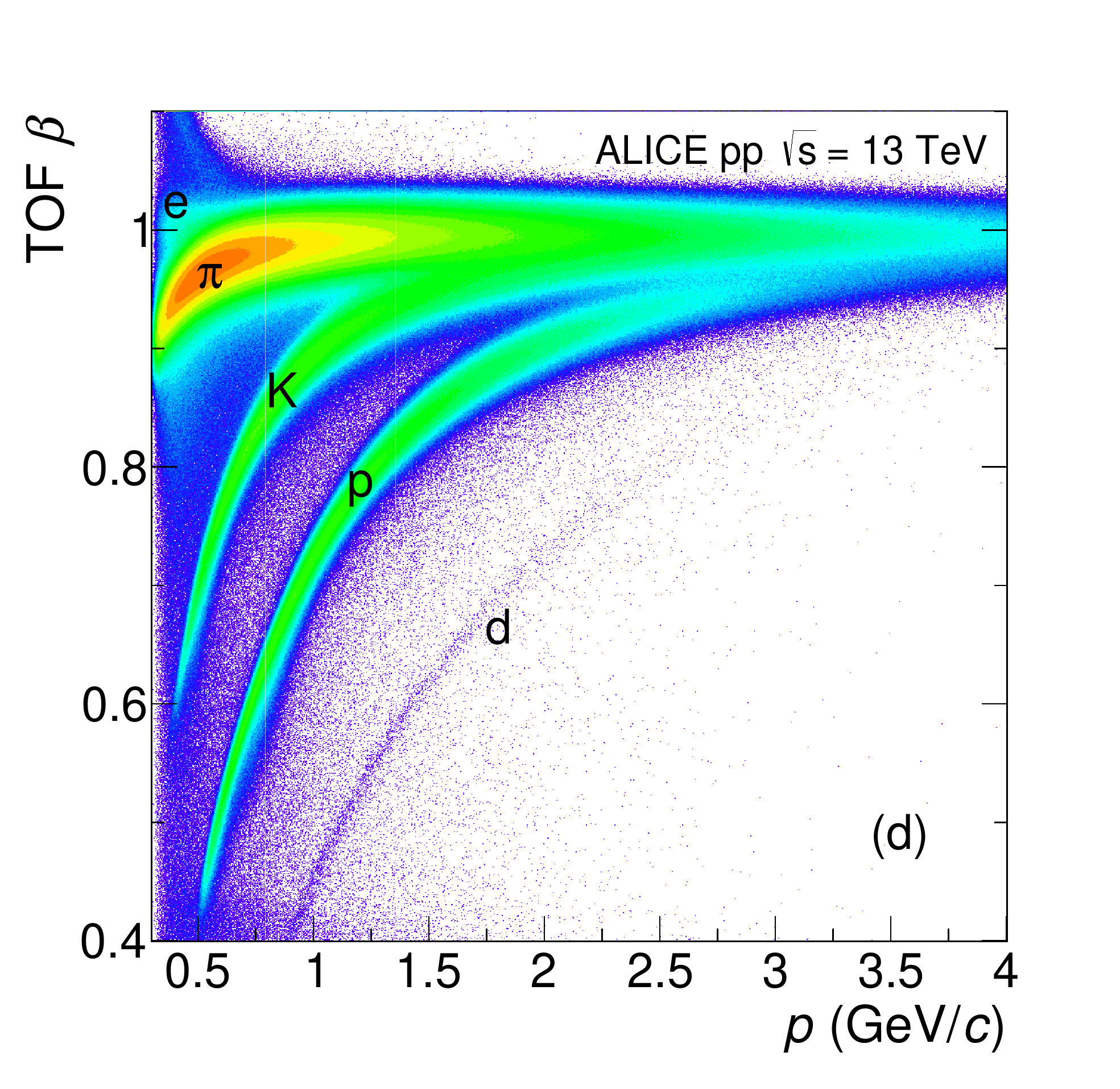}
      \caption[]{The performance during the LHC Run 2 period of the ALICE central barrel detectors used 
      for the measurements described in this paper.
      Panels a), b), c), and d) indicate the characteristic signal distributions of identified charged particles measured by 
      the ITS, TPC, TOF, and HMPID detector, respectively.
     }
      \label{fig:PidPerformance}
    \end{figure}
    \clearpage
    }

    \subsubsection{ITS stand-alone} \label{subsubsec:pidreco:pikp:itssa}

    In the ``ITS stand-alone'' analysis, both tracking and PID are performed based on information from the ITS detector only. For the present data sample, the contribution of tracks with wrongly 
    assigned clusters in the ITS is negligible due to the low pseudorapidity density of charged particles, $\dndetaCh_{|\eta|<0.5} = 5.31\pm0.18$, measured in the pseudorapidity region 
    $|\eta|<0.5$~\cite{Adam:2015pza}. The average \dedx signal in the four outer ITS layers used for PID is estimated by means of the truncated mean method~\cite{Aamodt:2011zj}. The measured \dedx 
    for the sample of ITS-sa tracks is shown in the top left panel of Fig.~\ref{fig:PidPerformance}, along with the Bethe--Bloch parametrization of the most probable values, which is the same as 
    the one used for the LHC Run 1 analyses~\cite{Adam:2015qaa}. Two identification strategies were used. In the main analysis, a unique identity is assigned to the ITS-sa track according to the 
    mass hypothesis for which the expected specific energy-loss value is the closest to the measured \dedx for a track with momentum $p$. The second analysis strategy uses the Bayesian PID 
    approach~\cite{Adam:2016acv}, based on likelihood parametrization with a set of iterative prior probabilities. The identification is based on the maximal probability method in which the species 
    with the highest probability is assigned to a track. For $\pt<\mevc{160}$, where the e$/\pi$ separation power in the ITS allows high-purity identification of electrons, four mass hypotheses (e/$\pi$/K/p) 
    are considered. For $\pt>\mevc{160}$, electrons and pions cannot be separated using their \dedx in the ITS detector, and the Bayesian approach is based on the $\pi$/K/p mass hypotheses only.
    
	For the ITS-sa analysis, the systematic uncertainties related to the PID procedure originate from the different techniques that are used (the truncated mean method and the Bayesian PID approach). 
	These range from about 1 to 5\% depending on particle species and \pt. The Lorentz force causes the migration of the cluster position in the ITS by driving the charge in opposite directions depending 
	on the polarity of the magnetic field of the experiment ($\mathbf{E}\times\mathbf{B}$ effect). The uncertainty related to this effect is estimated by analyzing data samples with opposite magnetic 
	field polarities, for which a difference at the level of ${\sim}2\%$ is observed.
    
    \subsubsection{TPC-TOF fits} \label{subsubsec:pidreco:pikp:tpctof}

    In the so-called ``TPC-TOF fits'' analysis, the distributions of the specific energy loss \dedx measured in the TPC and the velocity $\beta$ measured in the  TOF detector are fitted with functions 
    that describe the PID signals for different track momentum ($p$) intervals. The TPC provides a $3\stpc$ separation between pions and kaons up to $\pt{\sim}\mevc{600}$ and between kaons and protons 
    up to $\pt{\sim}\mevc{800}$. 
    
    Particle identification using this technique is possible in the \pt ranges $0.3<\pt<\gevc{0.5}$, $0.3<\pt<\gevc{0.6}$, and $0.4<\pt<\gevc{0.8}$ for \pix, \kx, and \px, respectively. The extraction 
    of the raw yield for a given species is done by integrating the $\mathrm{d}^{2}N/\mathrm{d}\pt\mathrm{d}n_{\sigma}$ distribution in these \pt intervals. In addition, in the transverse momentum ranges 
    $\pt < \gevc{0.4}$ for \pix, $\pt > \gevc{0.45}$ for \kx, and $\pt>\gevc{0.6}$ for \px, a Gaussian fit is used to remove the background contribution ($\text{e}^{\pm}$ for pions, \pix for kaons, and 
    $\kx+\text{e}^{\pm}$ for (anti)protons). The background contribution is small ($<1\%$) in all cases, except for \kx at $\pt>\gevc{0.55}$, where it reaches $\sim 13\%$. The TOF analysis uses the 
    sub-sample of global tracks for which the time-of-flight measurement is available. The procedure is performed in narrow regions of pseudorapidity, $|\eta|<0.2$ and $0.2<|\eta|<0.4$, 
    in order to achieve a sufficient level of separation and to strengthen the correlation between the total momentum and the transverse momentum. The TOF matching efficiency for the presented data sample 
    in the pseudorapidity region $|\eta|<0.2$ ($0.2<|\eta|<0.4$) increases rapidly with \pt up to around $50\%$ $(60\%)$ for pions at $\pt{\sim}\mevc{700}$, $45\%$ $(55\%)$ for kaons at $\pt{\sim}\gevc{1}$, 
    and $50\%$ $(65\%)$ for protons at $\pt{\sim}\mevc{800}$; it saturates at higher momenta~\cite{Akindinov:2013tea}. The PID performance of the TOF detector is shown in the bottom right panel of Fig.~\ref{fig:PidPerformance}, 
    where the velocity $\beta$ of the particles is reported as a function of momentum $p$. The raw particle yields are then obtained by fitting the measured $\beta$ distributions with a TOF response function, 
    where the contribution for a given species is centered around $\beta = p/E$. An additional template is added to account for wrongly associated (mismatched) hits in the TOF detector.
    
    The TPC-TOF fits analysis includes uncertainties related to the PID procedure from several sources. For the TPC part, uncertainties are estimated by integrating the ${\rm d}^{2} N/ {\rm d}p_{\rm T}{\rm d}n_{\sigma}$ 
    distribution for each particle species in $|n_{\sigma}|<2.5$ and $|n_{\sigma}|<3.5$, and comparing to the integral of the underlying distribution performed within $|n_{\sigma}|<3$; the larger resulting uncertainty 
    is used. Similarly, for TOF fits for $\pt<\gevc{1.5}$, the uncertainty related to PID is estimated by integrating the ${\rm d}^{2} N/ {\rm d}p_{\rm T}{\rm d}n_{\sigma}$ in the range of $|n_{\sigma}|<3$. 
    This results in uncertainties of up to $1\%$ for \pix, $5\%$ for \kx, and $2\%$ for (anti)protons. At higher \pt, where the separation between particle species becomes small, a better estimate of the 
    uncertainty of the method can be achieved by varying simultaneously the resolution \stof and the tail parameter~\cite{Adam:2015qaa,Akindinov:2013tea} of the fit function used to describe the PID signal 
    around their nominal values. An additional uncertainty is included to account for the TOF miscalibration, which becomes significant for $\pt>\gevc{1.5}$. This results in uncertainties of up to $8\%$ 
    for \pix, $14\%$ for \kx, and $7\%$ for (anti)protons. An uncertainty related to the matching of tracks to TOF hits arises from differences in the TPC-TOF matching efficiency in real data and simulations. 
    This uncertainty is $3\%$ for \pix, $6\%$ for \kx, and $4\%$ for (anti)protons, independent of \pt.
	
    \subsubsection{HMPID} \label{subsubsec:pidreco:pikp:hmpid}

    The HMPID analysis extends charged hadron identification into the intermediate-\pt region ($\gevc{2}\lesssim\pt\lesssim\gevc{10}$), combining the measurement of the emission angle of the Cherenkov photons 
    $\theta_{\rm Ch}$ and the momentum information of the particle under study. The Cherenkov photons are selected using the Hough Transform Method~\cite{DiBari:2003wy}. The measurement of the single photon 
    $\theta_{\rm Ch}$ angle in the HMPID requires the determination of the track parameters, which are calculated for tracks propagated from the central tracking detectors to the radiator volume where the 
    Cherenkov photons are emitted. Each track is extrapolated to the HMPID cathode plane and matched to the closest primary ionization (MIP) cluster. The distance within the cathode plane between the 
    extrapolated track and the matched cluster (denoted $d_{\rm{MIP-track}}$) is restricted to be less than 5~cm to reduce false matches. A mean Cherenkov angle $\left<\theta_{\rm Ch}\right>$, computed as 
    the weighted average of single photon angles, is associated to each particle track. The PID performance is shown in the bottom left panel of Fig.~\ref{fig:PidPerformance}, where the correlation between 
    the reconstructed Cherenkov angle and the track momentum is shown, indicating good agreement with the theoretically expected values. For yield extraction, a statistical unfolding technique is applied 
    by fitting the reconstructed Cherenkov angle distribution in a given momentum interval, which requires a precise knowledge of the detector response function. Yields are evaluated from the integral of 
    each of the three Gaussian functions, corresponding to the signals from pions, kaons, and protons. The HMPID allows pion and kaon identification in the momentum range $1.5 < p < \gevc{4}$, while 
    (anti)protons can be distinguished from pions and kaons up to $p=\gevc{6}$, with separation powers larger than $2\sigma$.
    
	Additionally, in the HMPID analysis, a data-driven correction for the selection criterion on the distance $d_{\rm{MIP-track}}$ has been evaluated by taking the ratio between the number of tracks that 
	pass the selection criterion on $d_{\rm{MIP-track}}$ and all the tracks in the detector acceptance. This matching efficiency correction is \pt dependent and it is about $20-40\%$; it is lower for particles 
	with velocity $\beta{\sim}1$~\cite{Adam:2015kca}. Negatively charged particle tracks have a distance correction ${\sim}3\%$ lower than positive ones due to a radial residual misalignment of the HMPID 
	chambers and an imperfect estimation of the energy loss in the material traversed by the track.
	
	In the HMPID analysis, the systematic uncertainty has contributions from tracking, PID, and track association~\cite{Adam:2015qaa,Adam:2015kca}. The PID uncertainties are estimated by varying the 
	parameters of the fit function used to extract the raw particle yield. This uncertainty is \pt-dependent and increases with \pt to a maximum value of $12\%$ for \pix and \kx, and $11\%$ for (anti)protons.
	Furthermore, the uncertainty of the association of the global track to the charged particle signal in the HMPID is obtained by varying the default value of the $d_{\rm{MIP-track}}$ distance criterion 
	required for the matching. The resulting uncertainty is \pt-dependent with a maximum value of about 4\% for (anti)protons at $\pt=\gevc{1.5}$.

    \subsubsection{TPC relativistic rise} \label{subsubsec:pidreco:pikp:tpcrr}

    In the TPC \dedx relativistic rise analysis charged pions, charged kaons, and (anti)protons can be identified up to $\pt=\gevc{20}$. The identification is achieved by measuring the specific energy loss 
    \dedx in the TPC in the relativistic rise regime of the Bethe--Bloch curve. The \dedx as a function of momentum $p$ is shown in the top right panel of Fig.~\ref{fig:PidPerformance}, indicating the \mdedx 
    response for charge-summed \pix, \kx, \px, and $e^{\pm}$. The separation power between particle species is about $4.5\stpc$ ($1.5\stpc$) for $\pi-\text{p}$ ($\text{K}-\text{p}$) at $\pt = \gevc{10}$, 
    and it is nearly constant at similar values for larger momenta. The results presented in this paper were obtained using the method detailed in Ref.~\cite{Adam:2015kca}. As discussed in 
    Refs.~\cite{Abelev:2014laa,Adam:2015kca,Adam:2015qaa,Adam:2016dau}, \dedx is calibrated taking into account chamber gain variations, track curvature and diffusion to obtain the best possible overall 
    performance, which results in a response that essentially only depends on $\beta\gamma$ ($=p/m$). The resolution is better at larger rapidities for the same \mdedx because of the longer integrated 
    track lengths. Hence, to analyze homogeneous samples, the analysis is performed in four equal-width intervals within $|\eta|<0.8$ ($|\eta|<0.2$, $0.2<|\eta|<0.4$, $0.4<|\eta|<0.6$ and $0.6<|\eta|<0.8$). 
    Samples of topologically identified pions from \kzs decays, protons from \lap decays, and electrons from photon conversions were used to parameterize the Bethe--Bloch response $\mdedx$ as a function of 
    $\beta\gamma$ and the relative resolution $\stpc/\mdedx$ as a function of \mdedx. The relative yields of pions, kaons, protons, and electrons are obtained as the \chpi, \chk, \chp, and \che yields 
    normalized to that for inclusive charged particles. They are obtained using four-Gaussian fits to \dedx distributions differentially in $p$ and $|\eta|$ intervals. The parameters (mean and width) of 
    the fits are fixed using the parameterized Bethe--Bloch and resolution curves. The relative yields as a function of \pt are found to be independent of $\eta$ and therefore averaged. Particle yields 
    are constructed using the corrected relative yields and the corrected charged particle yields~\cite{Adam:2015pza}. A Jacobian correction is applied to account for the pseudorapidity-to-rapidity conversion.
	
	In the TPC \dedx relativistic rise analysis, the pion and (anti)proton yields are corrected for secondary particles from weak decays using MC simulations for the relative fraction of secondaries. 
	The obtained fraction of secondary pions and (anti)protons are scaled to those extracted from $\text{DCA}_{xy}$ template fits to data. For $\pt\gtrsim\gevc{3}$, the correction is negligible for pions. 
	It is $\sim2\%$ for (anti)protons at $\pt=\gevc{3}$, decreases to $\sim1\%$ at $\pt=\gevc{10}$, and stays constant from that \pt onward. Moreover, at high \pt, there is a small contamination of primary 
	muons in the pion yields. Due to the similar muon and pion masses, the electron (fractional) yield is subtracted from the pion yield to correct for the muon contamination. This procedure gives a $<1\%$ 
	correction to the pion yield in the entire \pt range considered in the analysis. Furthermore, above $\pt=\gevc{3}$, both the contamination of kaons and the contamination of (anti)deuterons in the 
	(anti)proton sample are negligible.
    
    The tracking efficiency component is calculated as a relative correction factor. It is the ratio of the inclusive to identified charged particle efficiencies, and is applied to the relative yields. 
    At high \pt this correction is nearly constant, of the order of $3-6\%$, depending on the particle species.
	
	In the TPC \dedx relativistic rise analysis, the systematic uncertainties mainly originate from event and track selection and the PID procedure. The first component is based on the study of inclusive 
	charged particles~\cite{Adam:2015pza}, and it was recalculated to meet the event selection condition for inelastic events. Its value is estimated to be $7.3\%$ at high \pt (at which the value 
	is largest). It is the main contribution for pions. The second component was measured following the procedure explained in Ref.~\cite{Adam:2015kca}. Here, the largest contribution is due to the 
	uncertainties in the parameterization of the \dedx response, resulting in uncertainties of $6.5-15.0\%$ for \kx and $13.0-17.0\%$ for (anti)protons, depending on \pt.
	
    \subsubsection{Topological reconstruction of \kx kink decays} \label{subsubsec:pidreco:pikp:kinks}

\begin{figure}[t]
      \centering
	\mysidecaption{0.5}{%
	  \captionof{figure}{Kink invariant mass distribution for charge summed particles in 
	  the mother particle transverse momentum interval $0.2\,\text{GeV}/c < \pt < 6\,\text{GeV}/c$, before (open circle)
	    and after (full circle) topological selections.
	    \label{fig:InvMass_Kinks}
	    }%
            }{\includegraphics[keepaspectratio,width=\linewidth]{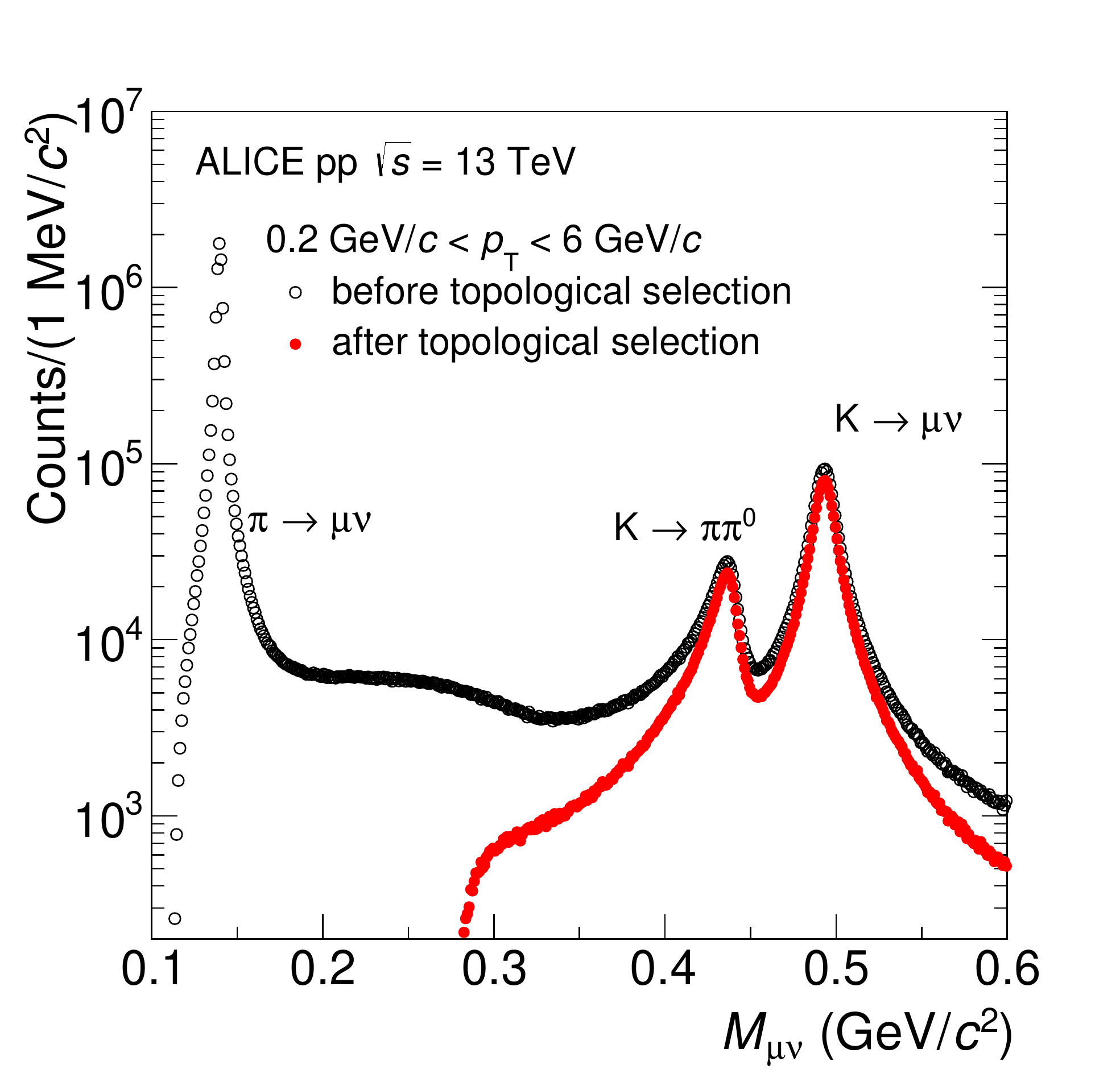}}[c]%
    \end{figure}

	Charged kaons are also measured by reconstructing the vertex of their weak decay in the TPC. The procedure extends the \pt reach of the identification of charged kaons on a track-by-track basis from 
	\gevc{4} (available with the HMPID) up to \gevc{7}. This method exploits the characteristic kink topology defined by the decay of a charged mother particle to a daughter with the same charge and a neutral 
	daughter~\cite{Adam:2015qaa}. Thanks to the two-body kinematics of the kink topology, it is possible to separate kaon decays from the background mainly caused by pion decays. For this purpose, a 
	topological selection is applied by imposing a selection criterion on the daughter track's momentum with respect to that of the mother track, and on the decay angle, defined as the angle between the 
	momenta of the mother and the charged daughter track. Furthermore, mother tracks are selected inside a $3.5\stpc$ band of the expected \dedx for kaons to enhance the purity of the sample. With the 
	assumption that the charged daughter track is a muon and the undetected neutral daughter particle is a neutrino, the reconstructed invariant mass $M_{\mu\nu}$ is calculated and is shown in 
	Fig.~\ref{fig:InvMass_Kinks}. The raw yield of the topologically selected kaons in a given \pt bin are obtained from the integral of the invariant mass distribution after the topological selection 
	criteria. The contamination due to fake kinks increases with \pt and saturates at $\pt{\sim}\gevc{1}$, reaching a maximum value of about 5\%.
    
    For the topological identification of charged kaons, the size of the correction related to contamination (arising from background or misidentification) was assigned as a \pt-dependent uncertainty 
    on the purity. The kink identification uncertainty is \pt dependent and it ranges from 2.5\% at low \pt to 2.2\% at high \pt. The systematic uncertainty on the efficiency for findable kink vertices 
    was estimated to be $3\%$, independent of \pt. The uncertainty due to contamination from fake kinks was estimated to be at most about 5\% around $\pt=\gevc{2}$, decreasing towards higher \pt.

    \subsubsection{Combination of \pix, \kx, and \px spectra from different analyses} \label{subsec:ptCombined}

\begin{figure}[t]
	\centering
	\includegraphics[keepaspectratio, width=0.49\columnwidth]{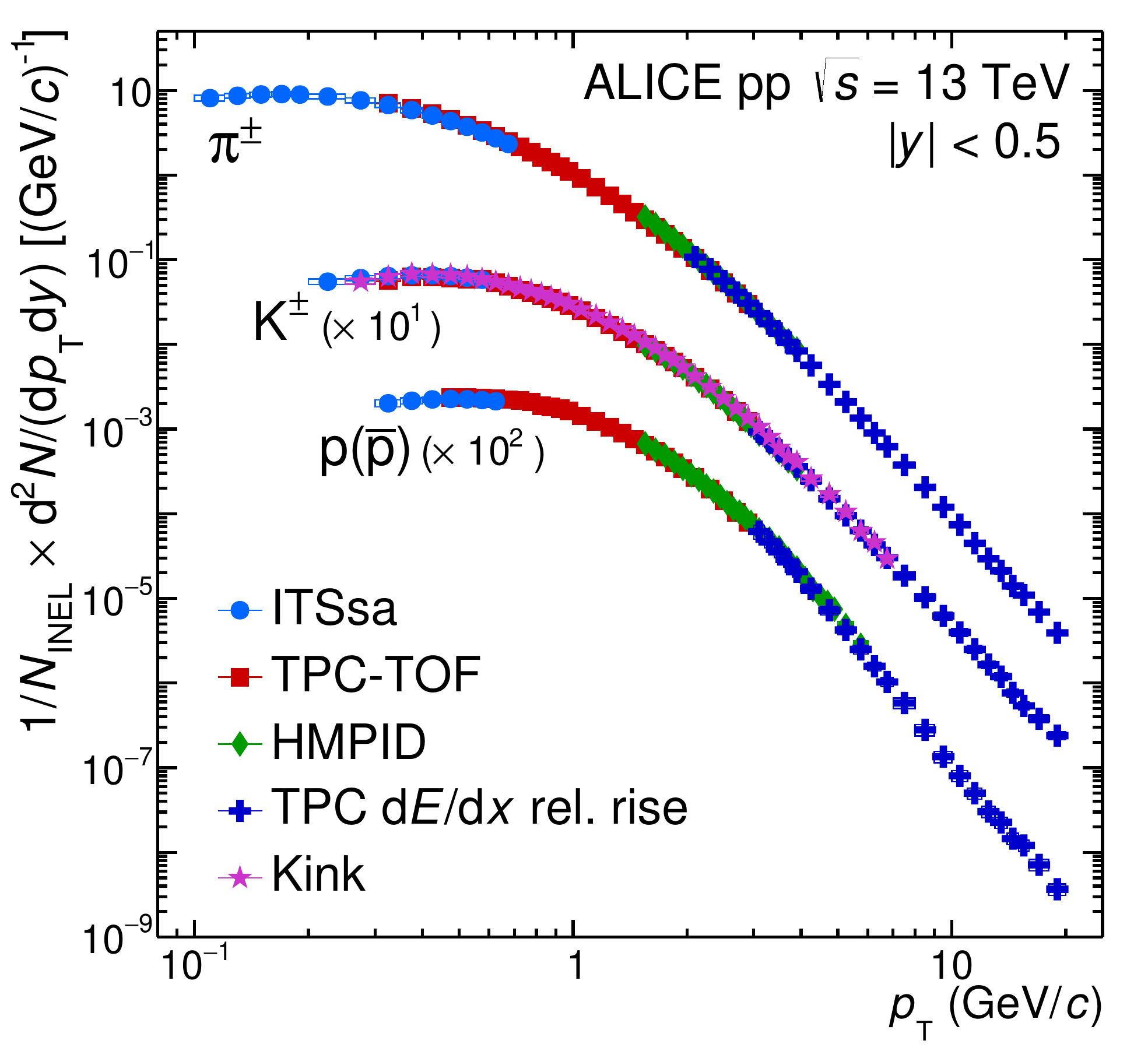}
	\includegraphics[keepaspectratio, width=0.49\columnwidth]{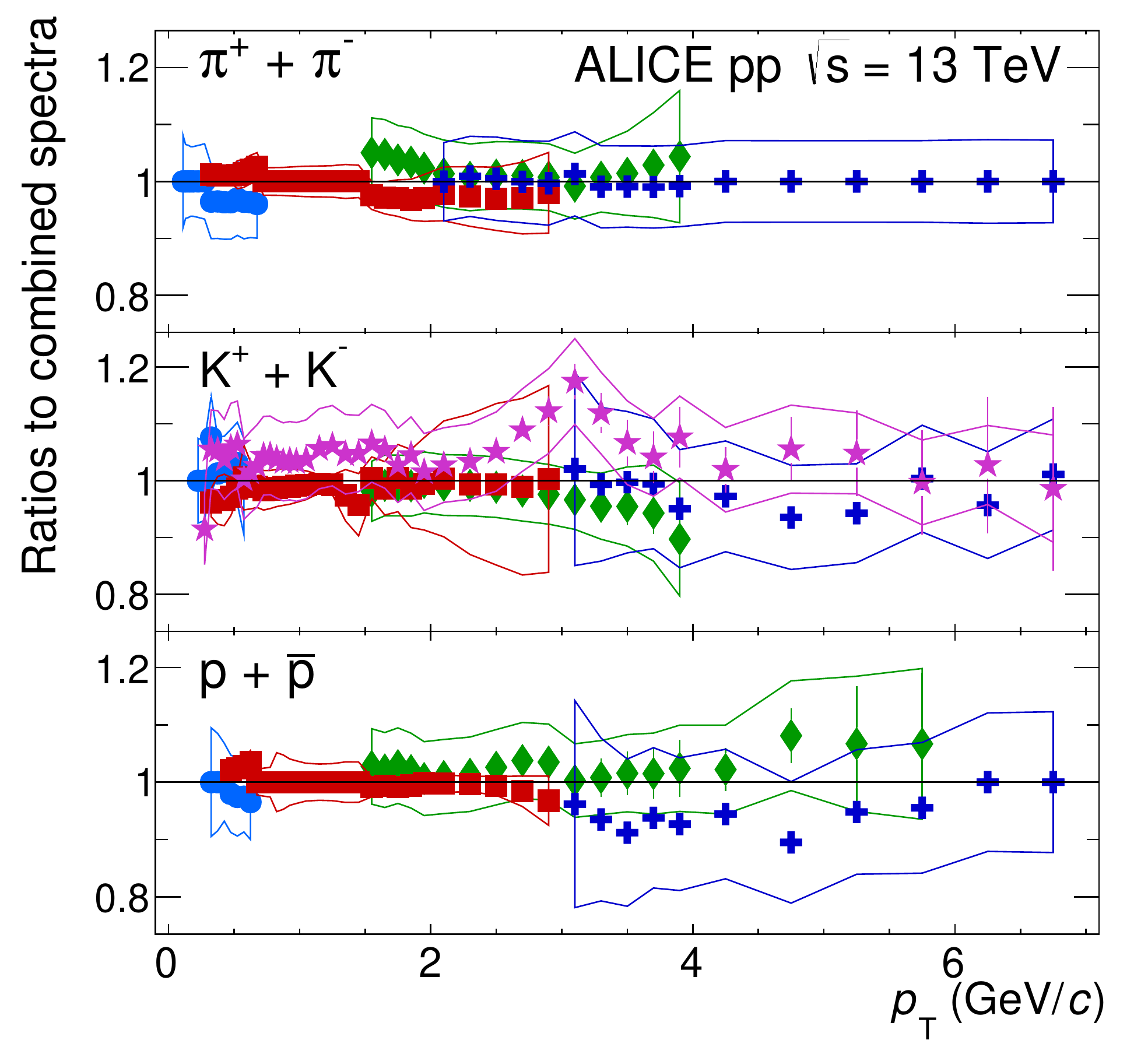}
	\caption{Left panel: \pt spectra of \pix, \kx, and \px measured at midrapidity ($|y|<0.5$) in \pp collisions at \sppt{13}
	using different PID techniques. The spectra are normalized to the number of inelastic collisions. 
	Right panel: The ratios of individual spectra to the combined spectra as a function of \pt for \pix (top), \kx (middle), and \px (bottom). 
	Only the \pt-range where the analyses overlap is shown. The vertical bars indicate statistical uncertainties while the bands show the 
	uncorrelated systematic uncertainties.
	}
	\label{fig:spectra_piKp_pp13_Combined}
	\end{figure}

    The charged pion, charged kaon and (anti)proton transverse momentum spectra were measured via several independent analyses as described in the preceding sections. To ensure the maximal \pt coverage, 
    the final \pt spectra were calculated as the average of all analyses weighted by the systematic uncertainties that are not shared between analyses, i.e. uncorrelated. The uncertainties related to the 
    ITS-TPC matching efficiency and the global tracking efficiency are largely correlated and were summed in quadrature with the uncorrelated part of the systematic uncertainties obtained after the averaging. 
    Only the TPC relativistic rise analysis is used above $\pt = 4, 7$, and \gevc{6} respectively for pions, kaons, and protons. To verify the validity of the procedure, the spectra obtained from the 
    individual analyses were compared to the final combined ones. The left panel of Fig.~\ref{fig:spectra_piKp_pp13_Combined} shows the \pix, \kx, and \px spectra obtained from the five analyses discussed 
    above, which are normalized to the number of inelastic collisions ($N_{\rm INEL}$). The right panel of Fig.~\ref{fig:spectra_piKp_pp13_Combined} shows the ratios of the individual spectra to the 
    combined spectra, which illustrates an excellent agreement in the overlapping \pt regions for every particle species within the uncorrelated part of the systematic uncertainties.    

	\subsection{Topological identification of weakly-decaying strange hadrons} \label{subsubsec:analyses:pidreco:weakdecaystrange}

    Primary strange hadrons \kzs, \lap, \lam, \xim, \xip, \omm, and \omp are reconstructed at midrapidity ($|y|<$0.5) via their characteristic weak decay topologies in the channels presented in Tab.~\ref{tab:analyses:reco}. 
    Single-strange hadrons \kzs, \lap and \lam decay into two oppositely charged daughter particles (\VZ decay). Multi-strange hadrons ($\Xi$ and $\Omega$) decay into a charged meson (bachelor) plus a \VZ decaying particle, 
    giving the two-step process known as a cascade. The identification methods for the \VZ (\kzs and \lap) and cascade-like ($\Xi$ and $\Omega$) candidates strictly follow those presented in earlier works~\cite{Aamodt:2011zza,Abelev:2012jp}. 
    An additional selection criterion on (anti)proton momentum ($p>0.3\gevc{1}$) measured at the inner wall of TPC was introduced because of the observed instability of track reconstruction for the lowest \pt bin of \lap\ 
    and \lam spectra. Several track, PID, and topological selection criteria are applied in order to find \VZ and cascade decay candidates. Charged tracks are selected using the standard criteria described in Sec.~\ref{subsubsec:tracksel}. 
    The identity of these daughter tracks is established with the requirement that the specific energy loss \dedx measured in the TPC is compatible with the expected mass hypothesis within $5\stpc$ ($4\stpc$) for the analysis 
    of \kzs and \lap ($\Xi$ and $\Omega$). These identified tracks are then combined to form invariant mass distributions, and fake combinations are reduced by applying selection criteria on topological variables.
    
	\begin{table}[t]
	\begin{center}
	\caption{Selection criteria for secondary and bachelor tracks as well as for \VZ and cascade candidates applied in the presented work.}	
	
\begin{threeparttable}
\begin{tabular}{l l l l l l}
\hline \\ [-10pt]
  Hadron species &&	\multicolumn{1}{c}{\kzs} & \multicolumn{1}{c}{\lap}  &\multicolumn{1}{c}{$\Xi$} & \multicolumn{1}{c}{$\Omega$} \\ [1pt]
\hline \\ [-10pt]
  \multicolumn{2}{l}{\textbf{Secondary track selections}} 		  	&			   &				&			 &				\\ [5pt]
  \multicolumn{2}{l}{Pseudorapidity range $|\eta|$}  					  	& $<$ 0.8		   & $<$ 0.8			& $<$ 0.8		 & $<$ 0.8			\\
  \multicolumn{2}{l}{DCA$_{xy}$ of \VZ\ daughter track} 			  	& 			   &				&			 & 				\\
  \multicolumn{2}{l}{to primary vertex (cm)}			  		& $>$ 0.06		   & $>$ 0.06 			& $>$ 0.04		 & $>$ 0.03 			\\
  \multicolumn{2}{l}{DCA$_{xy}$ of bachelor track} 				& 			   &				& 			 &				\\
  \multicolumn{2}{l}{to primary vertex (cm)}			  		& \multicolumn{1}{c}{---}			   & 	\multicolumn{1}{c}{---}			&  $>$ 0.05		 & $>$ 0.05			\\ [5pt]
  \multicolumn{2}{l}{TPC \dedx PID selection ($n_\sigma$)}			  	& $<$ 5 		   & $<$ 5  			& $<$ 4			 & $<$ 4			\\ [5pt]
\hline \\ [-10pt]
  \multicolumn{2}{l}{\textbf{\VZ\ selections}} 				  	&			   &				&			 &				\\ [5pt]
   \multicolumn{2}{l}{Rapidity range $|y|$} 	 					  	& $<$ 0.5 		   & $<$ 0.5			& \multicolumn{1}{c}{---} & \multicolumn{1}{c}{---}	\\
  \multicolumn{2}{l}{Transverse decay radius (cm)}  			  	& $>$ 0.5 		   & $>$ 0.5 			& $>$ 1.4 		 & $>$ 1.4			\\
  \multicolumn{2}{l}{DCA$_{xy}$ of \VZ\ to primary vertex (cm)}		  	& \multicolumn{1}{c}{---}   & \multicolumn{1}{c}{---}	& $>$ 0.07		 & $>$ 0.07			\\
  DCA between \VZ\ daughter tracks ($n_\sigma$)				  	& & $<$ 1		   	   & $<$ 1			& $<$ 1.5		 & $<$ 1.5			\\
  \multicolumn{2}{l}{Cosine of Pointing Angles}			 	& $>$ 0.97	 	   & $>$ 0.995  		& $>$ 0.97 		 & $>$ 0.97			\\
  \multicolumn{2}{l}{$|\Delta m|$ around nominal \lap\ mass (MeV/$c^{2}$)} 	& \multicolumn{1}{c}{---}   & \multicolumn{1}{c}{---} 	& $<$ 6			 & $<$ 6			\\  [5pt]
\hline \\ [-10pt]
  \multicolumn{2}{l}{\textbf{Cascade selections}} 			  	&			   &				& 			 &				\\  [5pt] 
  \multicolumn{2}{l}{Rapidity range $|y|$} 					& \multicolumn{1}{c}{---}  & \multicolumn{1}{c}{---}	& $<$ 0.5 		 & $<$ 0.5 			\\
  \multicolumn{2}{l}{Transverse decay radius (cm)} 			  	& \multicolumn{1}{c}{---}  & \multicolumn{1}{c}{---}	&  $>$ 0.8		 & $>$ 0.6			\\
  \multicolumn{2}{l}{DCA between \VZ\ and bachelor track (cm)}		  	& \multicolumn{1}{c}{---}  & \multicolumn{1}{c}{---}	&  $<$ 1.6		 & $<$ 1			\\
  \multicolumn{2}{l}{Cosine of Pointing Angle} 						 	& \multicolumn{1}{c}{---}  & \multicolumn{1}{c}{---}	& $>$  0.97		 & $>$  0.97			\\
  \multicolumn{2}{l}{$|\Delta m|$ around nominal $\Xi$\ mass (MeV/$c^{2}$)}	& \multicolumn{1}{c}{---}  & \multicolumn{1}{c}{---}	& $<$ 8 		 & $>$  8			\\  [5pt]
\hline 
\end{tabular}
  \end{threeparttable}
\label{Tab:V0AndCascadeselection}

	\label{tab:V0AndCascadeselection}
	\end{center}
	\end{table}

    Values for these selection criteria are summarized in Tab.~\ref{tab:V0AndCascadeselection} and a detailed description can be found in Ref.~\cite{Aamodt:2011zza}. \kzs (\lap) candidates compatible with the alternative \VZ hypothesis, 
    obtained by changing the mass assumption for the daughter tracks accordingly, are rejected if they lie within a fiducial window around the nominal \lap (\kzs) mass. A similar selection is applied for the $\Omega$ analysis where 
    candidates are rejected if the corresponding invariant mass, obtained by assuming the pion mass hypothesis for the bachelor track, is compatible within $\pm 8\ \mathrm{MeV}/c^2$ with the nominal $\Xi$ mass. A selection is also made 
    on the proper lifetime $c\tau=mL/p$, where $m$ is the particle mass and $L$ is the distance from the primary vertex to the decay vertex; $c\tau$ is required to be less than $20$~cm/$c$. A further selection is applied on the pointing 
    angle $\Theta$, the angle between the strange hadron's momentum vector and the position vector of its decay point with respect to the primary collision vertex. With requirements of $\cos\Theta>0.97$ for \kzs and $\cos\Theta>0.995$ 
    for \lap and \lam, about 1\% of secondary \lap and \lam generated in the detector material is removed.

    The particle yields are obtained as a function of \pt by extracting the signals from the relevant invariant mass distributions. Examples of the invariant mass peaks at \sppt{13} are shown in Fig.~\ref{fig:InvMass}; the distributions 
    of \kzs and \lap are very similar to those at \sppt{7}. The mean ($\mu$) and the width ($\sigma$) values of the distributions are found by fitting the distribution with a Gaussian for the signal plus a linear function describing the 
    background. The extracted $\mu$ values of the distributions both for {\VZ}s and cascades are in good agreement with the accepted values~\cite{PDG} and are well reproduced by MC simulations at $\sqrt{s}=7$ and \sppt{13} in all measured 
    \pt bins. The widths of the distributions evolve with \pt at $\sqrt{s}=13$ (7)~TeV by about 7 (14)~MeV/$c^{2}$ for \kzs and 2 (4)~MeV/$c^{2}$ for \lap and \lam, which agrees with MC simulations within $15-20\%$ at both reported energies.
    For the {\VZ}s (cascades) a region containing all the signal (signal region) is defined around the mean within $\pm~6\sigma$ ($3\sigma$), while a region to estimate the background (background region) is defined as side-bands from 
    -12$\sigma$ to -6$\sigma$ (-12$\sigma$ to -6$\sigma$) and from 6$\sigma$ to 12$\sigma$ (6$\sigma$ to 19$\sigma$). Given the flatness of the invariant mass distribution in the background region, the estimate of the background in the signal 
    region is obtained rescaling, by the ratio of the widths in the two regions, the sum of the entries of all the bins in the background region. An alternative method, used to estimate a possible systematic uncertainty, uses the fit of the invariant mass distribution in the background region to estimate the background contribution inside the peak region. In both cases, the signal is obtained subtracting the estimated background in the peak region from the integrated 
    counts in the peak region.
	
      \afterpage{%
    \begin{figure}[p]
      \centering
	\includegraphics[keepaspectratio,width=0.495\columnwidth]{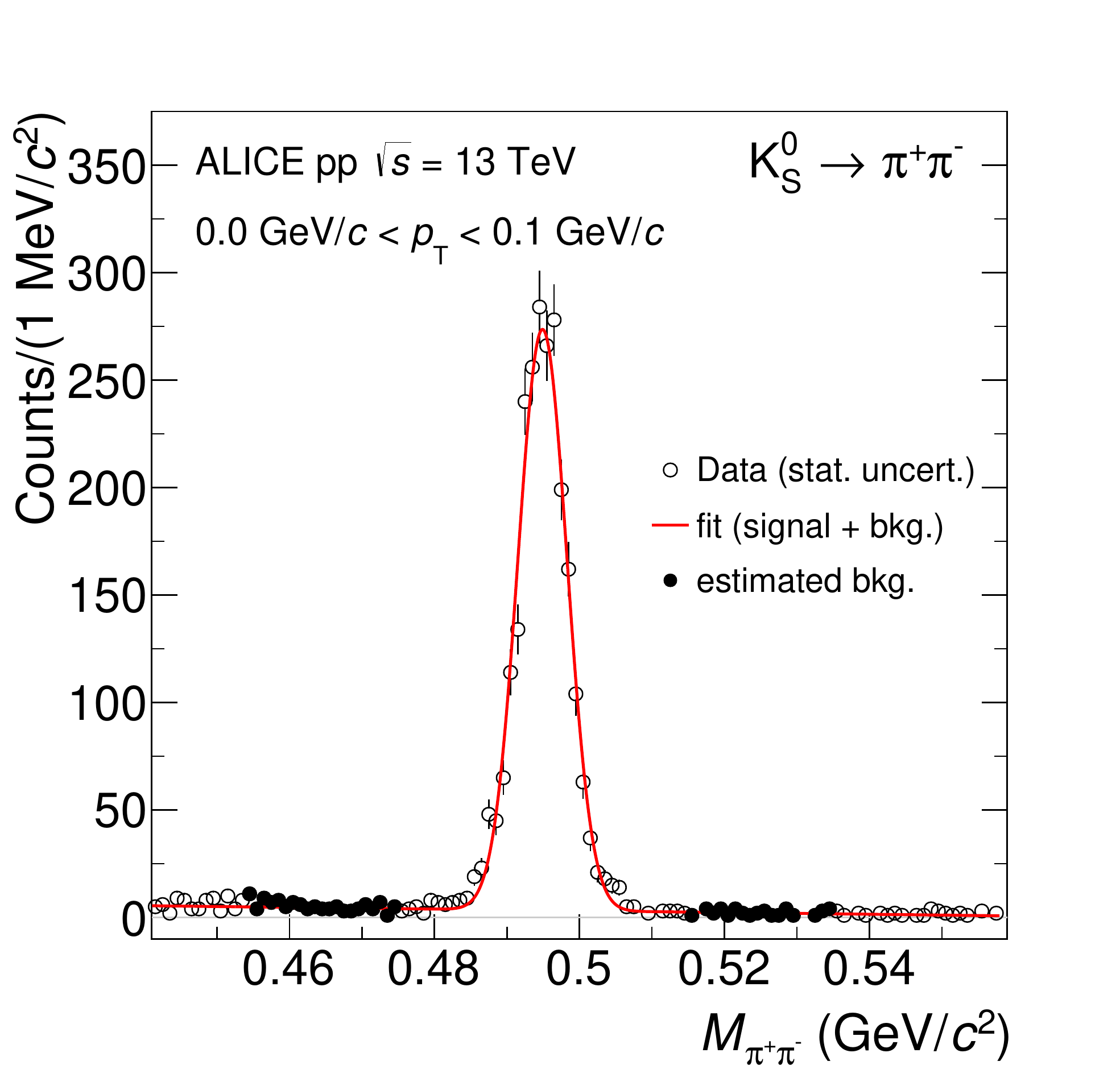}
	\includegraphics[keepaspectratio,width=0.495\columnwidth]{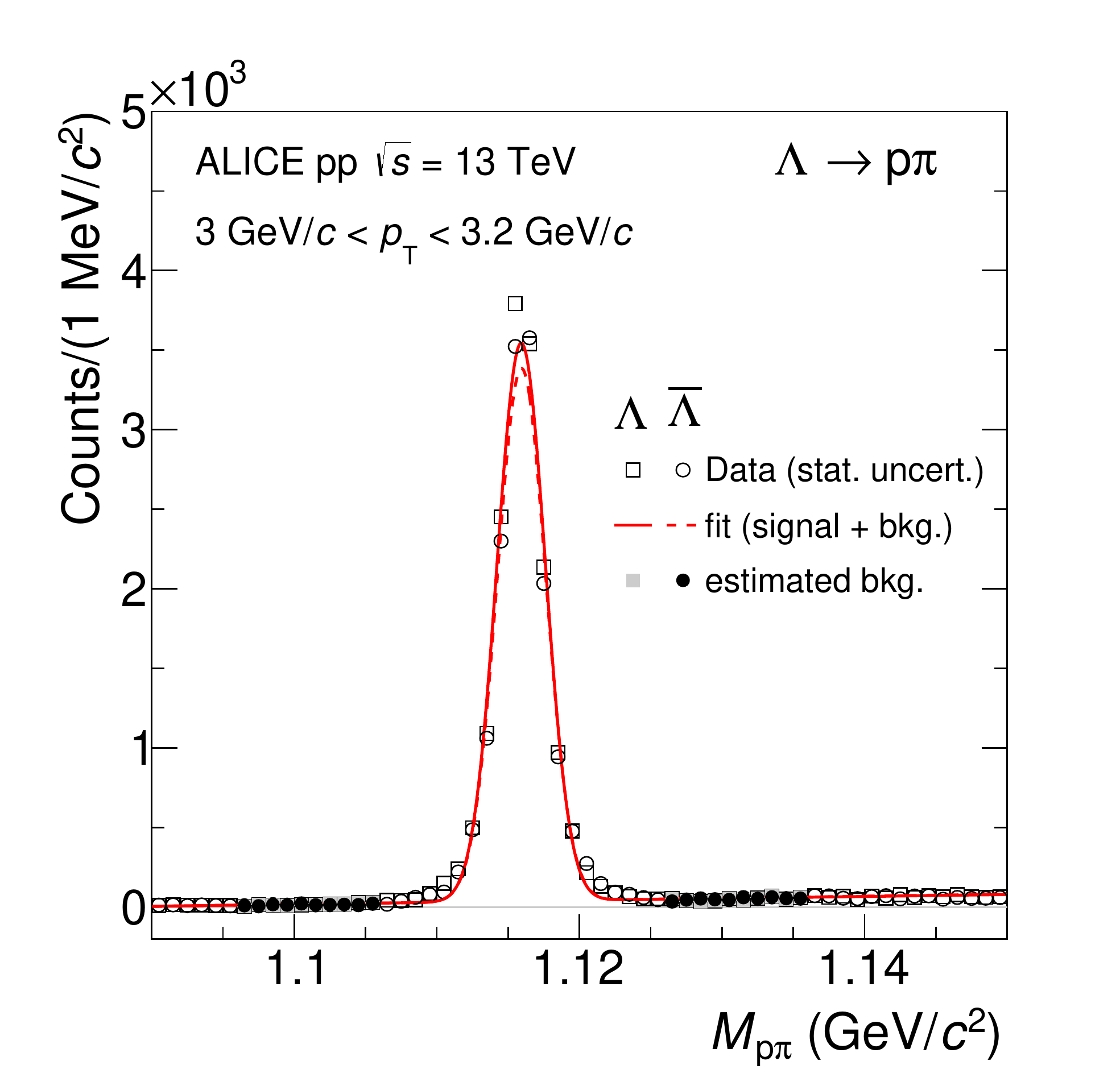}
	\includegraphics[keepaspectratio,width=0.495\columnwidth]{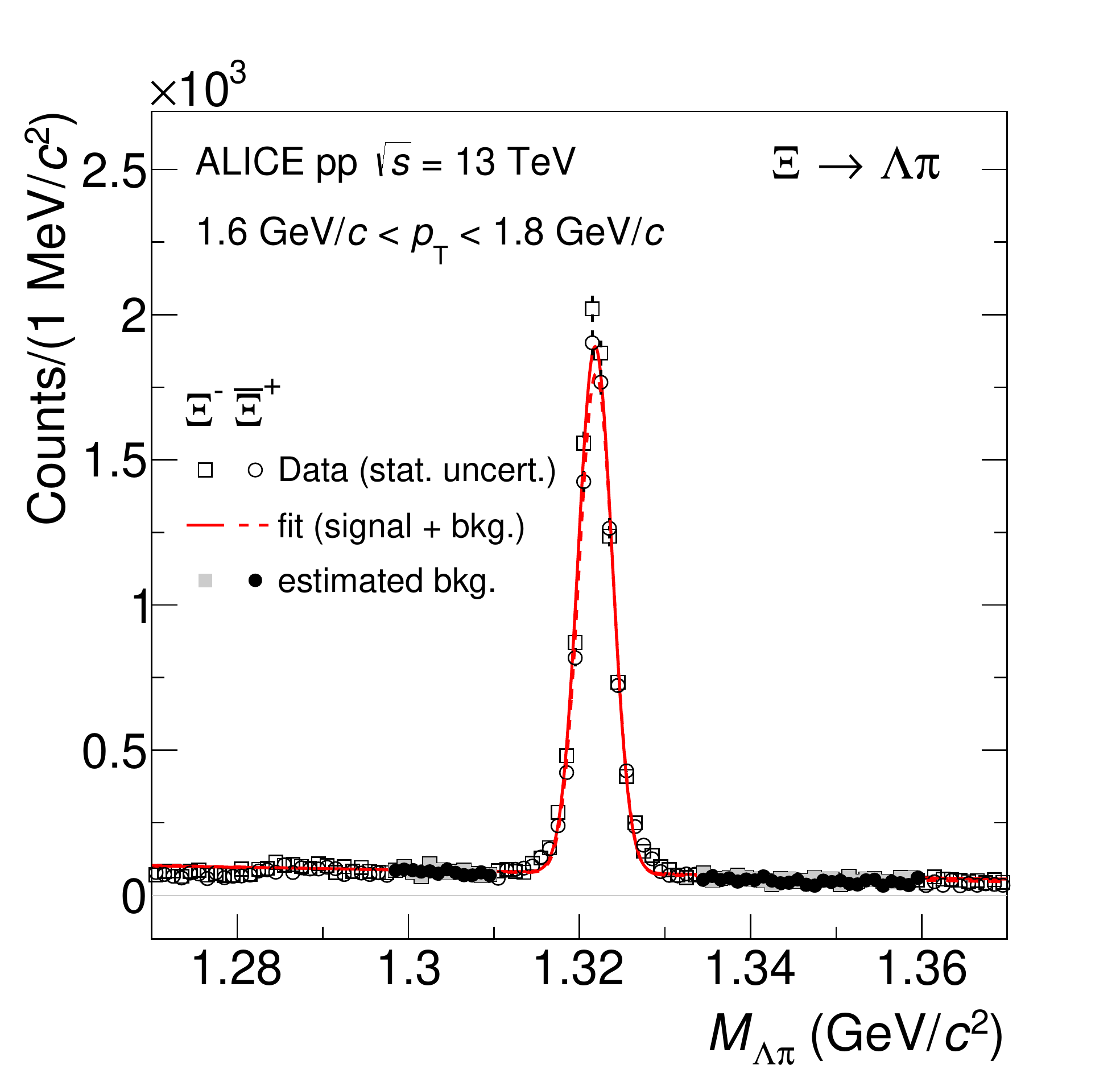}
	\includegraphics[keepaspectratio,width=0.495\columnwidth]{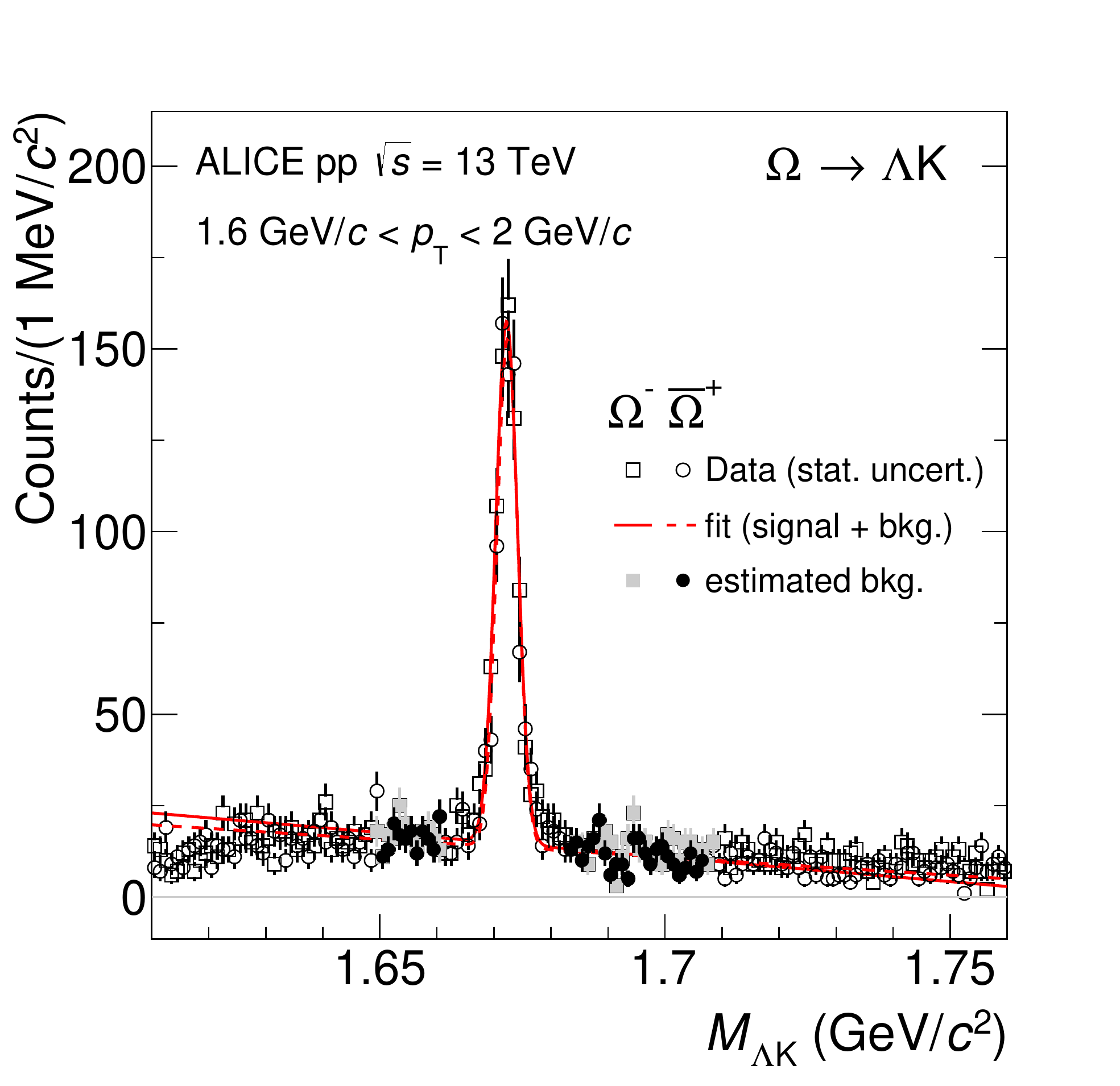}
	\includegraphics[keepaspectratio,width=0.495\columnwidth]{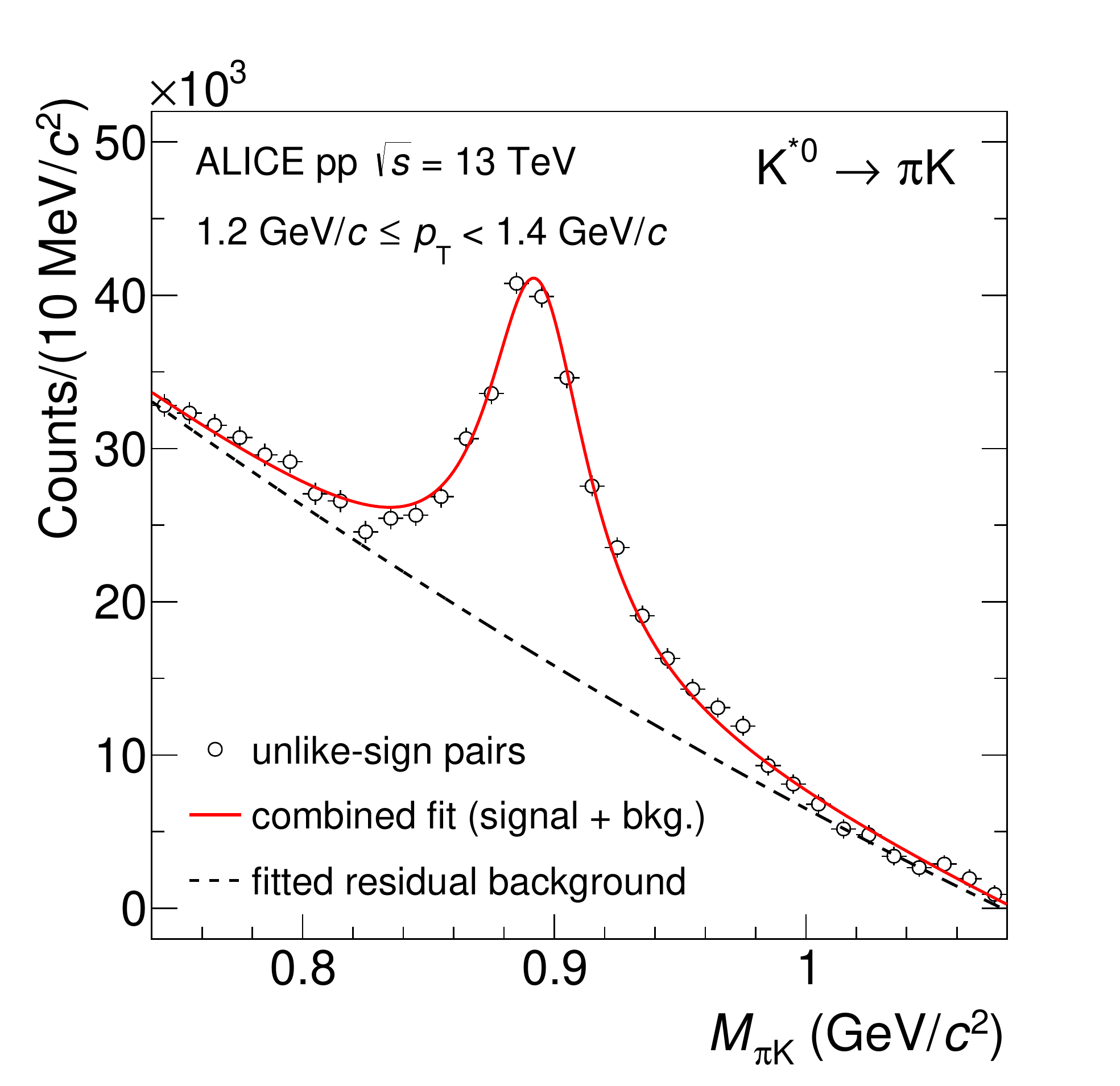}
	\includegraphics[keepaspectratio,width=0.495\columnwidth]{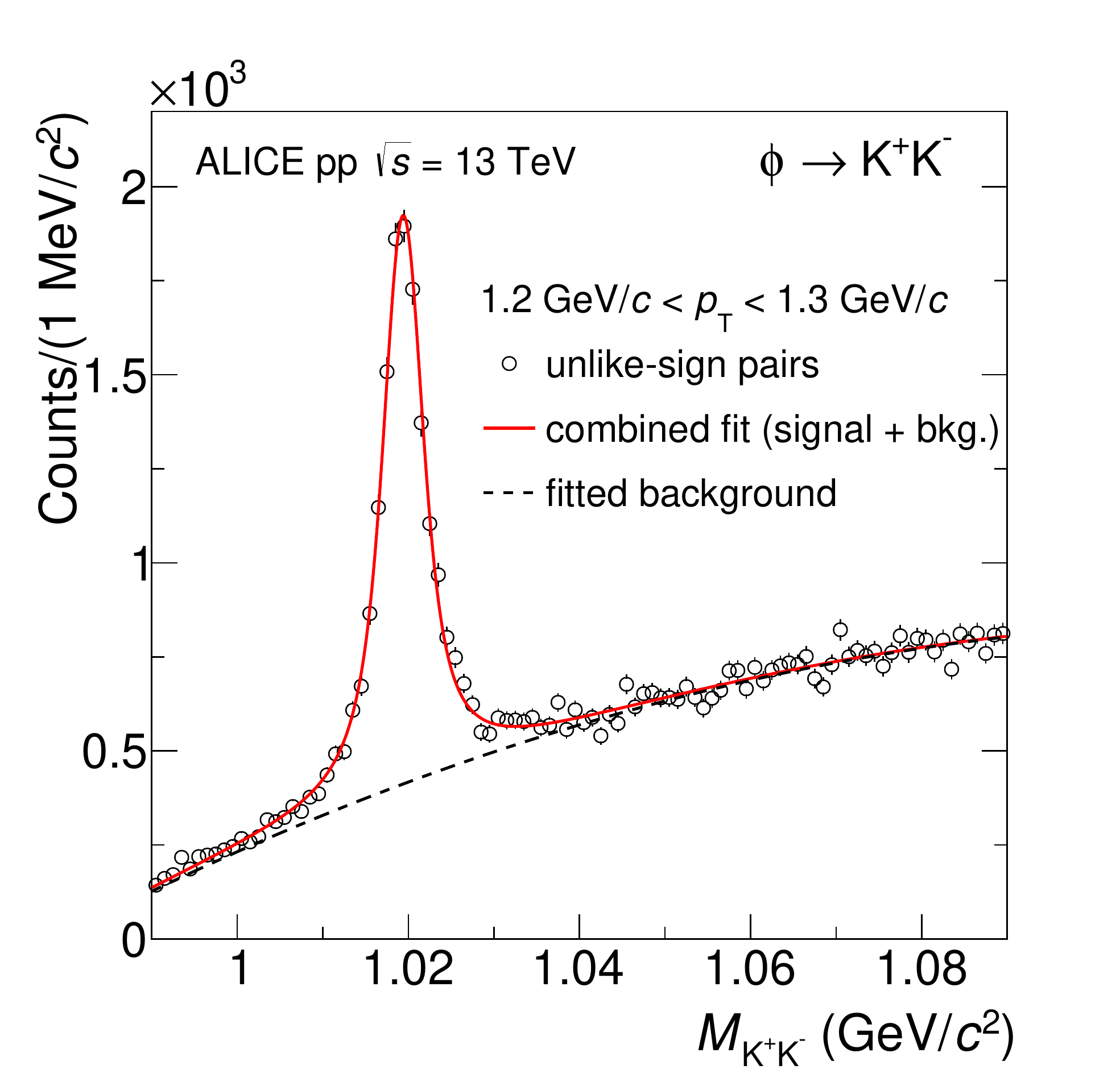}
      \caption[]{Invariant mass distributions of \kzs, \lap and \lam, \xim and \xip, and \omm and \omp, \ks and \ksb, and \ph. See the text for details.
      }
      \label{fig:InvMass}
    \end{figure}
    \clearpage
    }
    
    The yields for \lap (\lam) are significantly affected by secondary particles coming from the decays of \xim (\xip) and \xizero (\xizerobar). The feed-down fraction is computed for each \pt bin as the detection efficiency of \lap (\lam) 
    from $\Xi$ decays multiplied by the measured \xip (\xim) spectra, thereby assuming that the production rates of charged and neutral $\Xi$ are equal. The ratio of secondary {\lap} to the measured primary yield is about $6-22\%$, depending on \pt.

	For the \VZ and cascade analyses, the main sources of uncertainties are listed in Tab.~\ref{tab:syst:lambdak0s} and Tab.~\ref{tab:syst:xiomega}, respectively. The values of the selection criteria on the topological variables were varied 
	around their nominal values and the observed deviations for each component were summed in quadrature. Uncertainties related to signal extraction were estimated by varying the values of the width of the signal and background sampling regions 
	with respect to the default one, and adopting an alternative method to estimate the background in the peak region. The procedure resulted in a \pt-dependent uncertainty which rages from 0.2\% to 6.8\% for {\VZ}s and from 0.9\% to 2.4\% for 
	the cascades.

    For the measurement of \kzs, \lap and \lam at \sppt{13}, contributions from out-of-bunch pileup are removed as stated in Sec.~\ref{subsubsec:tracksel}. The applied correction reaches a maximum value of about $2\%$ for \kzs and about $3\%$ 
    for $\lap+\lam$ at high \pt. For \xip and \xim at \sppt{13}, a \pt-dependent correction factor, taken from Ref.~\cite{Acharya:2019kyh} is applied to remove the out-of-bunch pileup contribution; this correction is $0.5\%$ at low \pt and rises 
    to $2.1\%$ at high \pt. A similar correction for \omm and \omp is negligible and hence not applied. The out-of-bunch pileup contribution is found to be negligible for the measurement of \kzs, \lap, and \lam at \sppt{7}. A systematic uncertainty 
    to account for any residual effect is $0.3-1.1\%$ for the $\xip+\xim$, $1.2-4.6\%$ for $\lap+\lam$, depending on \pt, while it is negligible for \kzs.
    
    \afterpage{
	\begin{table}[t]
	  \centering
	  \caption{Summary of the main sources and values of the relative systematic uncertainties 
	  (expressed in \%) for the \kzs and $\lap+\lam$, \pt-differential yields. A single value 
	  between two or three columns indicates that no \pt dependence is observed. The values are 
	  reported for low, intermediate and high \pt. The abbreviation ``negl.'' indicates a negligible value.
	  }
	  \begin{threeparttable}
\begin{tabular}{K{0.28\textwidth} C{0.005\textwidth} C{0.005\textwidth} C{0.005\textwidth} C{0.005\textwidth} C{0.005\textwidth} C{0.001\textwidth} C{0.005\textwidth} C{0.005\textwidth} C{0.005\textwidth} C{0.005\textwidth} C{0.005\textwidth} C{0.001\textwidth} C{0.001\textwidth} C{0.001\textwidth}}
\toprule
{Hadron species}                                                        &  \multicolumn{6}{c}{\kzs}             &               & \multicolumn{6}{c}{$\lap+\lam$}               & \\
\cmidrule{1-15} \\[-12pt]
\multicolumn{15}{c}{Collision energy:~\sppt{13}} \\
\cmidrule{1-15} \\[-25pt] \\
{\pt (GeV/$c$)}                                                         & 0.05                  &               & 4.4                  &               & 14.25                 &               &               &       0.55                    &               & 4.2                 &               & 9.0       
        &       \\ [1pt]
\cmidrule{2-7} \cmidrule{9-15}                                                                                                          \\ [-10pt]
{Feed-down correction}                                          & \multicolumn{6}{c}{not applicable}       &       & 1.7   &       & 1.8   &       & 1.5                &               \\ [1pt]
{Hadronic interaction}                                          & \multicolumn{6}{c}{negligible}        &       & 1.7   &       & negl. &       & negl.                 &               \\ [1pt]
{Material budget}                                               & 4.7   &       & 0.5   &       & 0.5   &       &       & 6.7   &       & 0.8   &       & 0.8           &               \\ [1pt]
{Signal-Loss Correction}                                                & 1.9   &       & 0.2   &       & 0.4   &       &       & 2.0   &       & 0.2   &       & 0.2   &               \\ [1pt]
{Track selection}                                                       & negl.   &       & 3.5   &     & 9.5   &       &       & 1.0   &       & 2.2   &       & 2.1   &               \\ [1pt]
{Signal extraction}                                             & 0.7   &       & 0.2   &       & negl. &       &       & 0.3   &       & 1.7   &       & 3.8           &               \\ [1pt]
{Proper lifetime}                                               & \multicolumn{6}{c}{negligible}        &       & 0.7   &       & 1.3   &       & negl.                 &               \\ [1pt]
{Competing \VZ rejection}                                       & negl.   &       & 0.2   &     & negl. &       &       & negl. &       & 1.1   &       & 2.9           &               \\ [1pt]
{TPC \dedx}                                                     & \multicolumn{6}{c}{negligible}        &       & 0.6   &       & 0.2   &       & 3.3                   &               \\ [1pt]
{Topological selection}                                                 & 3.9   &       & 0.5   &       & negl. &       &       & 1.5   &       & 1.3   &       & 4.0   &               \\ [1pt]
{Out-of-bunch pileup rejection}                                 & \multicolumn{6}{c}{negligible}        &       & 1.2   &       & 4.6   &       & negl. &               &               \\ [1pt]
\cmidrule{1-15} \\[-12pt]
{Total}                 & 6.4   &       & 3.6   &       & 9.5   &       &       &       7.8     &       & 6.1   &               & 7.6                   &       \\ [1pt]
\midrule
\midrule
\multicolumn{15}{c}{Collision energy:~\sppt{7}} \\
\cmidrule{1-15} \\[-25pt] \\
{\pt (GeV/$c$)}                                                         & 0.05                  &               & 4.4                  &               & 14.25                 &               &               &       0.45                    &               & 4.2                  &               & 9.0          
        &       \\ [1pt]
\cmidrule{2-7} \cmidrule{9-15}                                                                                                          \\ [-10pt]
{Feed-down correction}                                          & \multicolumn{6}{c}{not applicable}       &       &       &  4.0  &       &       & 6.0           &               \\ [1pt]
{Hadronic interaction}                                          & \multicolumn{6}{c}{negligible}        &       &       &       & 1.0   &       &               &               \\ [1pt]
{Material budget}                                               &       &       & 4.0   &       &       &       &       &       &       & 4.0   &       &               &       \\ [1pt]
{Signal-Loss Correction}                                                & \multicolumn{6}{c}{negligible}        &       & \multicolumn{6}{c}{negligible}                \\ [1pt]
{Track selection}                                                       & 1.3   &       & 5.8   &       & 4.0   &       &       & 1.0   &       & 5.6   &       & 2.9           &               \\ [1pt]
{Signal extraction}                                             & 1.8   &       & 1.8   &       & 5.0   &       &       & 1.4   &       & 1.1   &       & 6.8           &               \\ [1pt]
{Proper lifetime}                                               & 0.5   &       & negl.   &       & 0.5   &       &       & 2.0   &       & 0.8   &       & 0.1           &               \\ [1pt]
{Competing \VZ rejection}                                       & negl.   &       & 0.5   &       & 3.2   &       &       & negl.   &       & 2.0   &       & 4.8           &               \\ [1pt]
{TPC \dedx}                                                     & negl.   &       & 2.3   &       & 2.6   &       &       & 4.9   &       & 0.8   &       & 6.0           &               \\ [1pt]
{Topological selection}                                                 & 1.9   &       & 1.1   &       & 5   &       &       & 1.3   &       & 1.2   &       & 4.8           &               \\ [1pt]
{Out-of-bunch pileup rejection}                                                & \multicolumn{6}{c}{negligible}        &       & \multicolumn{6}{c}{negligible}                \\ [1pt]
\cmidrule{1-15} \\[-12pt]
{Total}                 & 5.4   &       & 7.7   &       & 11.0  &       &       &       8.0     &       & 7.6   &               & 14.9                  &       \\ [1pt]
\bottomrule[1.5pt] \\ [1pt]
\end{tabular}
\end{threeparttable}

	  \label{tab:syst:lambdak0s}
	\end{table}
    \clearpage
    }

    \afterpage{
	\begin{table}[t]
	  \centering
	  \caption{Summary of the main sources and values of the relative systematic uncertainties 
	  (expressed in \%) for the $\xim+\xip$ and $\omm+\omp$ \pt-differential yields. A single value 
	  between two or three columns indicates that no \pt dependence is observed. The values are 
	  reported for low, intermediate and high \pt. The abbreviation ``negl.'' indicates a negligible value.
	  }
	  
\begin{threeparttable}
  \begin{tabular}{K{0.35\textwidth} C{0.005\textwidth} C{0.005\textwidth} C{0.005\textwidth} C{0.005\textwidth} C{0.005\textwidth} C{0.001\textwidth} C{0.001\textwidth} C{0.005\textwidth} C{0.005\textwidth} C{0.005\textwidth} C{0.005\textwidth} C{0.005\textwidth} C{0.001\textwidth} C{0.001\textwidth} C{0.001\textwidth}}
  \toprule
  {Hadron species}                &  \multicolumn{6}{c}{$\xim+\xip$} & & \multicolumn{6}{c}{$\omm+\omp$} & \\ 
  \midrule
  {\pt (GeV/$c$)}                               & 0.7  & & 1.9 & & 6          & & &   1.2     & & 2.56 & & 4.3  	&       \\ [1pt]
  \cmidrule{2-7}        \cmidrule{9-14}         \\ [-10pt]
  {Hadronic interaction}                    & 1.4   & & 0.1  	 & & negl.   & & & 0.8    & & negl. & & 1.0  	&       \\ [1pt]
  {Material budget}                         & 5.7   & & 1.9  	 & & 0.6     & & & 3.5    & &  		& 1.5 & &       \\ [1pt]
  {Signal-Loss correction}                  & 0.7   & & 0.2  	 & & 0.2     & & & 0.2    & & 0.1 	& & 0.1 &       \\ [1pt]
  {Track selection}                         & negl. & & 0.3  	 & & 2.9     & & &        & negl. & 	& & 4.1 &       \\ [1pt]
  {Signal extraction}                 		& 2.2   & & 0.9   	 & & 1.2     & & & 0.8    & & 1.7       & & 2.4 &       \\ [1pt]
  {Competing \VZ rejection}                 & 	\multicolumn{6}{c}{negligible} & &        & negl.&      & & 3.1 &       \\ [1pt]
  {TPC \dedx}                             	& 0.9   & & negl.    & & negl.   & & &\multicolumn{6}{c}{negligible} 	\\ [1pt]
  {Topological selection}                   & 1.7   & & 0.8   	 & & 1.8     & & & 2.5    & & 2.1 	& & 3.0 &       \\ [1pt]
  {$|y|$ cut at low $p_{T}$}                &       \multicolumn{6}{c}{negligible} & & 3.9    & &  & negl. &   	&       \\ [1pt]
  {Out-of-bunch pileup rejection}           & 0.3   & & 0.5      & & 1.1     & & & \multicolumn{6}{c}{not applicable}       \\ [1pt]
  \cmidrule{1-15} \\[-10pt]
  {Total}                                   & 6.6  & & 2.4 & & 3.7           & & & 6.0    & & 3.1 & & 6.7 		&      \\ [1pt]
    \bottomrule[1.5pt] \\ [1pt]
  \end{tabular}
\end{threeparttable}

	  \label{tab:syst:xiomega}
	\end{table}
	\begin{table}[t]
	  \let\center\empty
	  \let\endcenter\relax
	  \centering
	  \caption{Summary of the main sources and values of the relative systematic uncertainties 
	  (expressed in \%) for \ksa and \ph. A single value between two or three columns indicates 
	  the mean for all \pt bins, with little \pt dependence. The values are reported for low, 
	  intermediate and high \pt. The abbreviation ``negl.'' indicates a negligible value.
	  }
	  \begin{threeparttable}
\begin{tabular}{K{0.28\textwidth} C{0.005\textwidth} C{0.005\textwidth} C{0.005\textwidth} C{0.005\textwidth} C{0.005\textwidth} C{0.001\textwidth} C{0.005\textwidth} C{0.005\textwidth} C{0.005\textwidth} C{0.005\textwidth} C{0.005\textwidth} C{0.001\textwidth} C{0.001\textwidth} C{0.001\textwidth}}
\toprule
{Hadron species}                                &  \multicolumn{6}{c}{\ksa}             &               & \multicolumn{6}{c}{\ph}               & \\
\cmidrule{1-15} \\[-25pt] \\
{\pt (GeV/$c$)}                                	& 0.05                  &               & 3.0                  &               & 13.5                 &               &               &       0.5                    &               & 3.0                 &               & 8.0       
        &       \\ [1pt]
\cmidrule{2-7} \cmidrule{9-15}                                                                                                          \\ [-10pt]
  {Material Budget} 				& 3.4	& & 0.5 & & negl.& 	  & & 	2.2 	& & 0.7 & & negl.& 		  	\\ [1pt]
  {Hadronic Interaction} 			& 2.8	& & 1.2 & & negl.& 	  & & 	0.5 	& & 1.7 & & negl.& 		 	\\ [1pt]
  {Signal-Loss Correction} 			& 1.8	& & 0.3 & & 0.2	 & 	  & & 	1.7 	& & 0.5 & & 0.5	 & 		  	\\ [1pt]
  {Track Selection} 			& 8.2	& & 2.2 & & 6.8	 & 	  & & 	4.3 	& & 1.6 & & 4.1	 & 		  	\\ [1pt]
  {Tracking} 					& 	& & 2.0 & &  & 	  & & 	 	& & 2.0 & &  & 		 	\\ [1pt]
  {Signal Extraction} 				& 10.6  & & 5.4 & & 10.3 & 	  & & 	6.0 	& & 2.5 & & 4.9	 & 		  	\\ [1pt]
  {Branching Ratio} 				& \multicolumn{6}{c}{negligible}  & & 	 	& & 1.0 & & 	 & 		 	\\ [1pt]
  \midrule
  {Total} 					&  17.7	& & 9.9 & & 17.0&	  & & 	8.2 	& & 4.2 & & 6.5 &  	 	  	\\ [1pt]  
\bottomrule[1.5pt] \\ [1pt]
\end{tabular}
\end{threeparttable}

	  \label{tab:syst:phikstar}
	\end{table}
    \clearpage
    }
    
    The \VZ (topological selection) and track selection criteria are varied around their default values, producing up to $9.5\%$ ($5.8\%$) uncertainties in the \kzs and \lap yields at \sppt{13} (\sppt{7}). 
    The feed-down corrections for \lap\ and \lam\ carry uncertainties associated with the method and the uncertainty of the measured $\Xi$ spectrum. The values are estimated to be around $2\%$ ($4-6\%$) for 
    13 (7)~TeV, depending on \pt.
    The applied selection criteria for PID in the TPC, used for better discrimination between the combinatorial background and the signal for strange baryons, are varied in the range of $4-7\stpc$. For \lap and \lam, 
    this uncertainty is $\sim$0.8--6\% at $\sqrt{s}=7$ and 0.2--3.3\% at \sppt{13}. For $\Xi$ and $\Omega$, it is at most 1\% for both collision energies. For \kzs the difference was found to be negligible at 13~TeV 
    and at most 2.6\% at 7~TeV. The resulting total uncertainties from low to high \pt vary in the ranges 3.6--9.5\% for \kzs, 6.1--7.8\% for \lap, 6.6--3.7\% for $\Xi$, and 6.0--6.7\% for $\Omega$.

    In the invariant mass distribution (shown in Fig.~\ref{fig:InvMass}), one could observe for {\VZ}s and cascades an imperfect fit of the tails close to the peak; the excess of candidates comes from misidentified 
    daughters. The removal of this effect though application of further PID selection criteria would reduce significantly the number of candidates while not improving the signal extraction.

    \subsection{Reconstruction of resonances} \label{resonances}

    Using analysis techniques similar to those described in Ref.~\cite{ALICE_kstar_phi_PbPb}, \ks and \ksb mesons are identified through reconstruction of their decays to charged pions and kaons, while \ph mesons 
    are identified via their decays to pairs of charged kaons. Primary charged tracks are selected using the standard criteria described in Sec.~\ref{subsubsec:tracksel}. Pion and kaon candidates are identified using 
    their specific energy loss \dedx measured in the TPC and their velocity $\beta$ measured with the TOF. The specific energy loss for each pion (kaon) candidate with $p>\gevc{0.4}$ is required to be within 2\stpc of 
    the expected mean value for pions (kaons); a less restrictive selection is applied for lower momenta. In addition, if the charged track is matched to a hit in the TOF, $\beta$ must be within 3\stof of the expected 
    mean value.	For the analysis of the \ksa (\ph), all kaon candidates are paired with all oppositely charged pion (kaon) candidates from the same event and the pair invariant mass is calculated. The resulting distributions 
    are shown in Fig.~\ref{fig:InvMass}. The combinatorial background is estimated by calculating the invariant mass distribution of like-charge $\pi$K or KK pairs from the same event, by parameterizing the combinatorial 
    background with a simple function (for \ph only), or by pairing tracks from two different events (the ``mixed-event" technique). In order to ensure that the mixed events have similar characteristics, the $z$ positions 
    of their primary vertices are required to be separated by less than 1~cm and their charged particle multiplicities are required to differ by no more than 5 reconstructed tracks. The background-subtracted invariant mass 
    distributions are then fitted with a peak function added to a function that parameterizes the residual background contribution from correlated pairs. The \ph peak is described with a Voigtian function, the convolution 
    of a Breit--Wigner function and a Gaussian. The mass resolution for the \ks is much smaller than the width of that resonance and a Breit--Wigner function is used to describe the \ks peak. The yield of the \ksa (\ph) is 
    then calculated by integrating the invariant mass distribution within $0.8<\mpik<1.0$~\gvcc $(1.01<\mkk<1.03~\gvcc)$, subtracting the integral of the residual background in the same region, and adding the yield from 
    outside the peak region obtained from the peak fit function. The variations in the \ksa and \ph yields due to alternate treatments of the combinatorial background, residual background parameterizations, and peak fitting 
    functions are incorporated into the systematic uncertainties.
     
	The systematic uncertainties in the \ksa and \ph yields include the contributions listed in Tab.~\ref{tab:syst:phikstar}. The contribution due to variations in the PID, track, and event selection criteria 
	(``Track Selection" in the table) is 2--8\% for \ksa and 1-5\% for \ph. Variations in the combinatorial background construction, residual background parameterization, peak parameterization, and fit range 
	(``Signal Extraction" in the table) combine to give an uncertainty of 5--11\% for \ksa and 2--6\% for \ph, depending on \pt. The ITS-TPC matching uncertainty for single particles (pions and kaons) is 2\% for all \pt 
	intervals. The uncertainty in the branching ratio is $1\%$ for \ph and negligible for \ksa. The total systematic uncertainty for \ksa and \ph is estimated to be about $10-18\%$ and $8-13\%$, depending on \pt.
	

\section{Results} \label{sec:results}

  For the light-flavor hadrons discussed in this paper, the ratios of yields for particles and antiparticles are around one within the uncertainties, as expected at these collision energies in the midrapidity region. 
  Therefore, all the \pt spectra shown in the following are reported after summing particles and antiparticles, when a distinct antiparticle state exists. Unless explicitly stated, the sums of particles and antiparticles, 
  $\pip+\pim$, $\kp+\km$, $\ks+\ksb$, $\prp+\prm$, $\lap+\lam$, $\xip+\xim$, and $\omp+\omm$ are denoted as \pix, \kx, $\ks(\ksb)$, \px, $\lap(\lam)$, $\xim(\xip)$, $\omm(\omp)$ or simply as $\pi$, K, \ks, p, \lap, $\Xi$, 
  and $\Omega$, respectively, unless explicitly written.
  
  The uncertainty related to the overall normalization to inelastic (INEL) events is fully correlated between particle species and is not shown explicitly when plotting the results.

    \subsection{Transverse momentum distributions, integrated yields, and average transverse momenta} \label{subsec:ptspectra}
    
      \begin{figure}[t]
	\centering
	\includegraphics[keepaspectratio, width=\columnwidth]{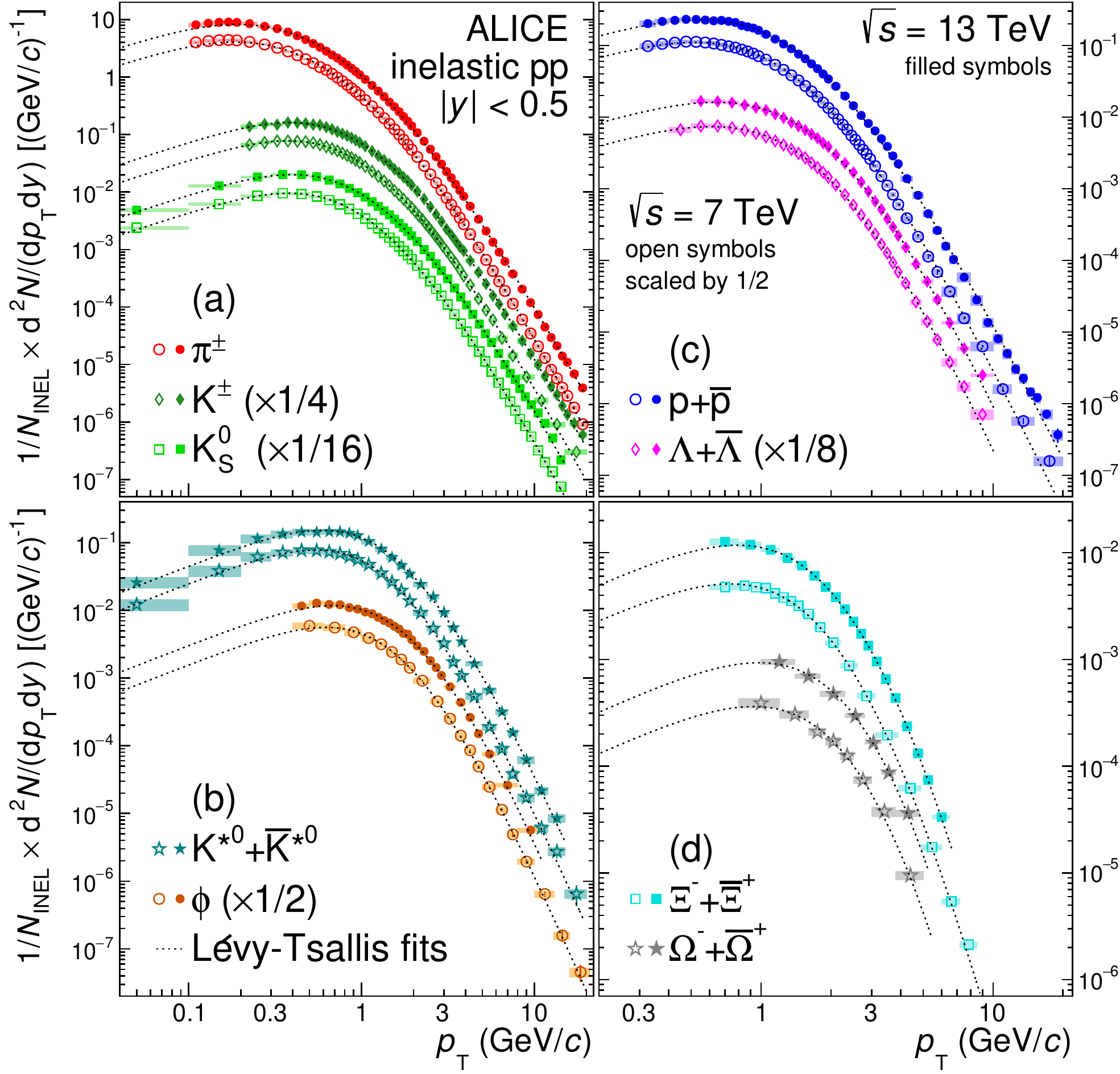}
	\caption{Transverse momentum spectra of light-flavor hadrons measured at midrapidity ($|y|<0.5$) in 
	inelastic \pp collisions at \sppt{13} (filled symbols) and \sppt{7} (open symbols, scaled by a factor 
	of $1/2$)~\cite{Adam:2016dau,Abelev:2012jp,Acharya:2019wyb}. Statistical and systematic uncertainties 
	are shown as vertical error bars and boxes, respectively. The data points are fitted using a L\'evy--Tsallis 
	function. The normalization uncertainty of $^{+7.3}_{-3.5}\%$ $(\pm2.6\%)$ for \pp collisions at 
	$\sqrt{s}=7$ (13)~TeV is common to all particle species and is excluded from the plotted uncertainties.
	}
	\label{fig:ptspectra}
      \end{figure}
    
    The \pt spectra of light-flavor hadrons measured at midrapidity in inelastic \pp collisions at $\sqrt{s} = 7$ and $13$\,TeV are given in Fig.~\ref{fig:ptspectra}. The \pt spectra of \kzs and \lap measured in this paper at 
    \sppt{7} are shown together with other particle species from previous ALICE measurements at the same center-of-mass energy~\cite{Abelev:2012jp,Adam:2016dau,Acharya:2019wyb}. For \sppt{7} INEL \pp collisions, the 
    reported \pt distributions of $\pi$, K, p~\cite{Abelev:2012jp,Adam:2016dau}, \ks, and \ph~\cite{Acharya:2019wyb} are from the updated measurements of ALICE, with extended \pt reach, and, for resonances, additionally 
    with an improved estimate of the systematic uncertainties. For clarity, some of the spectra have been scaled with the factors indicated in the legends.
    
    The \pt distributions are fitted with L\'evy--Tsallis functions~\cite{Tsallis,STAR_strange_pp_2007} in order to extrapolate the spectra to the unmeasured \pt regions, i.e. down to zero and up to high \pt, similar to what was 
    done in previous measurements~\cite{Adam:2015qaa,Aamodt:2011zza,Abelev:2012jp,Abelev:2012hy,Adam:2017zbf}. This procedure allows the \pt-integrated yields \dndyinline and the average transverse momenta \mpt to be extracted, for which 
    the measured as well as the extrapolated distributions are used. The obtained values are given in Tab.~\ref{tab:dydn_meanpt_allparticles}. The fit function describes well both the low-\pt exponential and high-\pt power-law nature of the \pt distribution, with $\chi^{2}/\rm{ndf}$ values in the 
    range $0.2-2.3$. The slope parameter of the L\'evy--Tsallis functions decreases for all particle species going from $\sqrt{s} = 7$ to $13$\,TeV. For example, for charged pions and (anti)protons it changes from $6.65\pm0.03$ to $6.42\pm0.03$ 
    and from $7.99\pm0.13$ to $ 7.71\pm0.11$.
    No extrapolation is needed for \kzs and \ks, as their yields are measured down to $\pt=\gevc{0}$. The fractions of extrapolated particle yields outside the measured \pt range at low \pt are given in the 
    last column of Tab.~\ref{tab:dydn_meanpt_allparticles}.

    Other fit ranges and other parameterizations (\mT exponential, Boltzmann distribution, Bose--Einstein distribution, Fermi--Dirac distribution, Boltzmann--Gibbs blast-wave function~\cite{BoltzmannGibbsBlastWave}) are also used and the resulting 
    variations in the \dndyinline and \mpt values are incorporated into the systematic uncertainties. The systematic uncertainties are 4--10\% for \dndyinline and 1--3\% for \mpt, similar to those estimated for measurements 
    during Run 1. For the present measurements, systematic uncertainties are dominant. The average yields for all particles species increase with collision energy. Compared to \pp collisions at \sppt{7}~\cite{Adam:2015qaa,Abelev:2012jp}, 
    the average increase of the \mpt and the yield per inelastic collision for all measured particle species is about $8\%$ and $11\%$, respectively. This is in agreement with the ${\approx}15\,\%$ increase of the average 
    pseudorapidity density of charged particles produced in $|\eta|<0.5$ as the collision energy increases from \sppt{7} to \sppt{13}~\cite{Adam:2015pza}.

	\begin{table}[t]
	  \let\center\empty
	  \let\endcenter\relax
	  \centering
	  \caption{Average transverse momentum \mpt and \pt-integrated yield \dndyinline values for light-flavor hadrons measured 
	  in \pp collisions at \sppt{13} and 7\,TeV. The first uncertainty 
	  is statistical and the second is systematic (including the low- and high-\pt extrapolation). The uncertainties due 
	  to the normalization to the number of inelastic events, which are $^{+7.3}_{-3.5}\%$ and $\pm2.6\%$ for \pp collisions 
	  at $\sqrt{s}=7$ and 13~TeV, respectively, are not included. The last column represents the fraction of extrapolated yield at low transverse momenta.
	  }
	  \resizebox{0.99\textwidth}{!}{\begin{threeparttable}
\begin{tabular}{c c c c}
\toprule
\toprule
\multirow{ 2}{*}{Hadron species}                                & \multirow{ 2}{*}{\dndyinline}                                         & \multirow{ 2}{*}{\mpt (GeV/$c$)}                                      & Extrapolated  \\
                                                                &                                                                       &                                                                       & fraction (\%) \\
  \midrule
\multicolumn{4}{c}{Collision energy:~\sppt{13}} \\
\midrule
  $\pi^{+}+\pi^{-}$                                     & $4.775 \pm 0.001 \pm 0.243$                                   & $(4.915 \pm 0.001 \pm 0.099)\times10^{-1}$                    & $8\pm1$   \\ [2pt]
  ${\rm K}^{+}+{\rm K}^{-}$                             & $(6.205 \pm 0.004 \pm 0.303)\times10^{-1}$                    & $(8.099 \pm 0.007 \pm 0.099)\times10^{-1}$                    & $9\pm0.1$   \\ [2pt]
  $\text{K}^{0}_{\text{S}}$                             & $(3.192 \pm 0.004 \pm 0.111)\times10^{-1}$                    & $(8.207 \pm 0.008 \pm 0.087)\times10^{-1}$                    & negl. \\ [2pt]
  \ksa                                                  & $(2.098 \pm 0.016 \pm 0.200)\times10^{-1}$                    & $1.121 \pm 0.005 \pm 0.030$                                   & negl. \\ [2pt]
  ${\rm p}+\overline{\rm p}$                            & $(2.750 \pm 0.002 \pm 0.188)\times10^{-1}$                    & $(9.659 \pm 0.008 \pm 0.144)\times10^{-1}$                    & $11\pm0.5$  \\ [2pt]
  \ph                                                   & $(3.734 \pm 0.040 \pm 0.213)\times10^{-2}$                    & $1.236 \pm 0.009 \pm 0.027$                                   & $13\pm1$  \\ [2pt]
  $\Lambda+\overline{\Lambda}$                          & $(1.807 \pm 0.005 \pm 0.102)\times10^{-1}$                    & $1.078 \pm 0.002 \pm 0.030$                                   & $22\pm12$  \\ [2pt]
  $\Xi^{-}+\overline{\Xi}^{+}$                          & $(1.980 \pm 0.012 \pm 0.082)\times10^{-2}$                    & $1.296 \pm 0.004 \pm 0.015$                                   & $20\pm4$  \\ [2pt]
  $\Omega^{-}+\overline{\Omega}^{+}$                    & $(1.846 \pm 0.046 \pm 0.122)\times10^{-3}$                    & $1.527 \pm 0.023 \pm 0.027$                                   & $34\pm7$  \\ [2pt]
\midrule
\multicolumn{4}{c}{Collision energy:~\sppt{7}} \\
\midrule
  $\text{K}^{0}_{\text{S}}$                             & $(2.802 \pm 0.002 \pm 0.149)\times10^{-1}$                    & $(7.731 \pm 0.006 \pm 0.100)\times10^{-1}$                    & negl. \\ [2pt]
  $\Lambda+\overline{\Lambda}$                          & $(1.523 \pm 0.002 \pm 0.110)\times10^{-1}$                    & $1.028 \pm 0.001 \pm 0.019$                                   & $16\pm1$  \\ [2pt]
\bottomrule
\end{tabular}
\end{threeparttable}
}
	  \label{tab:dydn_meanpt_allparticles}
       \end{table}


\section{Discussion} \label{sec:results}

    \begin{figure}[t]
      \centering
      \includegraphics[keepaspectratio, width=0.9\columnwidth]{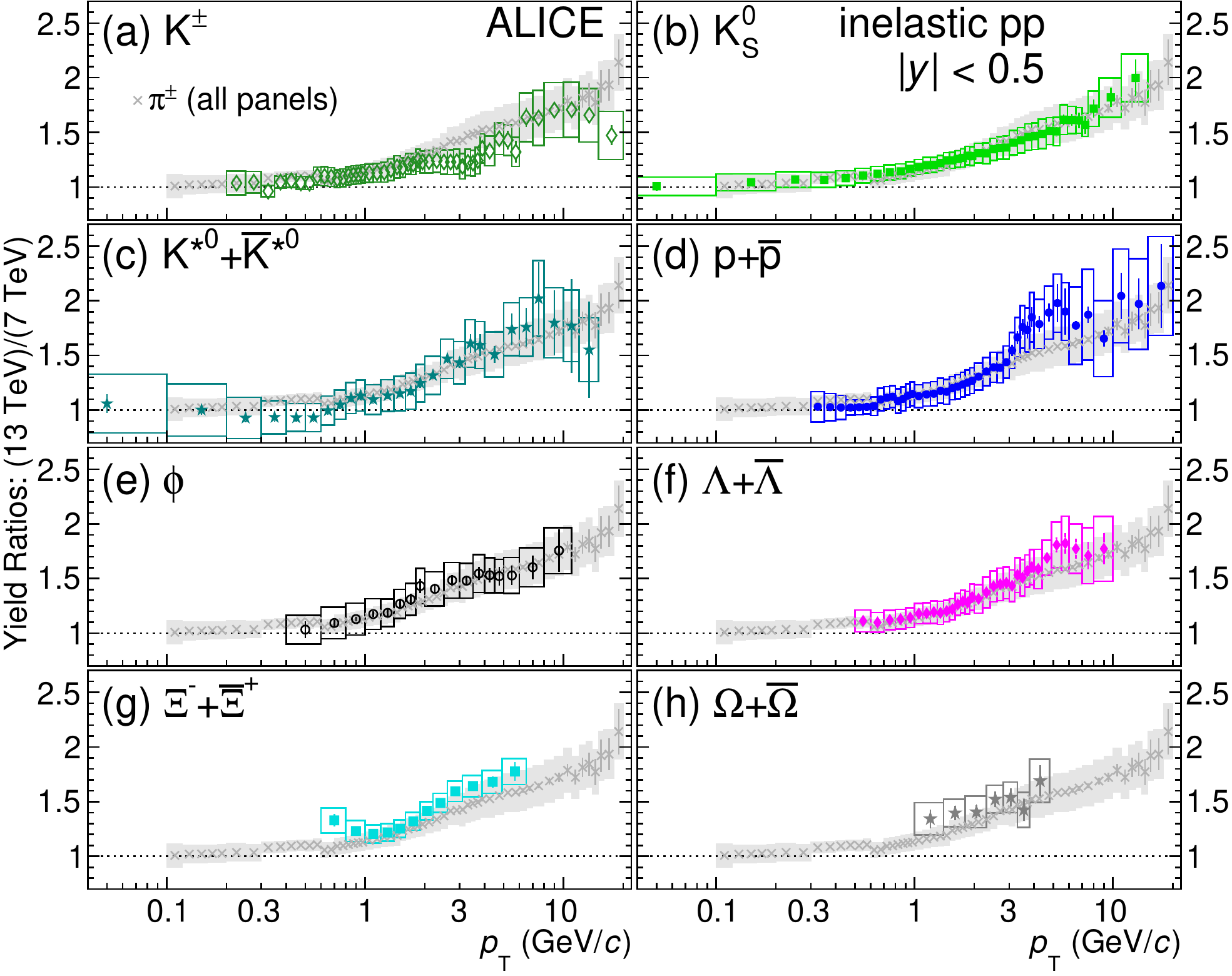}
      \caption{Ratios of the transverse-momentum spectra of light-flavor hadrons in inelastic \pp collisions 
      at \sppt{13} to those for \sppt{7}~\cite{Adam:2016dau,Abelev:2012jp,Acharya:2019wyb}. The ratio for 
      \pix is shown in each panel with grey crosses and boxes. Statistical and systematic uncertainties are shown as 
      vertical bars and boxes, respectively. The normalization uncertainty ($^{+10.8}_{-6.3}\%$) is excluded from the plotted uncertainties.
      }
      \label{fig:Fig_Yield_Ratio_to_276TeV}
    \end{figure}

  The transverse-momentum spectra reported in Fig.~\ref{fig:ptspectra} indicate a progressive and significant evolution of the spectral shapes at high \pt with increasing collision energy, which is similar for all particle 
  species under study. This behavior is better visualized in Fig.~\ref{fig:Fig_Yield_Ratio_to_276TeV} which shows the corresponding ratios of \pt spectra at \sppt{13} to those at \sppt{7}~\cite{Adam:2016dau,Abelev:2012jp,Acharya:2019wyb}. 
  The systematic uncertainties at both collision energies are largely uncorrelated and therefore their sum in quadrature is taken as the systematic uncertainty on the ratios. The uncertainty on the ratio due to normalization is $^{+10.8}_{-6.3}\%$.

  The ratios for all hadron species are above unity, which is consistent with the observed increase in the pseudorapidity density of inclusive charged particles with increasing collision energy~\cite{Adam:2015pza}. Furthermore, 
  all the ratios exhibit a clear increase as a function of \pt, indicating that hard processes become dominant in the production of high-\pt particles. The \pt dependence demonstrates that the spectral shapes are significantly 
  harder at \sppt{13} than at \sppt{7}, which is also evident in the reduction of the slope parameter of the L\'evy--Tsallis functions mentioned above. A universal shape --- independent of \pt within uncertainties --- can be observed 
  for most species (excluding $\Xi$ and $\Omega$) in the soft regime, $\pt\lesssim\gevc{1}$. There is a hint that the ratio for \px may be enhanced above the one for \pix in the \pt region ${\sim}3-\gevc{6}$, although the enhancement 
  is barely significant given the uncertainties. Such an enhancement would be consistent with the appearance of the baryon anomaly, an increased baryon-to-meson production ratio at intermediate transverse momenta ($\gevc{2}\lesssim\pt\lesssim\gevc{10}$), 
  observed in previous ALICE measurements of light-flavor hadron production~\cite{Abelev:2014laa,Acharya:2019yoi,Acharya:2018orn,Adam:2016dau,Abelev:2013xaa}.
  It is worth noting that the hardening of the \pt spectra with increasing collision energy has been reported in our earlier work for inclusive charged particles~\cite{Adam:2015pza}, although with different event selection criteria. 
  There, a requirement of at least one charged particle with $\pt>\gevc{0}$ in $|\eta|<1$ was imposed, selecting events corresponding to 75\% of the total inelastic cross section. In Ref.~\cite{Adam:2015pza}, the observed trend was 
  found to be well captured by the \py and \eplhc MC generators. In Sec.~\ref{subsec:models} the ratios of \pt spectra of light-flavor hadrons at \sppt{13} to those at \sppt{7} are compared to results from these common event generators.

    \subsection{Scaling properties of hadron production}

    Two kinds of universal scaling of identified particle production have been observed in high energy \pp collisions: transverse mass (\mT) scaling, which was originally seen in the lower \pt region of hadron spectra, and \xt 
    scaling~\cite{Brodsky:1973kr,SIVERS19761,PhysRevD.18.900,Arleo:2009ch,Arleo:2010kw}, observed in the higher \pt region. New studies of the \mT and \xt scaling properties of light-flavor hadrons in pp collisions at $\sqrt{s}=7$ and 13~TeV are discussed below.


    \subsubsection{Transverse mass (\mT) scaling}\label{subsec:mTscaling}

    At ISR energies~\cite{Alper:1974rw,Alpgard:1981ke} it has been observed that hadron \mT spectra in \pp collisions seem to follow an approximately universal curve after scaling with arbitrary normalization factors, an effect known as \mT scaling. 
    The STAR and PHENIX collaborations observed the breaking of \mT-scaling in \pp collisions at \sppg{200}~\cite{Abelev:2006cs,Adare:2010fe}, where a clear separation between baryon and meson spectra was observed for $\mT\geq2$~\gvcc. In \pp 
    collisions at \sppg{200}, the separation between the baryon and meson spectra seems to increase over the measured \mT range. We, the ALICE collaboration, recently reported a similar breaking in \pp collisions at \sppt{8}\cite{Acharya:2017tlv}. 
    The measurements in this paper allow this study to be extended to higher \mT and to the highest LHC energies.

\begin{figure}[hbtp]
    \centering
    \includegraphics[keepaspectratio,width=0.49\columnwidth]{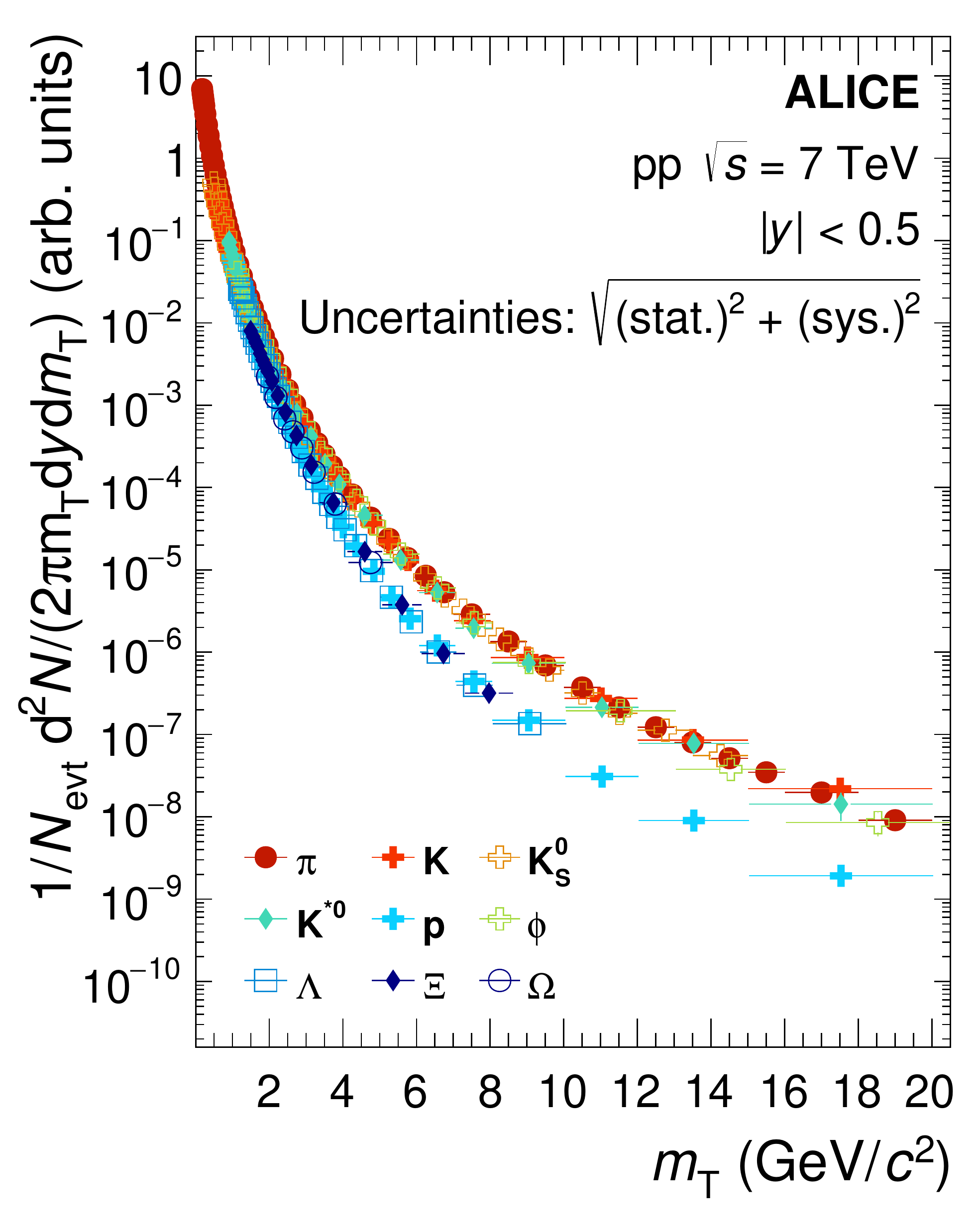}
    \includegraphics[keepaspectratio,width=0.49\columnwidth]{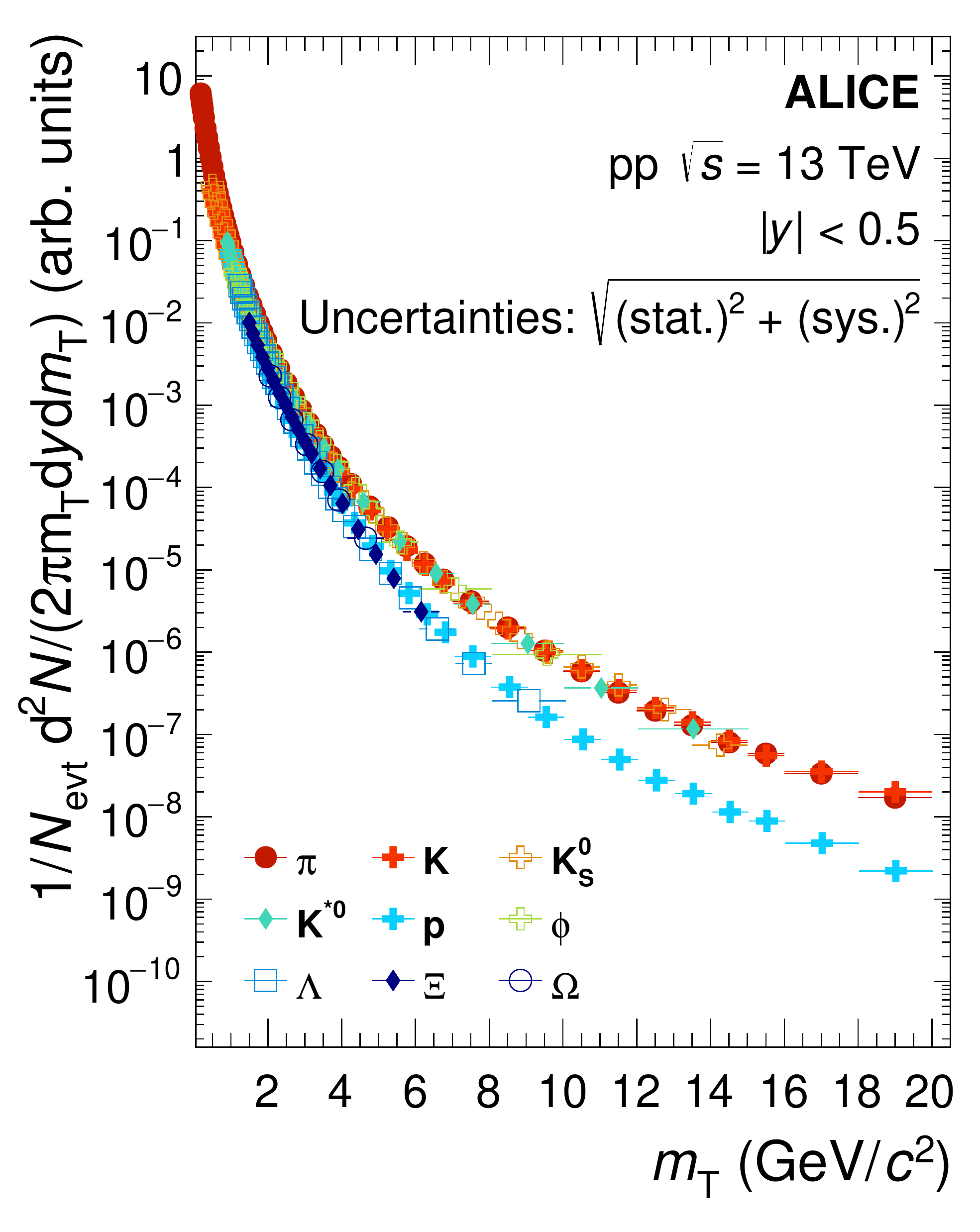}
    \includegraphics[keepaspectratio,width=0.49\columnwidth]{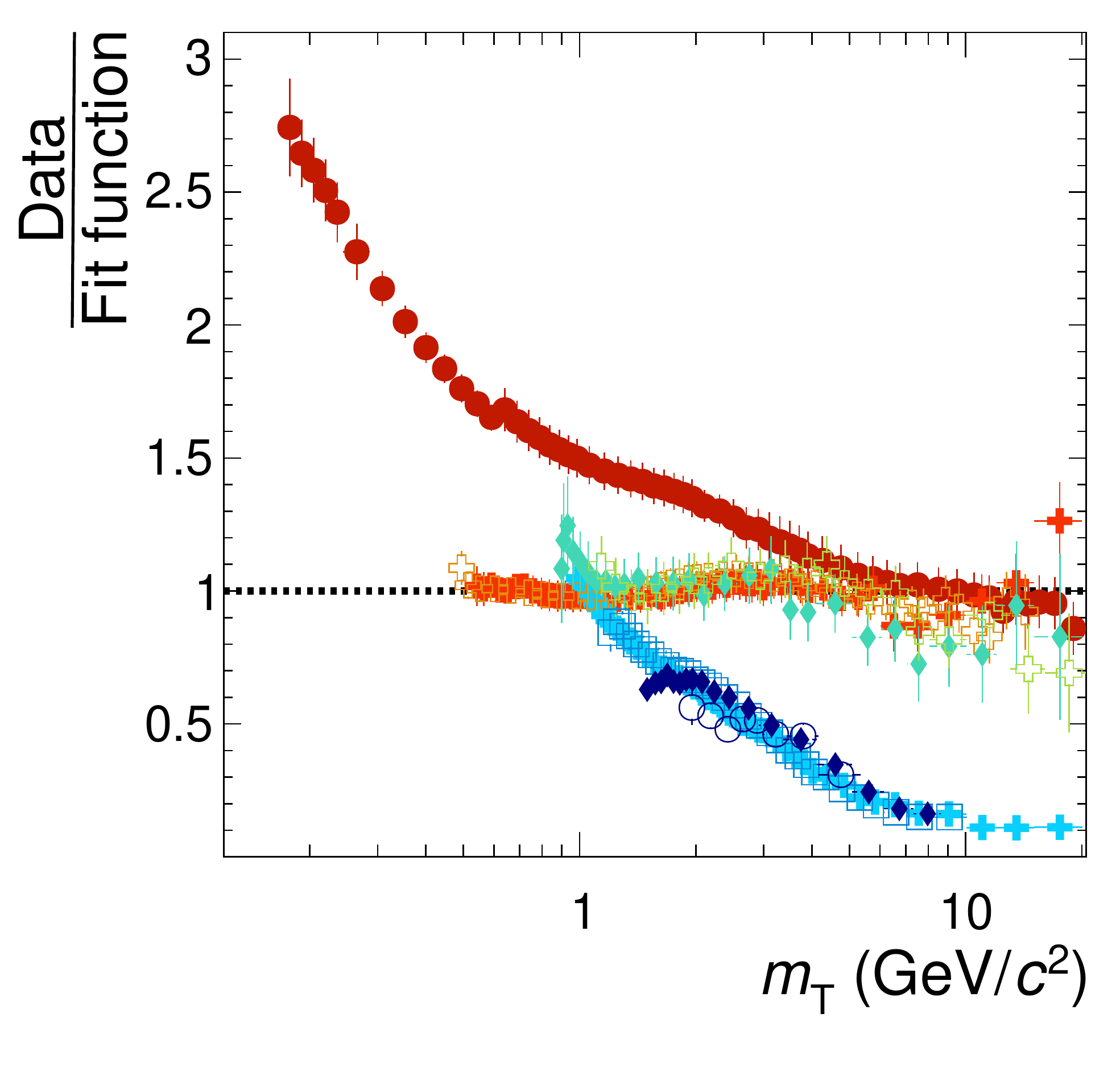}
    \includegraphics[keepaspectratio,width=0.49\columnwidth]{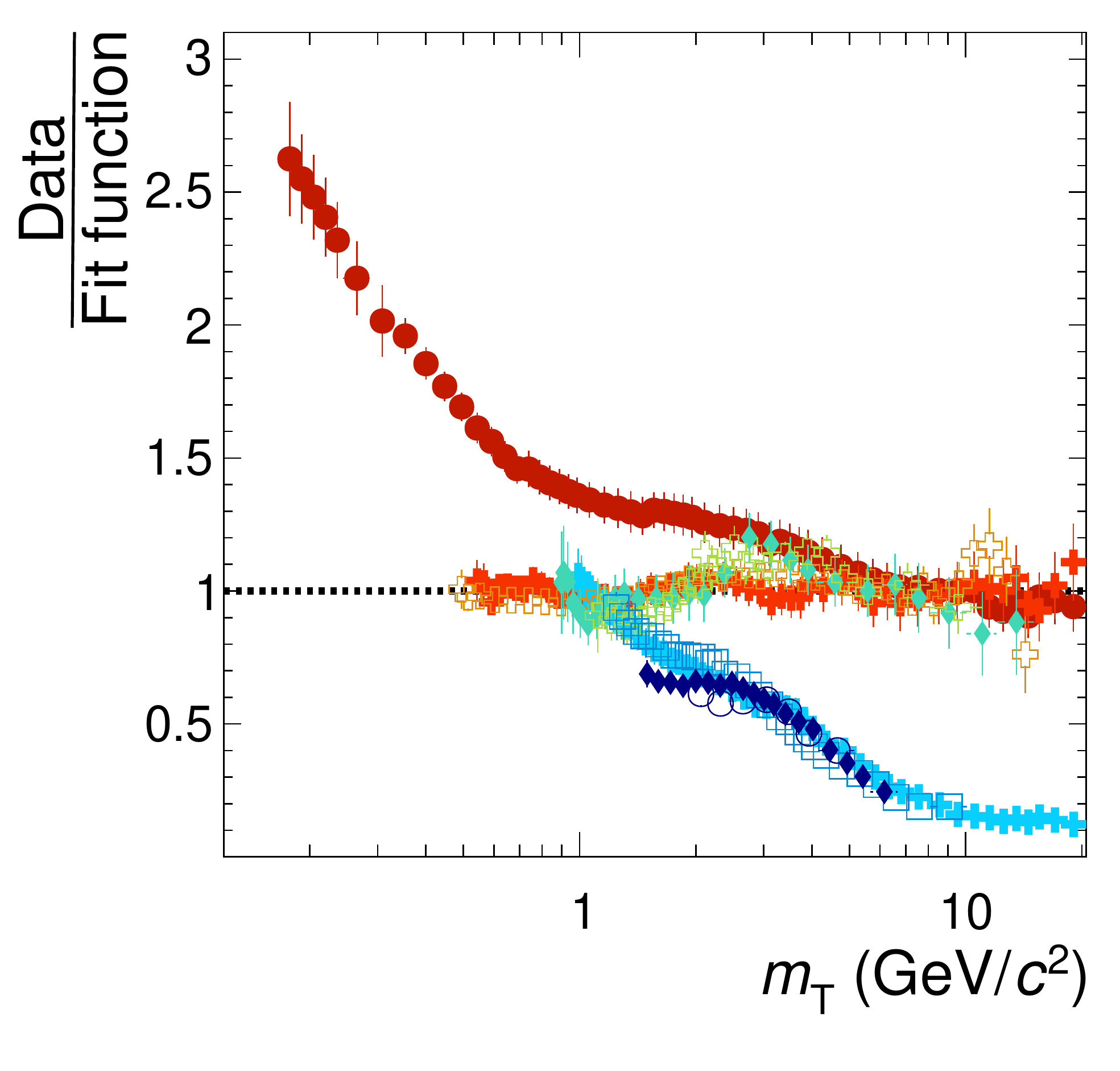}
    \caption{Upper panels: Scaled \mT spectra for identified particles in pp collisions at \rs[7~TeV] (left) and 13~TeV (right). 
    Lower panels: Ratios of the scaled \mT spectra to the \kx fit functions. The reference data for \pp at \sppt{7} are from 
    Refs.~\cite{Adam:2016dau,Abelev:2012jp,Acharya:2019wyb}
    }
    \label{fig:mtscale1}
    \end{figure}

    The charged kaon and (anti)proton \mT spectra are fitted separately with a modified Hagedorn function of the form $A\times(e^{-a\mT}+\mT/b)^{-n}$~\cite{Khandai:2012xx}. The meson \mT spectra are then scaled by multiplicative factors so that 
    their integrals over the measured \mT ranges (or $\mT>4$~\gvcc for the pions) match the integral of the kaon fit function. Similarly, the baryon \mT spectra are scaled so that their integrals match the integral of the proton fit function over 
    their measured \mT ranges. The baryon spectra are further scaled so that all fits to the \mT spectra have the same value at $\mT=1$~\gvcc. Figure~\ref{fig:mtscale1} shows the result of this study for \pp collisions at 
    \sppt{7}~\cite{Adam:2016dau,Abelev:2012jp,Acharya:2019wyb} and 13~TeV, with the lower panels showing the ratios of the various scaled \mT spectra to the \kx fit function.

    The pion \mT spectra deviate from the trend followed by the other mesons for $\mT\lesssim2$~\gvcc, which is likely due to feed-down from resonance decays~\cite{Jiang:2013gxa}. The measured primary \pix yield contains a significant contribution 
    mostly from $\rho$ and $\omega$ decays, which, according to a recent study~\cite{Altenkamper:2017qot}, affects the low-\pt ($\lesssim\gevc{1}$) part of the spectrum with increasing importance towards higher collision energies. This is the reason 
    why the \kx \mT spectrum is used as the reference for the other mesons. A clear difference in the slope is observed between the baryon and meson spectra for $\mT\gtrsim2$~\gvcc. The separation between the meson and baryon \mT spectra may be a 
    reflection of the fact that, according to the Lund model of hadronization, meson formation via the fragmentation of strings requires the breakup to only a (quark, anti-quark) pair, while a baryon--antibaryon pair can be formed by the (diquark, anti-diquark) 
    break up of the string~\cite{ANDERSSON198245,Andersson:1984af}. The separation between the (anti)proton and meson \mT spectra becomes approximately constant for $\mT>10$~\gvcc. 

    The breaking of \mT scaling for pions at low \mT as shown in Fig.~\ref{fig:mtscale1} serves as a motivation to quantify the effect for different mesons. Following Refs.~\cite{Altenkamper:2017qot,SchaffnerBielich:2001qj}, instead of using the \mT spectra, 
    the scaling law can be better studied practically as a function of \pt, (note that the invariant yields are equal in terms of these variables). This requires changing the functional form of the invariant yield parameterization through the substitution 
    $\pt\to\sqrt{\mT^{2}-m^{2}}$, where $m$ is the particle mass. In doing so, the \pt-differential invariant yield $Y_{s\textprime}$ of a particle species $s\textprime$ can be obtained by scaling the parameterization of the yield $Y^{\rm ref}_{s}$ of 
    reference particle species $s$. When both spectra are evaluated at the same transverse mass, $(p_{\mathrm{T},s}^{2}+m_{s}^{2}) = (p_{\mathrm{T},s\textprime}^{2}+m_{s\textprime}^{2})$, the yield for species $s\textprime$ is 
    $Y_{s\textprime}(p_{\mathrm{T},s\textprime}) = c\times Y^{\rm ref}_{s}\left(\sqrt{p^{2}_{\mathrm{T},s\textprime} + m^{2}_{s\textprime} - m^{2}_{s}}\right)$. Here, $c$ is a constant offset between the yields for species $s$ and $s\textprime$, determined in the 
    \pt region where the spectral shapes for the two species are the same (i.e. at high \pt).

\begin{figure}[ht]
      \centering
	\includegraphics[keepaspectratio, width=0.95\columnwidth]{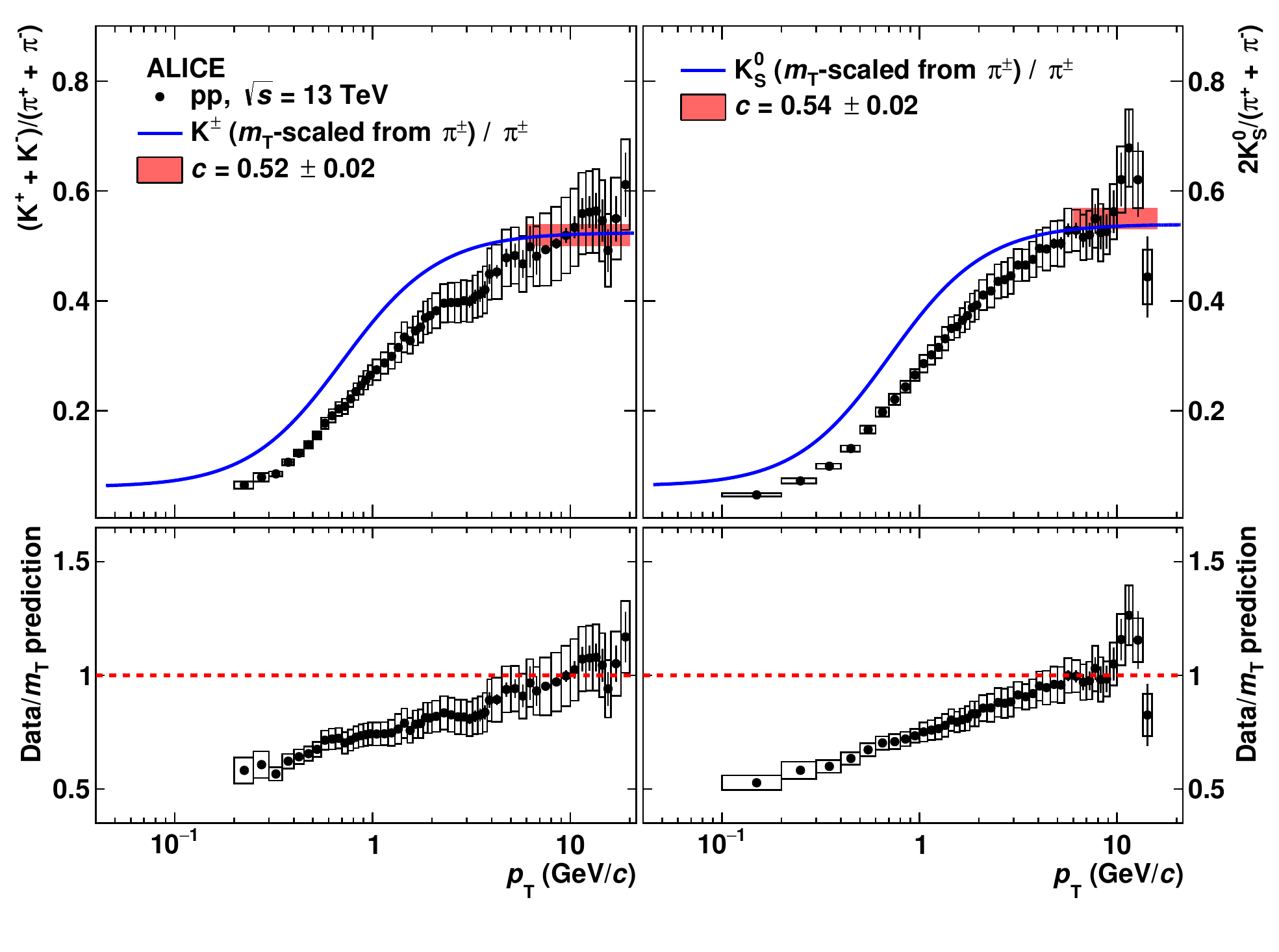}\\ \vspace{-5mm}
	\includegraphics[keepaspectratio, width=0.95\columnwidth]{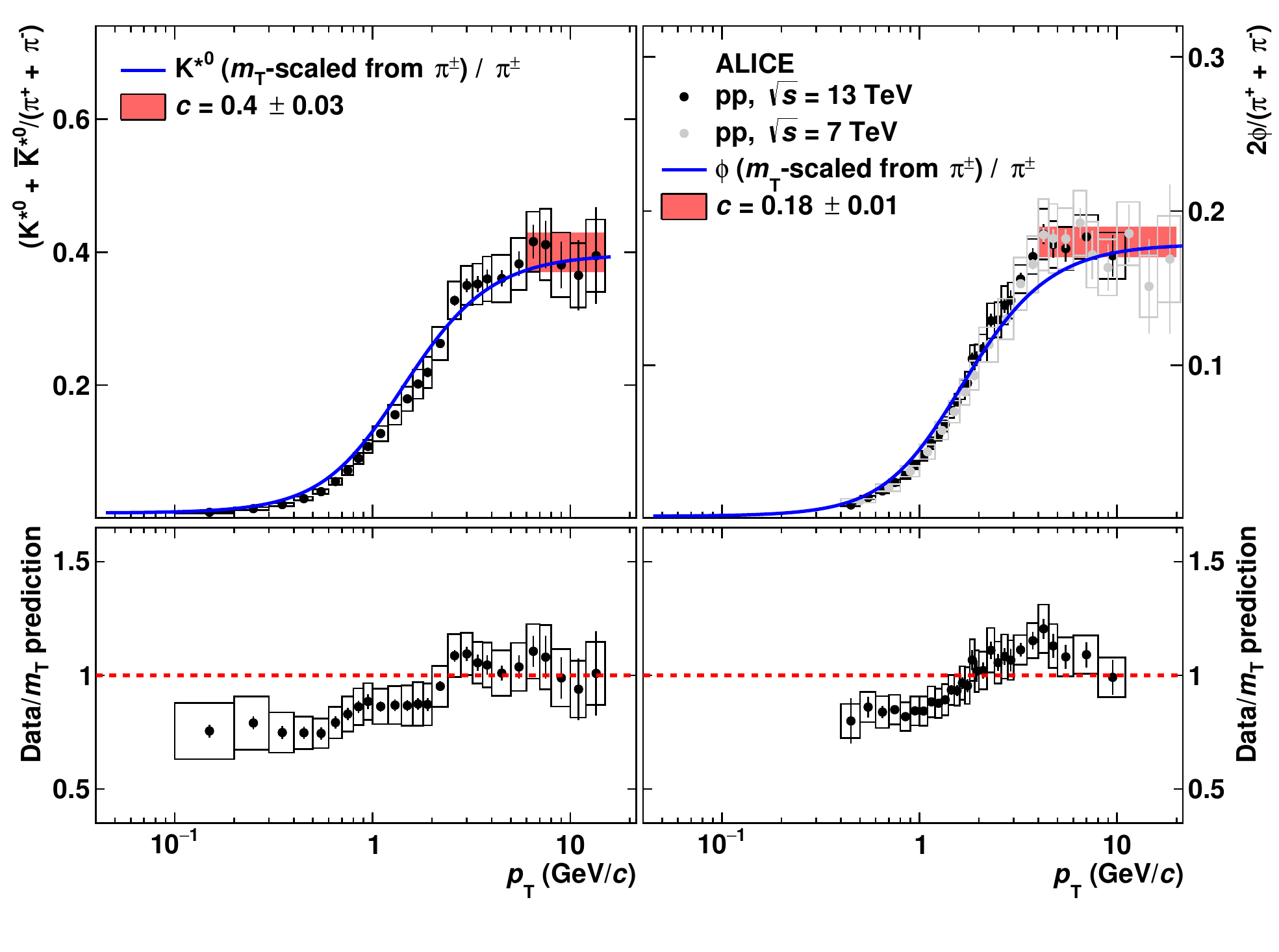} \vspace{-5mm}
	\caption{Particle ratios \kx/\pix, \kzs/\pix (top) and \ks/\pix, $\phi$/\pix (bottom) as a function of \pt measured 
	in \pp collisions at \sppt{13}. The measured ratio is reported together with that obtained from transverse 
	mass (\mT) scaling of charged pions shown as a solid line. The shaded red band indicates the constant fit, in 
	the \pt region where it was performed, which is used to determine the constant offset parameter $c$. See the 
	text for details. The \ph spectrum in \pp at \sppt{7} is from Ref.~\cite{Acharya:2019wyb}}
	\label{fig:mtscale2}
    \end{figure}

    Charged pions are used as the reference species to test the \mT scaling of the \kx/\pix, \kzs/\pix, \ks/\pix and $\phi$/\pix particle ratios. The reference \pt spectrum of \pix is parameterized with a L\'evy--Tsallis function, which is also used for 
    the extraction of the \pt-integrated yields and is found to describe the data within $15\%$ over its entire \pt range. The \mT scaling relation is then applied to the parametrized pion yield. The appropriate offset parameter $c$ is determined by 
    fitting the measured particle ratios separately in the high-\pt region where they saturate. The measured \kx, \kzs, and \ks (\ph) yield in pp collisions at \rs[13] (7)~TeV are fitted with a constant linear function in the range $\pt>\gevc{6}$ ($\pt>\gevc{4}$) 
    with statistical and systematic uncertainties summed in quadrature. The resulting $c$ values for the various particle ratios are given in the legends of Fig.~\ref{fig:mtscale2} and are shown as shaded bands in the \pt region from which they are 
    extracted. The reported widths of the bands correspond to the statistical uncertainties on $c$ obtained from the fits.
    
    Along with the measured particle yield ratios, Fig.~\ref{fig:mtscale2} shows the ratios of the \mT-scaled parameterizations to the reference parametrization (solid blue lines) for the particle species in question. A significant deviation between the 
    parameterized curves and the measured data is observed in the low-to-mid \pt region ($\pt\lesssim 5-\gevc{6}$) for the \kx/\pix and \kzs/\pix ratios, which indicates the breaking of empirical \mT scaling. In contrast, the \mT-scaling predictions for 
    the $\ks/\pix$ and $\ph/\pix$ ratios are notably closer to the measurements. In order to quantify the level of the scale breaking, the double ratios between the measured yield ratios and those based on the \mT-scaling relation are evaluated. The double 
    ratio values for \pt below the threshold of $\gevc{6}$ (i.e., below the region used for the determination of the offset parameter $c$) decrease towards lower \pt, deviating beyond $16\%$ for $\pt\lesssim\gevc{2}$. The significance of the deviation 
    from the \mT scaling hypothesis is $4.7\sigma$ at $\pt=\gevc{1}$, far from the threshold (note that adjacent \pt bins have fully uncorrelated uncertainties). This further confirms the breaking of the empirical \mT scaling relation for the quoted \pt region.

    \subsubsection{\xt scaling}\label{subsec:xTscaling}

    The validity of empirical \xt scaling is tested using the \sppt{13} \pp measurements reported here and those obtained at \sppt{2.76}~\cite{Abelev:2014laa,Adam:2017zbf} and \sppt{7}~\cite{Adam:2016dau,Acharya:2019wyb}. Due to the lack of a complete 
    set of measurements of hadron yields at high \pt at these three collision energies, only \pix, \kx, \ks, and \px are used.

    The invariant cross sections are determined from the measured invariant yields as $E\,\mathrm{d}^{3}\sigma/\mathrm{d}p^{3} = \sigma_{\rm inel}\times\,E\,\mathrm{d}^{3}N/\mathrm{d}p^{3}$, where $\sigma_{\rm inel}$ is the inelastic cross section in \pp 
    collisions at \sppt{13}~\cite{Loizides:2017ack}. The logarithm of the ratio of the invariant cross sections at two different collision energies, scaled by the logarithm of the ratio of the two collision energies is calculated. This quantity, denoted as $n$ 
    in the following, depends on \xt and \cme~\cite{PhysRevD.11.1199}. It increases with \xt in the low \xt region, where particle production is dominated by soft processes, and appears to saturate in the high \xt region. 
    
    The \xt scaling exponents $n$ as a function of \xt are extensively studied in Ref.~\cite{Arleo:2009ch} for unidentified hadrons and pions using the data from the E706~\cite{Apanasevich:2002wt}, PHENIX/ISR~\cite{Adare:2008qb,Arleo:2008zd}, PHENIX~\cite{Adare:2008qb,Adare:2007dg}, 
    UA1~\cite{Albajar:1989an} and CDF~\cite{Abe:1988yu,Acosta:2001rm} collaborations. 
    The value of $n$ decreases towards the high-\xt region. The measured mean exponent value was about $8.2$ in the range $\xt = 10^{-1}-4\times10^{-1}$ and reaches a value of $\approx5.2$ in the range $\xt = 10^{-2}-7\times10^{-3}$. 
    In our measurements at \sppt{13}, we are able to probe an even lower \xt region.
    
\begin{figure}[t]
	\centering
	\includegraphics[scale=0.8]{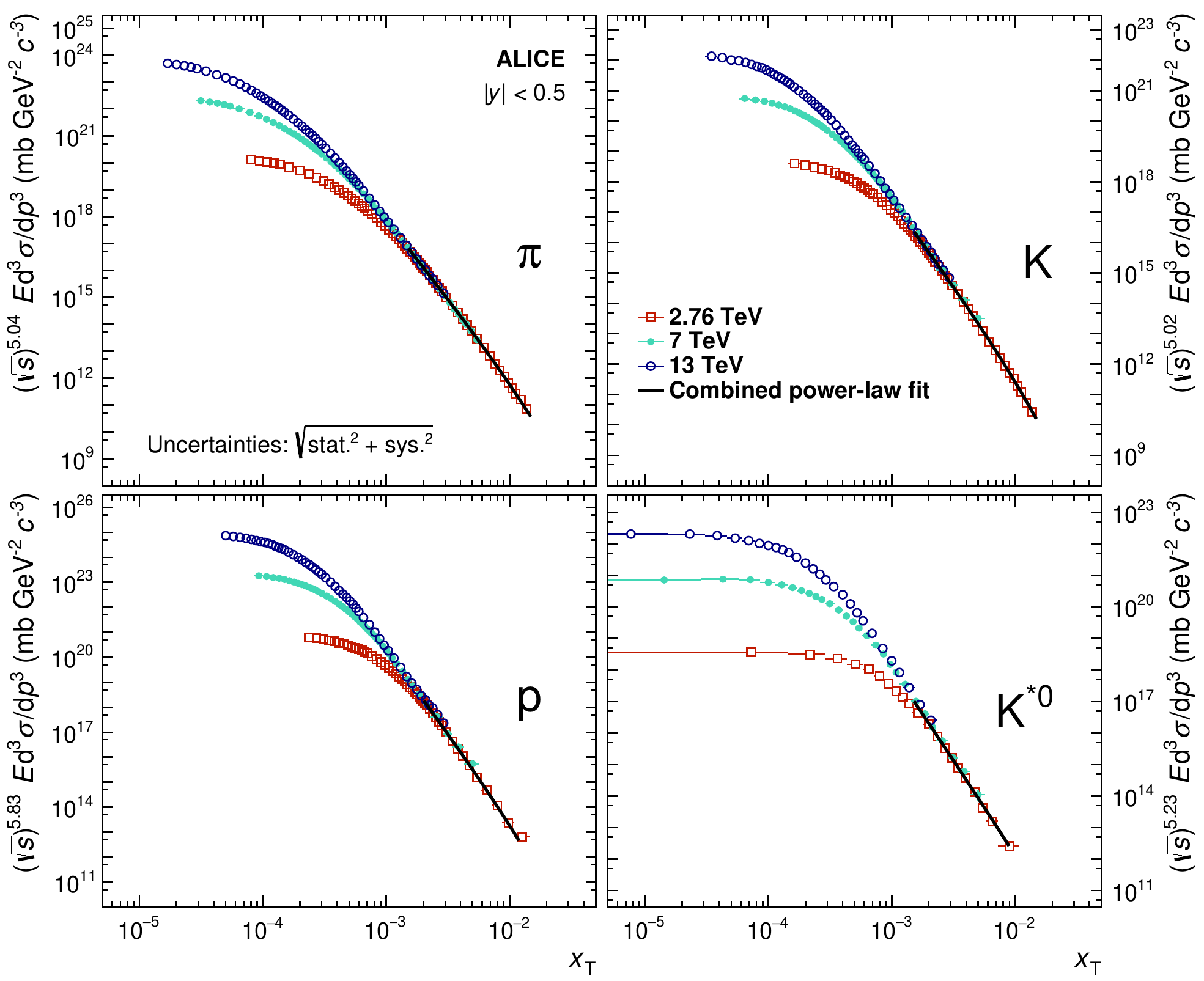}
	\caption{Scaled invariant yields of \pix, \kx, \px, and \ks as a function of $\xt=2\pt/\sqrt{s}$ 
	at different collision energies of \sppt{2.76}~\cite{Abelev:2014laa,Adam:2017zbf}, 
	\sppt{7}~\cite{Adam:2016dau,Acharya:2019wyb}, and \sppt{13}.
	The solid line represents a combined power-law fit in the high-\xt region where the distributions 
	show a scaling behavior.
	}
	\label{fig:xtscale}
	\end{figure}

    Each $n(\xt,\cme)$ distribution is 
    fitted with a constant in the range $2\times10^{-3}\leq\xt\leq6\times10^{-3}$ to obtain the respective $n$ values for different energy combinations; these are then averaged to obtain the mean value $\langle n\rangle$ for each particle species. The \xt spectra 
    for different particle species are scaled by $(\sqrt{s}/\text{GeV})^{\langle n\rangle}$. Within the quoted \xt range, the best scaling is achieved with the exponents $\langle n\rangle= 5.04\pm0.02$ for \pix, $\langle n\rangle = 5.02^{+0.21}_{-0.25}$ for 
    \kx, $\langle n\rangle = 5.83^{+0.13}_{-0.21}$ for \px, and $\langle n\rangle= 5.23\pm0.15$ for \ks. The uncertainties on the $\langle n\rangle$ values are the maximum observed deviations from the mean. The scaling exponents for the meson spectra are 
    found to be consistent within systematic uncertainties, indicating that ratios of meson spectra at high \pt are constant and attain similar values for all beam energies. Figure~\ref{fig:xtscale} shows the \xt-scaled spectra for \pix, \kx, \px, and \ks 
    at collision energies of \sppt{2.76}~\cite{Abelev:2014laa,Adam:2017zbf}, 7\,TeV~\cite{Adam:2016dau,Acharya:2019wyb} and 13\,TeV.
    
    These results suggest that identified particle yields at LHC energies follow \xt scaling above $x_{\mathrm{T}}{\sim}10^{-3}$. The measured \xt scaling exponent at LHC energies is consistent with the \xt-dependent trend of the exponent values~\cite{Arleo:2009ch}. 
    The value of the parameter $n$ is found to be reduced at LHC energies compared to lower collision energies~\cite{Adare:2008qb,Arleo:2008zd,Apanasevich:2002wt,Adare:2007dg,Abe:1988yu,Acosta:2001rm,Albajar:1989an,Alper:1975jm,Adler:2003pb,Antreasyan:1978cw,Banner:1982zw,Adams:2006nd}, 
    which is attributed to the increasing importance of hard scattering processes at higher \cme, as suggested by Fig.~\ref{fig:Fig_Yield_Ratio_to_276TeV}. 
    
    It is also interesting to note that the exponent $n$ takes on larger values for baryons than for mesons in the investigated \xt range. On the one hand, this is connected to the decrease of the \ptopisum ratio with increasing \pt (see Fig.~\ref{fig:BaryonToMeson_vs_pT}), 
    as opposed to the constant behavior of the \ktopisum and \kstopisum ratios (see Fig.~\ref{fig:mtscale2}), which suggests that meson spectra are harder than baryon spectra in the corresponding \pt range. On the other hand, as discussed in Ref.~\cite{Arleo:2009ch}, 
    the NLO pQCD predictions including higher-twist processes, i.e. in which the detected hadron can be exclusively produced in the hard subprocess reaction, there is evidence for the larger value of the exponent for baryons than for mesons. This is in contrast to 
    the observations based on the leading-twist processes, where the exponent $n$ has only a weak dependence on hadron species.

    The quality of the scaling behavior is verified by combined fitting the differential cross sections with a power-law function of the form $a\times\xt^{b}\times(1+\xt)^{c}$. Here, $a$, $b$, and $c$ are free parameters and the region below 
    $\xt=1.5\times10^{-3}$ ($\xt=2\times10^{-3}$ for protons) is excluded to avoid the dominant contribution from soft particle production, which does not follow \xt scaling. The fits are of good quality with $\chi^{2}/\rm{ndf}$ values in the range $0.4-1.5$. 
    In spite of the naive assumption of a power law function and the expected non-scaling behaviors discussed in Ref.~\cite{Sassot:2010bh}, the measurements agree with the global power law fits within the region of overlap ($2\times10^{-3}\lesssim\xt\lesssim6\times10^{-3}$) 
    within roughly $40\%$, depending on particle species. The measurements from ALICE performed at \sppt{13} are consistent over the accessible \xt range ($2\times10^{-3}\leq\xt\leq6\times10^{-3}$) with empirical \xt scaling and with measurements from \pp collisions at \rs[2.76] and 7~TeV.

	\begin{figure}[t]
	\centering
	\includegraphics[keepaspectratio, width=\linewidth]{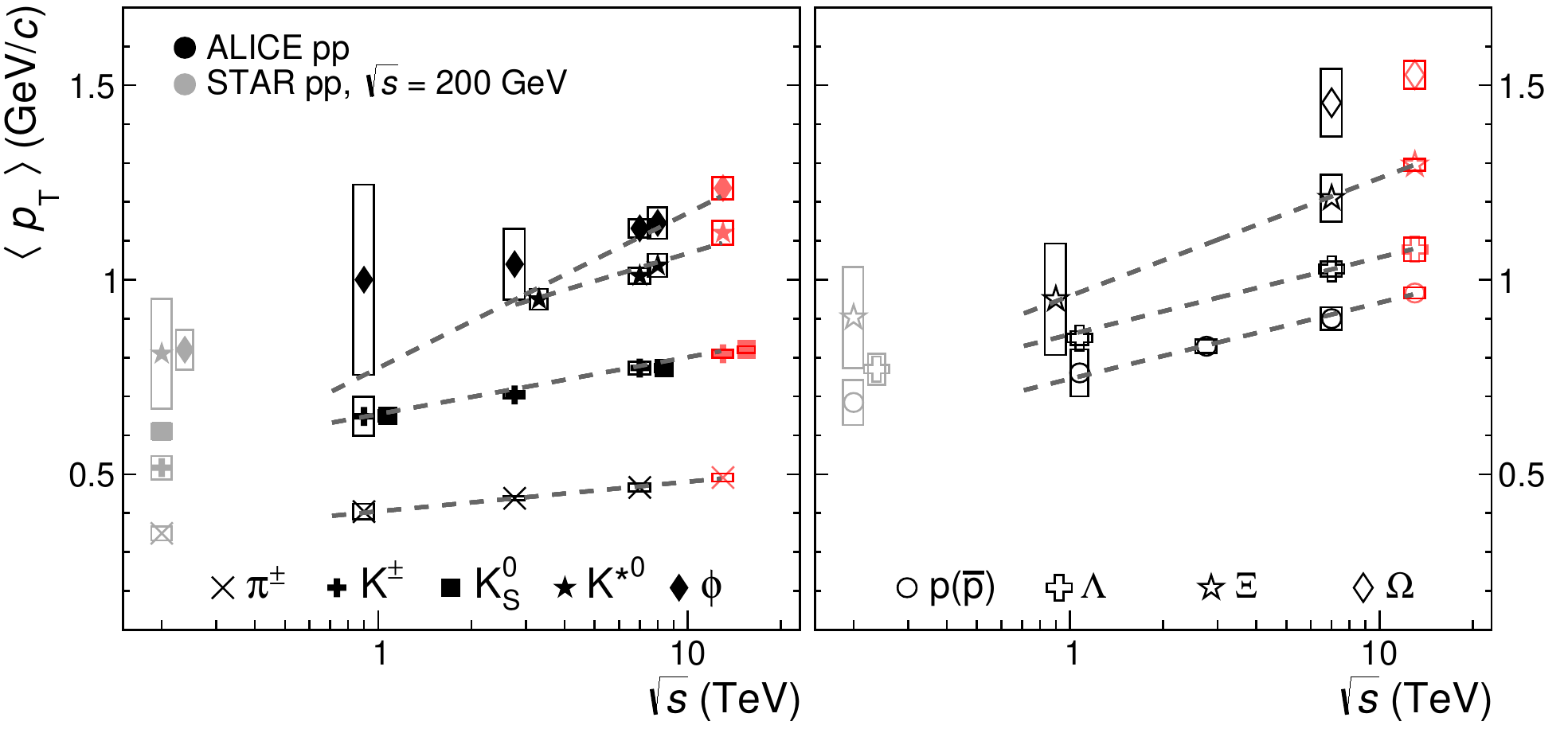}
	\caption{
	Average transverse momentum \mpt as a function of the center-of-mass energy. Open boxes indicate the statistical 
	and systematic uncertainties (when available) summed in quadrature. Results from ALICE~\cite{Aamodt:2011zj,Abelev:2014laa,Adam:2015qaa,Aamodt:2011zza,Adam:2017zbf,Abelev:2012jp,Abelev:2012hy,Acharya:2019wyb} 
	are compared with those from STAR measured at \sppg{200}~\cite{Abelev:2008ab}. Some data points are slightly offset 
	from their true energy for better visibility. Dashed curves show linear fits in $\ln{s}$. Note that the data points 
	of \kzs are not fitted due to their very similar values of \mpt to those of charged kaons. Red data points are for \sppt{13}.
	}
	\label{fig:AVGPT_vs_collisionEnergy}
	\end{figure}

	\subsection{Excitation functions} \label{subsec:excitationFunc}

    Figure~\ref{fig:AVGPT_vs_collisionEnergy} compiles the excitation functions of the average transverse momenta \mpt for light-flavor hadrons in inelastic \pp collisions; the focus is on the LHC energy regime, but some RHIC results are also shown. 
    As discussed in Ref.~\cite{dEnterria:2016oxo}, the measurement of \mpt as a function of the collision energy is particularly useful in probing the saturation scale of the gluons inside the proton.

	Results at midrapidity are presented from \sppg{200} up to the top LHC energy \sppt{13}, spanning nearly two orders of magnitude in center-of-mass energy. The average \pt increases with \cme; it rises steeper for heavier particles, as seen in 
	our earlier measurements at lower collision energies~\cite{Adam:2015qaa,Abelev:2012jp,Abelev:2012hy,Acharya:2019wyb}. The growth of \mpt with \cme is attributed to the increasing importance of hard processes for higher collision energies. 
	For single- and multi-strange hadrons, this observation is equivalent to the hardening of \mpt as the collision energy increases from \rs[7] to 13~TeV for event classes with a similar \dndetainline, as reported in Ref.~\cite{Acharya:2019kyh}.
    
	\begin{figure}[t]
	\centering
	\includegraphics[keepaspectratio, width=\columnwidth]{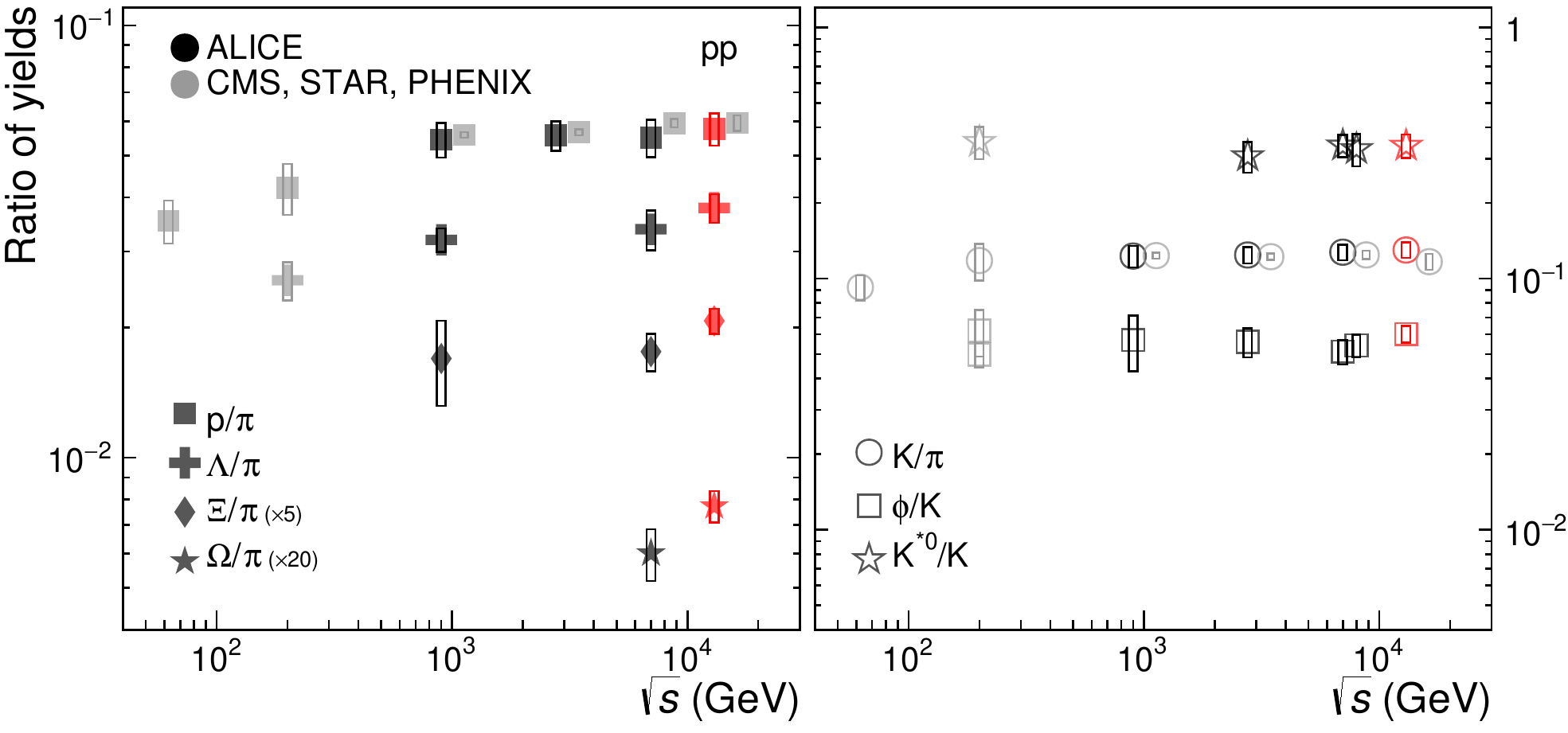}
	\caption{Ratio of particle yields to \pix and \kx as a function of the collision energy \cme (GeV). 
	Results from ALICE~\cite{Aamodt:2011zj,Abelev:2014laa,Adam:2015qaa,Aamodt:2011zza,Adam:2017zbf,Abelev:2012jp,Abelev:2012hy,Acharya:2019wyb} are 
	compared with those from CMS~\cite{PhysRevD.96.112003} at the same energies, and with those from STAR~\cite{Abelev:2008ab,Adams:2004ux,Adams:2003xp} 
	at \sppg{200} and PHENIX~\cite{Adare:2011vy,Adare:2010fe} at \sppg{62.4} obtained at BNL RHIC. The CMS data points are slightly shifted horizontally for clarity. 
	Open boxes represent statistical and systematic uncertainties summed in quadrature. Red data points are ALICE measurements for \sppt{13}.
	}
	\label{fig:RatioToPion_vs_Energy}	
	\end{figure}	
	
    It is worth noting that the proton and \lap have different \mpt values than the \ks and \ph, despite the similar masses of these particles. In particular, regardless of the considered \cme, the average transverse momenta are greater for those 
    resonances than for protons or \lap, with the \ph having the largest value. This clearly indicates a violation of mass ordering among these particles. It is noteworthy that in our measurements at \sppt{7} as a function of charged-particle 
    multiplicity, the difference in \mpt between \px and \ks (and \ph) increases with multiplicity~\cite{Acharya:2018orn}. 
    
    It is of particular importance to study the \pt-integrated particle ratios as a function of collision energy, which might provide more information on hadron production mechanisms and their dependence on the collision energy or charged particle 
    multiplicity. Given the observation of an increase in (multi-)strange hadron production as a function of multiplicity in \pp collisions at \sppt{7}~\cite{ALICE:2017jyt}, an increase in these yields with collision energy would also be expected.
    Figure~\ref{fig:RatioToPion_vs_Energy} shows the ratios of particle yields to the yields of pions and kaons as a function of the collision energy; results from ALICE are compared with those from CMS~\cite{PhysRevD.96.112003} and with 
    STAR~\cite{Abelev:2008ab,Adams:2004ux,Adams:2003xp} and PHENIX~\cite{Adare:2011vy,Adare:2010fe} results from RHIC measured at \rs[200] and 62.4~GeV. All the ratios except $\Omega/\pi$ and $\Xi/\pi$ appear to saturate in the LHC energy regime.
    For the $\Omega/\pi$ and $\Xi/\pi$ ratios, the relative increases are ${\sim}29\%$ and ${\sim}18\%$, respectively; these are larger than the ${\sim}12\%$ increase of the $\lap/\pi$ ratio, indicating that the strangeness content may control 
    the magnitude of the increase in the yield ratios as the energy changes from \rs[7] to 13~TeV. 
    
    The ratios of the \ph and \ks yields to those of charged kaons ($\ph/\text{K}$ and $\ks/\text{K}$) do not exhibit any dependence on the collision energy. It is worth noting that there may be a decrease in the $\ks/\text{K}$ ratio in 
    high-multiplicity \pp collisions at \sppt{7}~\cite{Acharya:2018orn}. In nuclear collisions, such a modification of resonance yields is often described as a consequence of scattering processes during the hadron gas phase of the collision 
    system evolution.

    \subsection{Mass and baryon number effects on \mpt}
    
	\begin{figure}[t]
	\centering
	\includegraphics[keepaspectratio, width=\columnwidth]{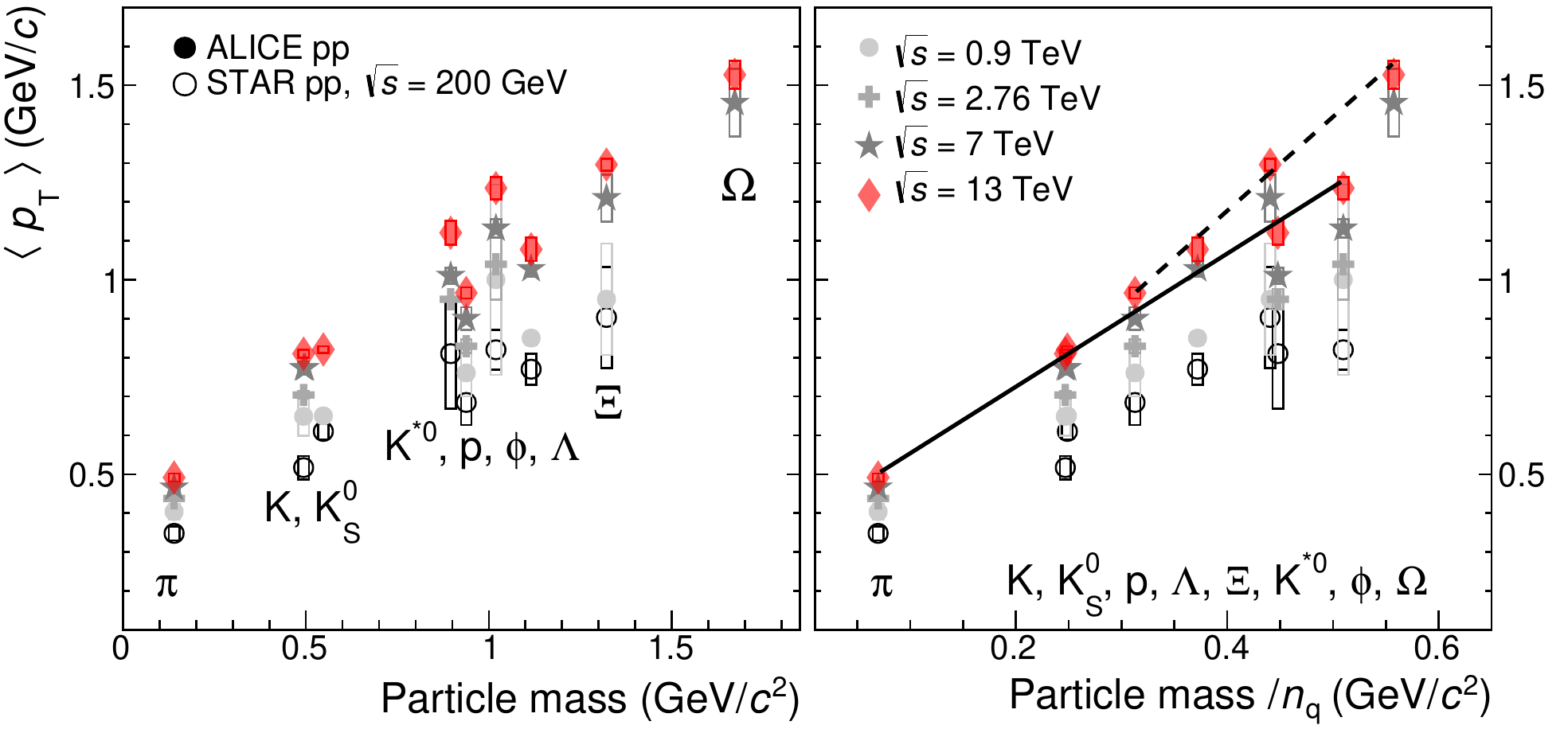}
	\caption{Average transverse momenta of light-flavor hadrons as a function of hadron mass (left) and 
	as a function of hadron mass normalized to the number of constituent quarks (right) for different collision energies (\cme). 
	Results reported from ALICE~\cite{Aamodt:2011zj,Abelev:2014laa,Adam:2015qaa,Aamodt:2011zza,Adam:2017zbf,Abelev:2012jp,Abelev:2012hy} 
	are compared with those measured by STAR at RHIC at \sppg{200}~\cite{Abelev:2008ab,Adams:2004ux,Adams:2003xp}. 
	Uncertainties that are not visible are smaller than the size of the symbol. Solid and dashed lines are drawn as visual aids and 
	represent separate linear fits to meson and baryon data at \sppt{13}. The data points for \kzs are slightly shifted horizontally for clarity. 
	Red data points are for \sppt{13}.
	}
	\label{fig:AVGPT}	
	\end{figure}	
    
    The average transverse momenta \mpt as a function of the particle mass are reported in the left panel of Fig.~\ref{fig:AVGPT} for all light-flavor hadrons under study. The \sppt{13} \pp results are compared to our earlier measurements 
    reported at lower collision energies~\cite{Aamodt:2011zj,Adam:2015qaa,Adam:2017zbf,Aamodt:2011zza,Abelev:2012jp,Abelev:2014laa,Abelev:2012hy}, and to those measured at \sppg{200} by the STAR Collaboration at RHIC~\cite{Abelev:2008ab,Adams:2004ux,Adams:2003xp}. At \sppt{13}, two different 
    linear trends can be observed for mesons and baryons separately, reflecting the violation of mass ordering in \mpt.
    
    The observation of scale breaking between mesons and baryons as a function of the transverse mass, as discussed in Sec.~\ref{subsec:mTscaling}, might lead to a violation of mass ordering in the average transverse momenta of the produced 
    particles. For particles with similar masses (like \ks, p, \ph, and \lap), the meson spectra will be harder and the \mpt values greater in comparison to the baryons. Furthermore, the softer pion \mT spectra could lead to a deviation of 
    the pion \mpt values from the trend observed for the other mesons.    
    
    For $\cme<1$~TeV all the considered hadrons appear to exhibit scaling of \mpt with the reduced hadron mass $m/n_{q}$, i.e. the mass normalized by the number of constituent quarks $n_{q}$. This is observed in \pp collisions at \sppg{900} 
    at ALICE and is shown in the right panel of Fig.~\ref{fig:AVGPT}. Reference~\cite{Ortiz:2015cma} suggests that such scaling holds even for multi-strange baryons ($\Xi$, $\Omega$) when the average pseudorapidity density of charged particles 
    measured at mid-pseudorapidity is small ($\dndetainline=3.81\pm0.01~(\text{stat.})\pm0.07~(\text{syst.})$~\cite{Aamodt:2010pp}). At higher collision energies, the scaling behavior is obviously broken. Two separate trends are observed at 
    \sppt{13}: one for mesons (slope $1.71\pm0.06$) and one for baryons (slope $2.42\pm0.15$).

    \subsection{Yield ratios} \label{subsec:partratio}

\begin{figure}[t]
	\centering
	\includegraphics[keepaspectratio, width=\columnwidth]{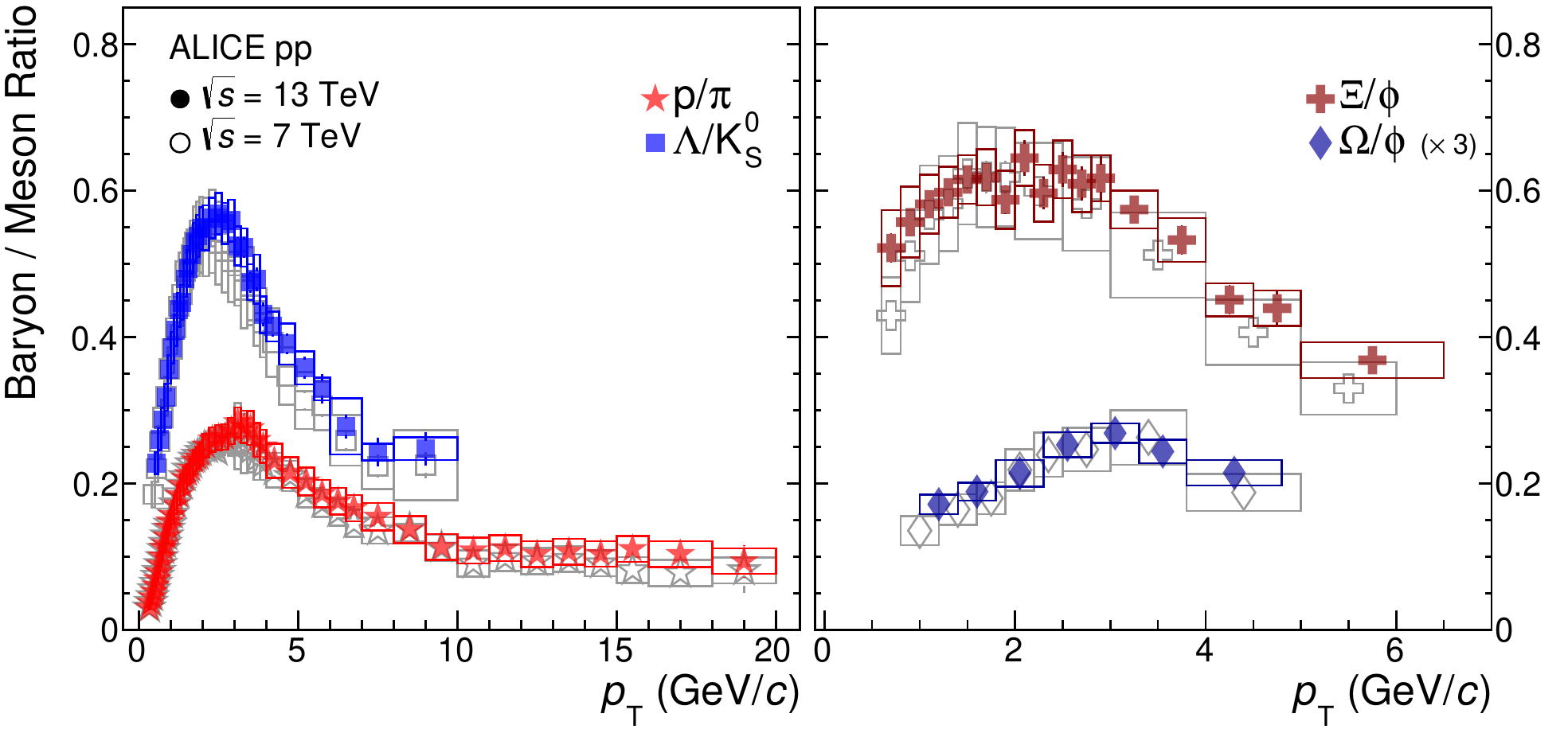}
	\caption{Baryon-to-meson ratios as a function of transverse momentum (\pt) measured in \pp collisions at 
	\rs[13] (full symbols) and 7~TeV~\cite{Adam:2016dau,Abelev:2012jp} (open symbols). Note the different scale of 
	the horizontal axis in the right panel.
	}
	\label{fig:BaryonToMeson_vs_pT}	
	\end{figure}

    The ratios of hadron yields are investigated as a function of transverse momentum. This allows the \pt spectra of different particle species, characterized by their unique mass and quark content, to be compared. Additionally, 
    measurements at different \cme are included, which helps to quantify any change in spectral shapes with \cme. The uncertainties related to normalization cancel in these ratios. Figure~\ref{fig:BaryonToMeson_vs_pT} shows the 
    \ptopisum, \ltokzssum (left panel), \xitophisum, and \omtophisum (right panel) baryon-to-meson ratios as a function of \pt at $\sqrt{s}=7$~\cite{Adam:2016dau,Abelev:2013xaa,Abelev:2012jp} (open symbols) and \sppt{13} (full symbols).
    The left panel includes particle ratios with baryons containing zero (\ptopisum) and one (\ltokzssum) strange valence quark, whereas the right panel collects ratios for hadrons with two (\xitophisum) and three (\omtophisum) strange quarks. 
    The \omtophisum ratio compares hadrons that consist entirely of strange valence (anti)quarks: three for $\Omega$ in and two for \ph.    
    
    At low \pt, all of the ratios increase with \pt as expected from the higher \mpt observed for higher mass particles. In this \pt regime, all of the ratios at \sppt{13} show good agreement with those at \sppt{7} within their systematic 
    uncertainties, suggesting that the collision energy has no observable effect on the magnitude or shape of these yield ratios. This observation remains valid in the higher \pt region. Within the systematic uncertainties the 
    \ptopisum, \ltokzssum, \xitophisum, and \omtophisum ratios are consistent for the two collision energies. It is noteworthy that the \ptopisum ratio at $\sppt{13}$ has a hint of enhancement at intermediate \pt with respect to that at \sppt{7}, 
    however this is barely significant given the quoted uncertainties. From $\pt>\gevc{10}$ onward, the \ptopisum ratio becomes fairly constant for both collision energies.

    \subsection{Comparison to theoretical models} \label{subsec:models}

    The high-precision measurements of the \pt spectra reported in this paper are crucial inputs for the further tuning of Monte Carlo event generators and to improve the understanding of particle production mechanisms at the investigated 
    collision energies. The measurements of the light-flavor hadron species presented above are compared with Monte Carlo model predictions based on general-purpose event generators: \py~6, \py~8, and \eplhc. 
    
    The \py MC model contains a rigorous description of hard scatterings through pQCD, combined with phenomenological models for semi-hard/soft processes. The Perugia-2011 tune~\cite{Skands:2010ak} for \py~6 and the Monash 2013 tune~\cite{Skands_Monash_Tune} 
    for \py~8, which have different sets of parameters, are used. The \py~8 model uses an updated parameter set for Lund hadronization for light (and heavy) quarks. The widely-used Monash 2013 tune has an improved description of diffractive 
    processes with respect to \py~6. Both sets of parameters have been obtained from recent (2011 and 2013) analyses of minimum bias, underlying event, and/or Drell--Yan measurements in \pp collisions at \sppt{7}; the Monash 2013 tune was 
    optimized to describe early data collected by the LHC experiments as well as lower energy data. Moreover, both versions of the model have strong final-state parton interactions implemented through different color reconnection
    models~\cite{Skands:2007zg,Sjostrand:2014zea}. As a consequence of the different energy evolution of the cutoff for multiple parton interactions, the Monash tune has larger multiple parton interaction activity at a given collision energy than the Perugia tune.
    
    Conversely, the \eplhc event generator (used with the CRMC package, version 1.5.4) invokes Gribov's Reggeon Field Theory~\cite{Schuler:1993wr} for multiple scatterings; this formalism features collective hadronization with the core-corona 
    mechanism~\cite{Werner:2007bf}. After multiple scattering, the final-state partonic system consists mainly of longitudinal flux tubes that fragment into string segments. If the energy density from string segments is high enough, they fuse into 
    the so-called ``core'' region, which then evolves hydrodynamically and eventually hadronizes to form the bulk part of the system. On the other hand, in the low-density region the strings expand and eventually break via the production of 
    quark-antiquark pairs, which hadronize using the unmodified string fragmentation and form the ``corona'' region. The \eplhc model uses recent data available from LHC, which helps in reproducing minimum bias results with transverse momenta up 
    to a few GeV/$c$. 
    
    Figure~\ref{fig:models} shows the ratios of \pt spectra extracted from the \py~8, \py~6, and \eplhc models to the spectra measured at \rs[7] and 13~TeV; the models give similar descriptions of the data at both energies. For the comparison of 
    the measured data to the MC generators, the total fractional uncertainties of the data are shown, i.e. the statistical and systematic uncertainties of the measurement have been summed in quadrature. Both \py versions can generally describe 
    the shapes of particle \pt spectra with reasonable accuracy at intermediate \pt ($\gevc{2}\lesssim\pt\lesssim\gevc{10}$), but generally give softer \pt spectra than observed at low \pt ($\lesssim\gevc{2}$), although pions show the opposite trend, 
    with a harder spectrum. This low-\pt behavior is due to the fact that the MC generators are known to have difficulties at describing diffractive processes that play a role at very low \pt~\cite{Abelev:2012sea}. In contrast, at high \pt \py 
    predicts harder spectra.
    
    \afterpage{%
    \begin{figure}[t]
      \centering
      \includegraphics[keepaspectratio, width=\columnwidth]{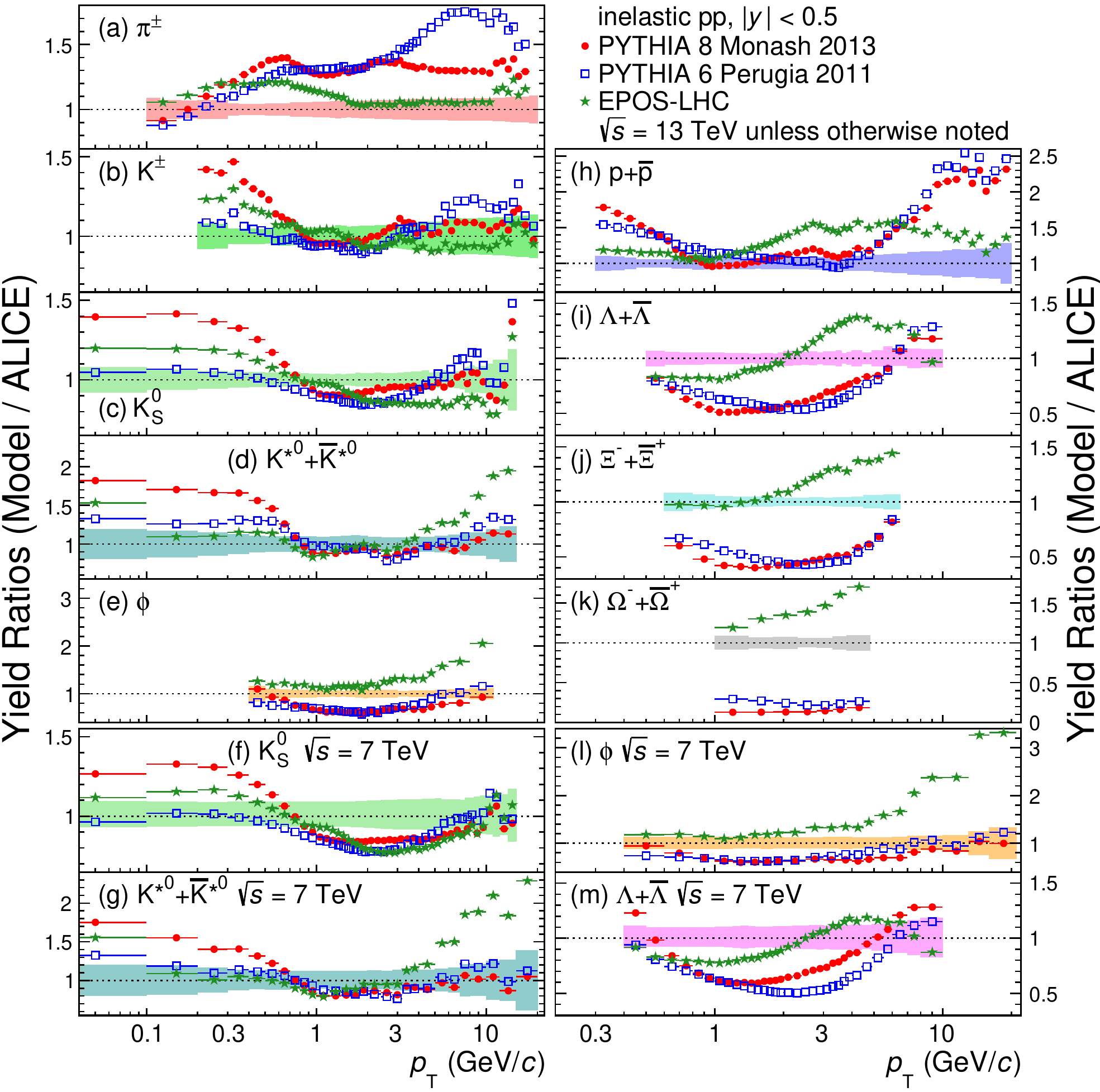}
      \caption{Ratios of \pt spectra from model calculations to the \pt spectra 
      measured by ALICE in \pp collisions at \rs[13] and 7~TeV. The total fractional uncertainties 
      of the ALICE data are shown as shaded boxes. The measured \pt spectra of \ks and \ph in 
      \pp collisions at \rs[7~TeV] are from Ref.~\cite{Acharya:2019wyb}
      }
      \label{fig:models}
    \end{figure}
    \clearpage
    }

    For the most abundant pions and charged kaons, the spectral shapes above ${\sim}\gevc{1}$ are described significantly better by \py~8 than \py~6. For both \kx and \kzs, the model-to-measured ratios for both \py versions are qualitatively similar, 
    although \py~8 provides a better agreement with the data at mid-to-high \pt than \py~6. These \py tunes describe the (anti)proton \pt spectra best for $0.8<\pt<\gevc{4}$. For single-strange (\lap) and multi-strange ($\Xi$, $\Omega$) baryons,
    both of the \py tunes underestimate the amount of produced particles in almost the entire \pt range. A similar discrepancy is observed for the case of \py~6 with the Perugia 2011 tune at \sppt{7}~\cite{Abelev:2012jp} and with older tunes 
    (D6T, ATLAS-CSC, Perugia 0) for the $\Xi$ as reported for \sppg{900} in Ref.~\cite{Aamodt:2011zza}. It is worth noting that a discrepancy is also observed when the yields of \lap, $\Xi$, and $\Omega$ measured as a function of the charged-particle 
    multiplicity~\cite{Acharya:2019kyh} are compared to results from \py and \eplhc. In all cases, the observed deviation becomes larger for hadrons with higher strange valence quark content. In Fig.~\ref{fig:models}, the model underestimates the 
    measured $\Xi$ $(\Omega)$ yields by a factor of two (four), although the discrepancy weakens for the $\Xi$ towards higher \pt values. For the \ph meson (with zero net strangeness content), \py predicts yields within $10-20\%$ for $\pt{\sim}\gevc{8}$, 
    while at higher \pt the measured yields and the \py values agree within uncertainties.

    For \pix, \kx, and \px, the \eplhc model predicts the spectral shape and normalization better than both of the \py tunes in the entire \pt range. For the resonances and multi-strange baryons, \eplhc gives harder \pt spectra than the measured ones 
    and performs better than \py at describing the yields of the multi-strange baryons. The \eplhc model gives an accurate description for the $\Xi$ baryon at $\pt\lesssim\gevc{2}$ both in shape and normalization, but deviates at higher \pt. The model 
    also describes the shape of the \pt spectrum of $\Omega$ baryons, which have higher strangeness content, but does not reproduce the yield. For the \kzs and \lap (reported here for the first time at both \rs[7] and 13~TeV), as well as \ks and \ph, 
    the models give similar descriptions at both collision energies. However, at \sppt{7} the agreement in normalization worsens for \kzs at intermediate \pt, while \eplhc predicts harder spectra at higher \pt. For the \ph meson, the \eplhc model 
    shows a marginal agreement with the data at low \pt, and monotonically deviates from the measured spectrum as the \pt increases.
    
\begin{figure}[p]
      \centering
      \includegraphics[keepaspectratio, width=\columnwidth]{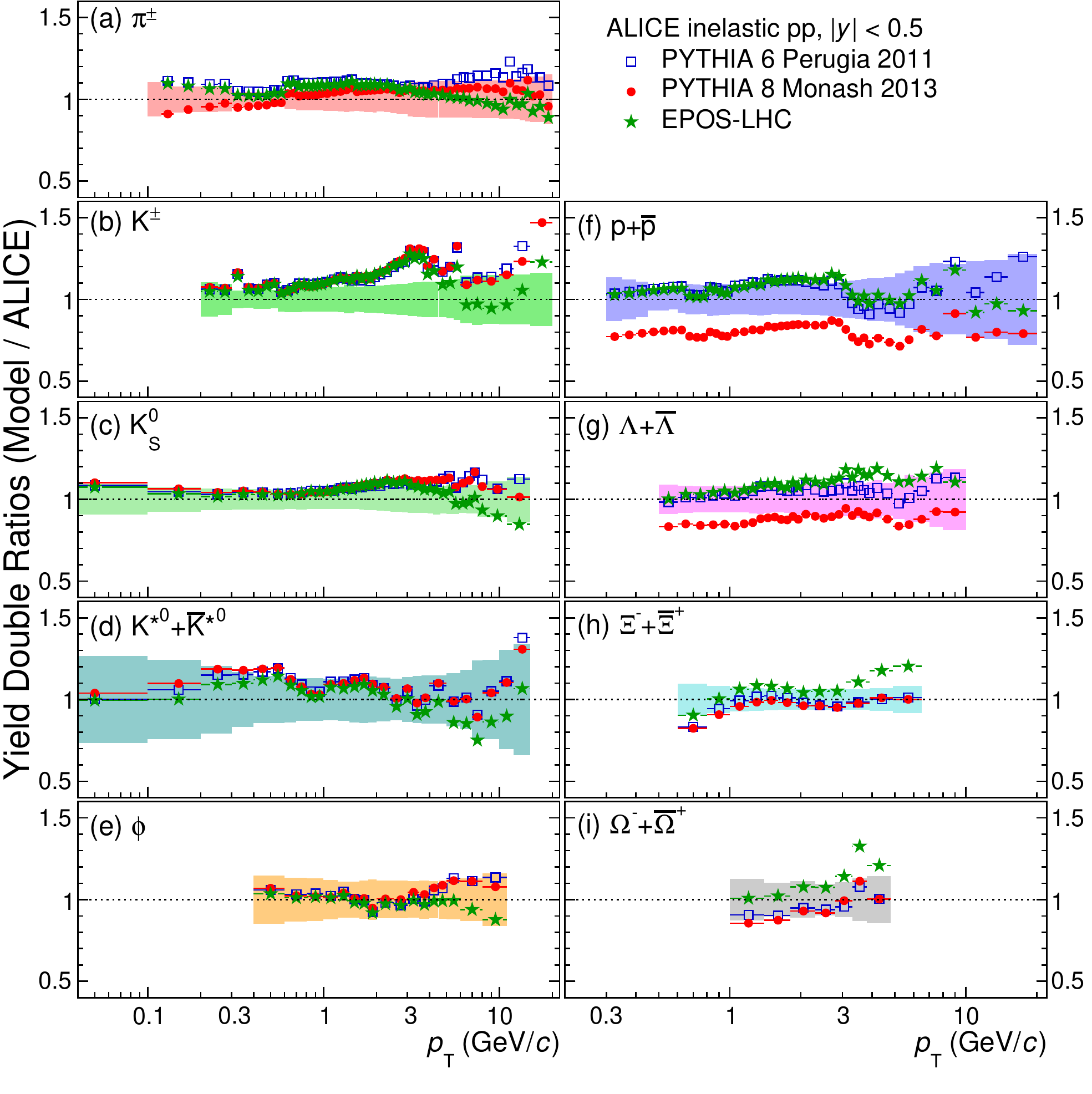}
      \caption{The measured 13-to-7~TeV ratios of hadron \pt spectra are compared with model calculations: where $Y$ is the particle yield, the plotted double ratio is 
      $(Y^{13\;\mathrm{TeV}}_{\mathrm{model}}/Y^{7\;\mathrm{TeV}}_{\mathrm{model}})/(Y^{13\;\mathrm{TeV}}_{\mathrm{measured}}/Y^{7\;\mathrm{TeV}}_{\mathrm{measured}})$. 
      The measured \pt spectra at \sppt{7} are from Refs.~\cite{Adam:2016dau,Abelev:2012jp,Acharya:2019wyb}. 
      The total fractional uncertainties of the measured yield ratios are shown as shaded boxes.
      }
      \label{fig:ratios_13vs7_MC}
    \end{figure}

    Generally, the deviations of these models from the ALICE measurements are similar to those observed at lower \cme, which were reported in Refs.~\cite{Adam:2017zbf, Abelev:2012hy} for \ks and \ph and in Ref.~\cite{Adam:2015qaa} for \pix, \kx, and \px, 
    although with a restricted \pt reach for the latter three particle species. To study how the models follow the changes in spectral shapes and normalization as a function of \pt as the energy increases from \rs[7] to 13~TeV, Fig.~\ref{fig:ratios_13vs7_MC} 
    shows the double ratios: the ratios of the measured \pt spectra at \sppt{13} to \sppt{7} (see Fig.~\ref{fig:Fig_Yield_Ratio_to_276TeV}) are divided by the same ratios obtained from MC models. The comparisons indicate that the models capture well the 
    increase of the yields with \cme for the mesons shown in panels (a)\,--\,(e); for \pix, \kx, and \kzs, such an observation between two distinct collision energies in this \pt regime is reported here for the first time. For charged kaons, the models 
    predict a more pronounced hardening with collision energy at high \pt. The tension seen in the range $2\lesssim\pt\lesssim\gevc{6}$ originates from the different analysis techniques which were used to obtain the combined spectra for \sppt{7} and at \sppt{13}. 
    Note that the normalization uncertainties at both energies are not included in the reported fractional uncertainties in the figure. For both \ks and \ph mesons, the similarity of the deviations of the \py model (though with a different tune) from 
    ALICE measurements was seen in our earlier measurements at \sppt{2.76} and \sppt{7}, as reported in Ref.~\cite{Adam:2017zbf}. Panels (f) and (g) demonstrate that the \py~6 and \eplhc models agree with the measurements of \px and \lap, in contrast, 
    \py~8 systematically underestimates the measured data for $\pt\lesssim\gevc{8}$. For the multi-strange baryons (panels (h) and (i)), \eplhc predicts the \cme evolution of the \pt spectra above $\pt=\gevc{1}$ the best, however the \py models also agree 
    with the measured data for the $\Omega$ baryons above $\pt{\sim}\gevc{3}$. For baryons, the model description for \py~8 improves as the strangeness content increases.

    \begin{figure}[t]
      \centering
      \includegraphics[keepaspectratio, width=\columnwidth]{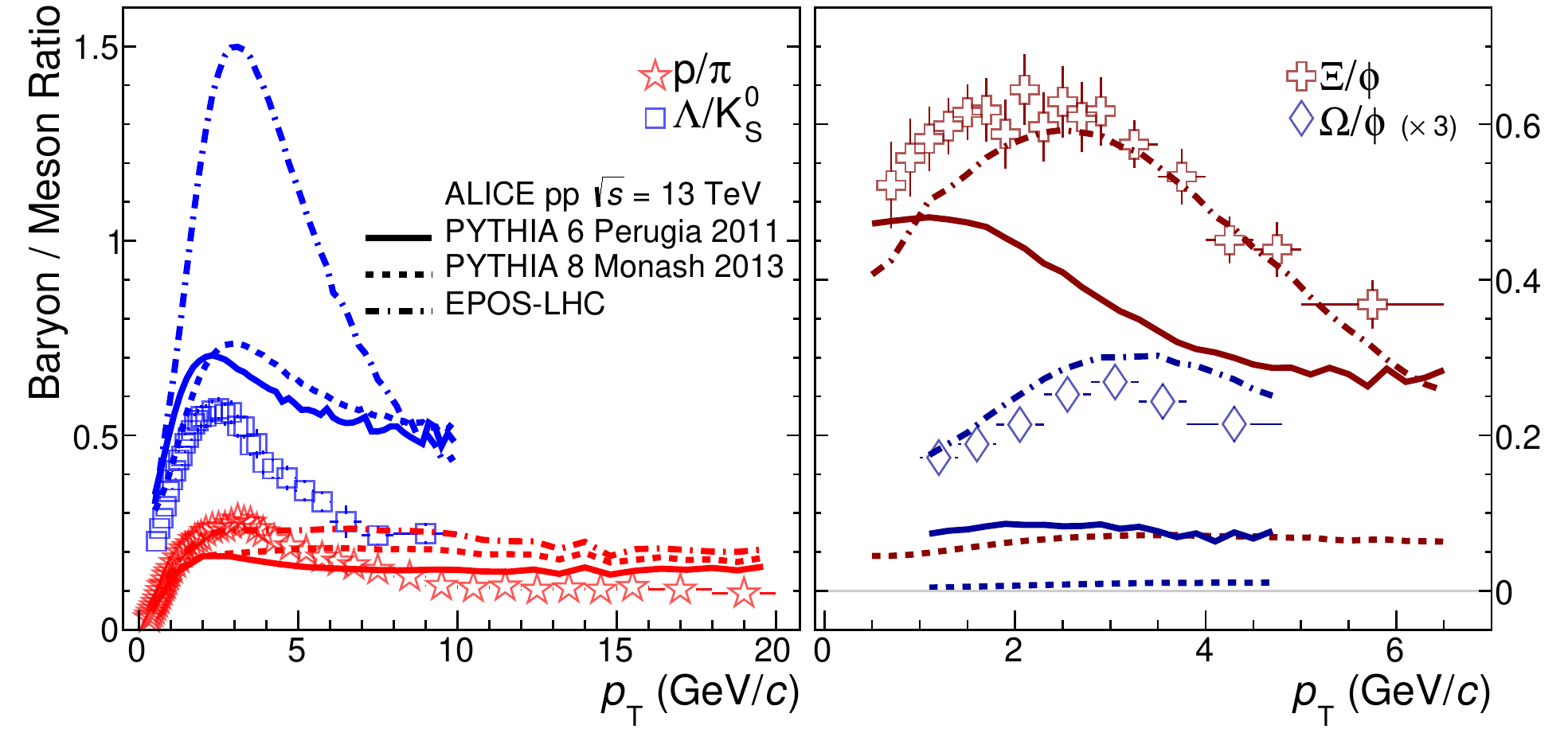}
      \caption{Baryon-to-meson particle ratios as a function of transverse momentum measured in \pp collisions at \sppt{13}. 
      The data are compared with several general-purpose Monte Carlo models. Statistical and systematic uncertainties are summed 
      in quadrature, and are shown as vertical error bars.
      }
      \label{fig1:Baryon_to_Meson_vs_pT_Canvas3x2}
      \end{figure}

    The relative impact of the hardening observed in the \pt spectra in Fig.~\ref{fig:Fig_Yield_Ratio_to_276TeV} for the various light-flavor hadrons is better seen in terms of the \pt-dependent yield ratios of different particle species. 
    The comparison of these ratios to MC models allows the different hadronization mechanisms implemented in the event generators to be tested. In Fig.~\ref{fig1:Baryon_to_Meson_vs_pT_Canvas3x2} the \pt-dependent baryon-to-meson ratios 
    measured in \pp collisions at \sppt{13} and discussed in the previous section are shown as a function of \pt and compared with the same MC models  discussed above. The ratios for multi-strange baryons are better approximated by \py~6 
    than \py~8. As discussed for multi-strange baryon measurements performed by ALICE in \pp collisions at \sppt{7}~\cite{Abelev:2012jp}, this is a consequence of the removal of the ``popcorn'' mechanism~\cite{Eden:1996xi} in the Perugia 2011 
    tune which in turn suppresses baryon production by favoring soft quark-antiquark pairing.

    Although the Perugia 2011 tune reproduces the shape of the $\Omega$ baryon \pt spectrum within roughly $10\%$ (as seen in Fig.~\ref{fig:models}), the $\Omega/\phi$ ratio given by this model has large deviations from the data for 
    $\pt\lesssim\gevc{4}$, the same \pt region where the model \ph meson spectrum has its largest disagreement with the measured data (about $40\%$). The Perugia 2011 tune better characterizes the \pt evolution of the $\Xi/\ph$ ratio than 
    the Monash 2013 tune of \py~8 for $\pt>\gevc{2}$. The \eplhc model gives a good description of the shapes of the multi-strange baryon-to-meson ratios in the entire \pt range of the measurements. The model predicts quantitatively the magnitude 
    of the $\Xi/\ph$ ($\Omega/\phi$) ratio with reasonable accuracy above (below) $\pt=\gevc{2}$.
    All three models approximate the \pt evolution of the \ltokzssum ratio qualitatively, but systematically over-predict its magnitude for the full \pt range. The \py~8 and \eplhc models predict the 
    maximum of the ratio at higher \pt values than measured. \eplhc fails to reproduce the magnitude of the \ltokzssum ratio, deviating from the measured ratio by about a factor of three at the maximum of the peak. The \ptopisum ratio is 
    well described by the models in the low-\pt region ($\lesssim\gevc{2}$), however they fail to follow the enhancement and depletion behavior seen in the \pt region between $\gevc{2}$ and $\gevc{6}$. All three models indicate a flattening 
    behavior above ${\sim}\gevc{10}$, similar to what is seen in the measured data.

    \subsection{Comparison to pQCD calculations} \label{subsec:pqcd}

    The measured invariant cross sections are compared to next-to-leading order (NLO) perturbative QCD calculations using CT10NLO proton PDFs~\cite{Lai:2010vv} with the DSS (de Florian, Sassot, and Stratmann) FF set~\cite{deFlorian:2007aj,deFlorian:2007ekg}. 
    For charged pions a new version of the DSS FFs is available: the DSS14 FF~\cite{deFlorian:2014xna} set. The NLO calculations are based on Ref.~\cite{Helenius:2012wd} which applies the same factorization scale value, $\mu=\pt$ for the factorization, 
    renormalization and fragmentation scales. The variation of the scales to $\mu=\pt/2$ and $\mu=2\,\pt$ gives an estimate of the theoretical uncertainty; the PDF uncertainties are negligible in comparison to the scale uncertainty. The rather 
    large scale uncertainty observed at lower \pt ($2<\pt<\gevc{10}$) stabilizes at $\pm20-30\%$ for $\pt\simeq\gevc{10}$, which is the region where the NLO calculations are trustworthy and free from non-perturbative effects.

\begin{figure}[t]
     \centering
      \includegraphics[keepaspectratio,width=0.49\columnwidth]{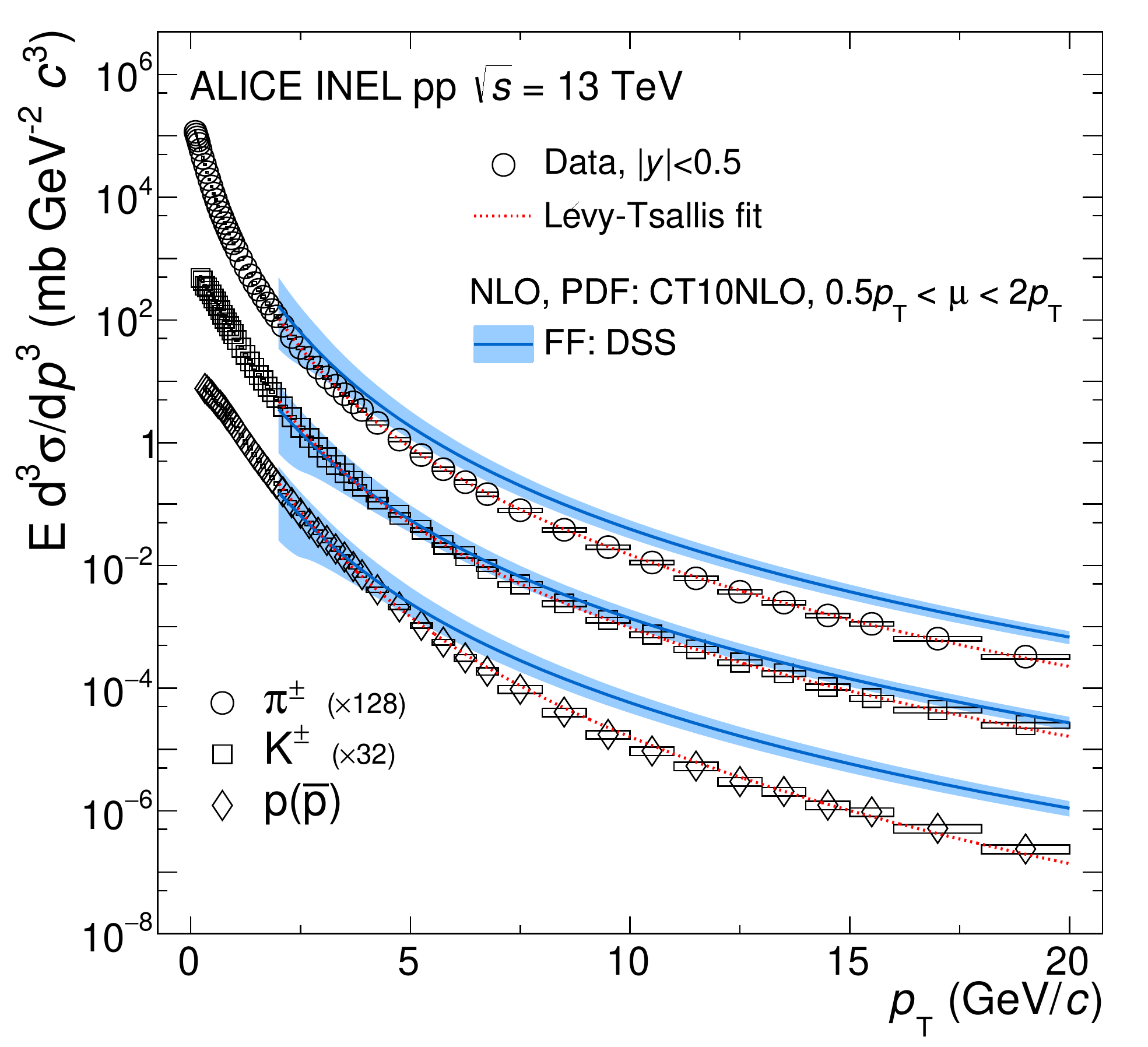}
      \includegraphics[keepaspectratio,width=0.49\columnwidth]{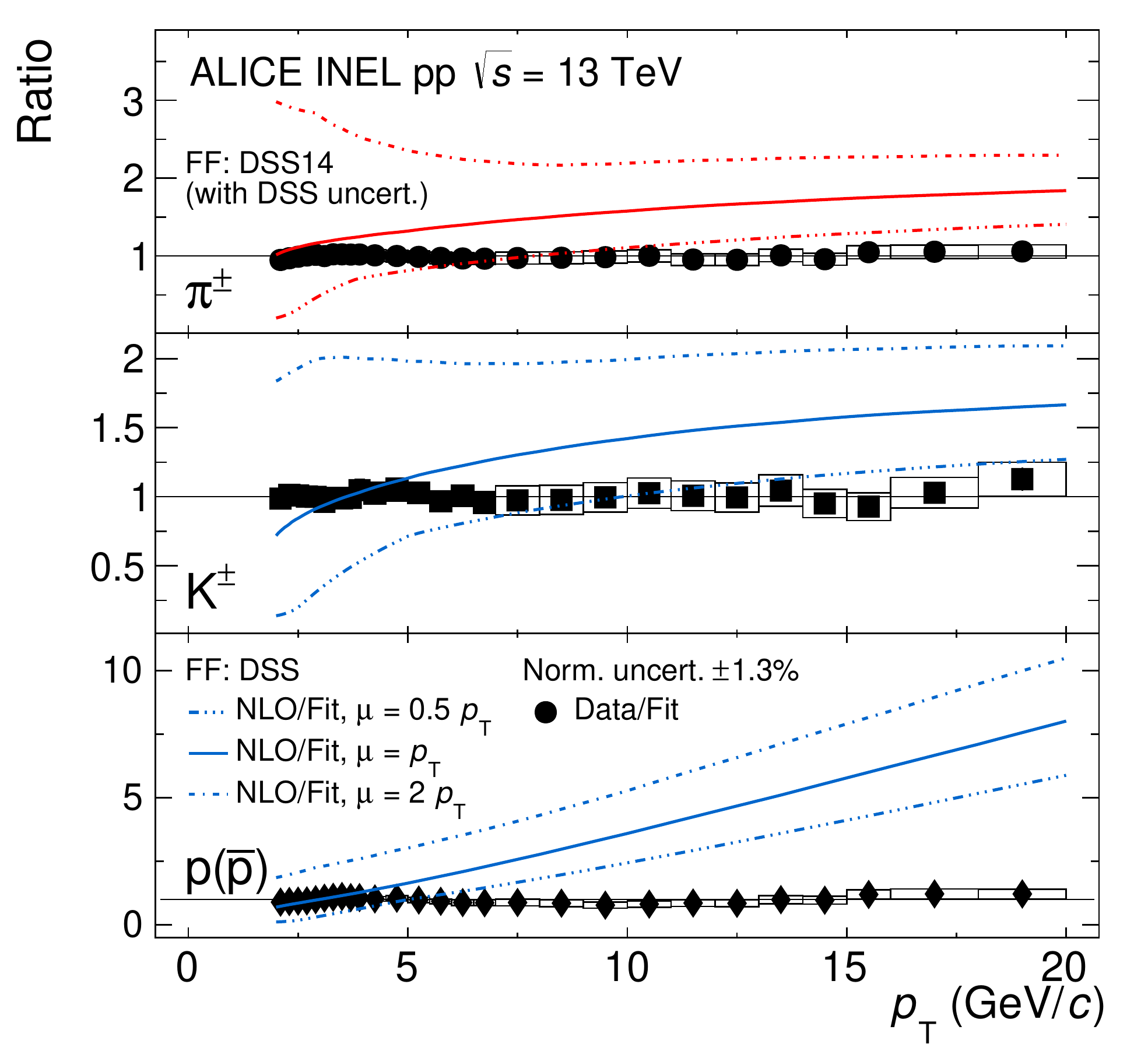}
      \caption[]{(Left) Invariant differential cross sections for \pix, \kx, and \px production, compared with NLO pQCD calculations 
      using CT10NLO PDFs~\cite{Lai:2010vv} with DSS14 FFs~\cite{deFlorian:2014xna} for \pix and DSS FFs~\cite{deFlorian:2007aj,deFlorian:2007ekg} 
      for \kx and \px. The shaded bands around the NLO calculations correspond to the scale uncertainty. (Right) Ratio of measured data (points) 
      or NLO calculations (lines) to a L\'evy--Tsallis function that is fitted to the data. The NLO-to-fit ratio is shown for the scale $\mu=\pt$ 
      and the variations $\mu=\pt/2$ and $\mu=2\,\pt$. The fully correlated normalization uncertainty is indicated in the legend.
      }
     \label{fig:YieldRatios_vs_pt_Models_Experiments_Thesis_2017}
    \end{figure}

    The production of \pix, \kx, and \px from hard scattering becomes dominated by gluon fragmentation with increasing collision energy in the \pt range of the measurement~\cite{deFlorian:2007ekg}. The presented identified charged-hadron spectra 
    can therefore help to constrain the gluon-to-charged-hadron fragmentation function~\cite{dEnterria:2013sgr} which is of crucial importance to a better description of the LHC charged-hadron data with NLO pQCD (see e.g. Ref.~\cite{Abelev:2013ala}). 
    The invariant differential cross sections for \pix, \kx, and \px are shown in the left panel of Fig.~\ref{fig:YieldRatios_vs_pt_Models_Experiments_Thesis_2017} in comparison to NLO pQCD calculations. Since, to date, no calculation exists for the 
    scale uncertainties of the DSS14, only the DSS FFs with the corresponding scale uncertainties are reported in the figure. In the right panel of Fig.~\ref{fig:YieldRatios_vs_pt_Models_Experiments_Thesis_2017} the ratios of the measured data and 
    the NLO pQCD calculations to the L\'evy--Tsallis fits of the \pix, \kx, and \px cross sections are shown. For $\pt>\gevc{10}$ the NLO pQCD calculations, employing the DSS14 FFs and using the DSS scale uncertainties, over-predict the measured 
    pion cross section by up to a factor of approximately two, but describe the shape of the \pt spectrum rather well. Similar discrepancies between NLO pQCD calculations and the measured cross sections have also been reported for the measurements 
    of neutral pions ($\pi^{0}$) at $\sqrt{s}~=~7$ and 8\,TeV~\cite{Abelev:2012cn,Acharya:2017tlv} from ALICE, leaving room for future improvements in the calculations. It is worth noting that the published $\pi^{0}$ measurement at \sppt{7}~\cite{Abelev:2012cn} 
    adds important constraints for gluon FFs, which would help reduce the FF uncertainties. The NLO calculations describe charged kaons better than pions, which is reflected in the better agreement between the calculated cross section and with the 
    measured data-to-fit ratio within the quoted uncertainties. The deviations between the NLO calculations and the data generally increase with \pt. This increase is significantly stronger for protons, which deviate the most from the measured values for all $\mu$ scale choices. The NLO calculations significantly overestimate the measured data at high \pt. 
        
    \begin{figure}[t]
      \centering
      \includegraphics[keepaspectratio, width=\columnwidth]{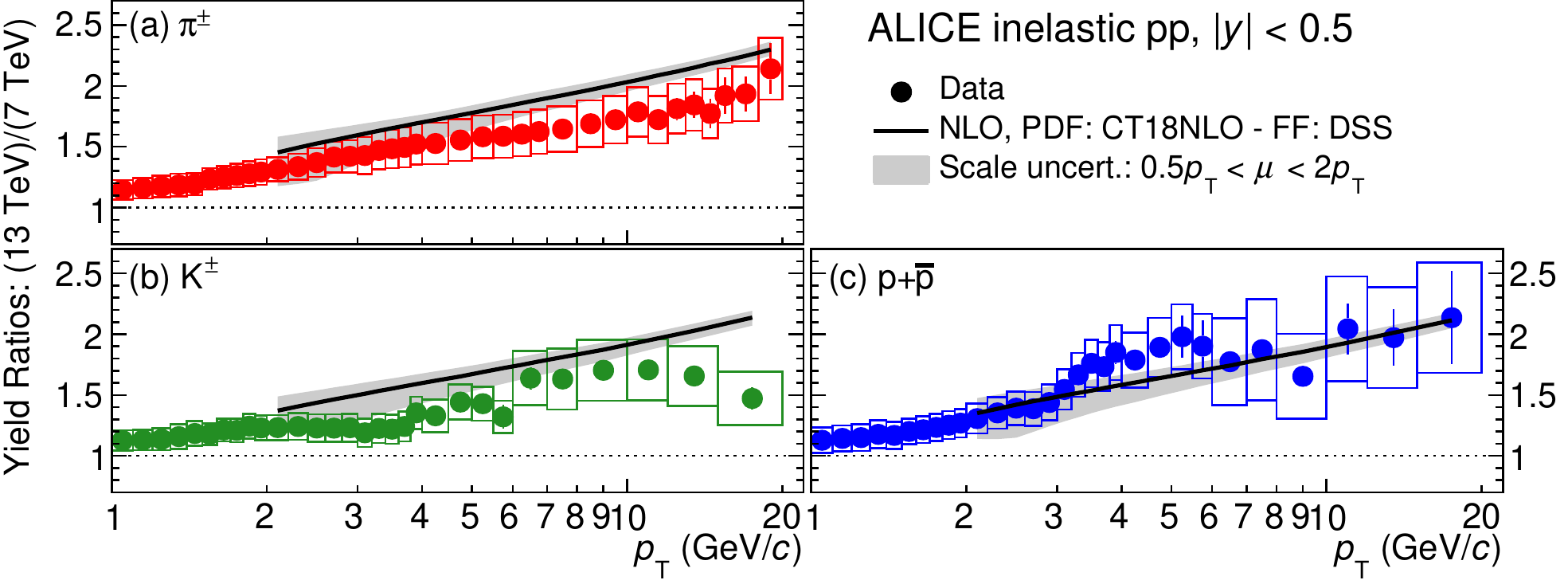}
      \caption{Ratios of transverse-momentum spectra of \pix, \kx, and \px in minimum bias inelastic \pp collisions 
      at \sppt{13} to those at \sppt{7}~\cite{Adam:2016dau}. Solid lines show predictions from a next-to-leading order 
      (NLO) pQCD calculation using DSS fragmentation functions~\cite{Helenius:2012wd}. Statistical and systematic 
      uncertainties are shown as vertical error bars and boxes, respectively.
      }
      \label{fig:ratios_13vs7_pQCD}
    \end{figure}

    At lower \pt ($<\gevc{10}$) the discrepancy between the NLO calculations and the measured data is reduced for all particle species, but at the cost of an increase of the scale uncertainties. In this \pt regime, soft parton interactions and 
    resonance decays dominate particle production, which cannot be described within the framework of pQCD. This is also reflected in the shape of the distributions, which is not described by the calculation.  The presented results show that independent 
    fragmentation works rather well for mesons, in particular for $\pt>\gevc{6}$, however for (anti)protons the spectral shape deviates towards the high-\pt region. 

    Though the \pt dependence of the cross sections at a given collision energy is not described well by the NLO calculations, the \pt dependence of the ratios of the \pt spectra at \rs[7]~\cite{Adam:2016dau} and 13~TeV is described better. Such a 
    comparison is shown in Fig.~\ref{fig:ratios_13vs7_pQCD}, where the ratio of invariant yields in inelastic \pp collisions at \rs[13~TeV] to those at 7~TeV is compared to the same ratio calculated using NLO pQCD. The agreement between the data and 
    NLO calculations is notably improved compared to the NLO results for the spectra themselves. Taking a double ratio, where the ratios of the measured spectra at \rs[13] and 7~TeV are divided by the ratios from the NLO pQCD calculations, the observed 
    difference is below $10\%$ ($20\%$) for pions (kaons and protons).


    \section{Summary} \label{sec:conclusions}

    The production of light-flavor hadrons at midrapidity was measured in inelastic \pp collisions at \sppt{13}. Additionally, single-particle \pt distributions of \kzs, \lap, and \lam were also measured in inelastic \pp collisions at \sppt{7}.
    The presented measurements complement the existing ones at lower collision energies, allowing particle production to be studied over a wide range of \cme. All \pt distributions are observed to become harder for $\pt>\gevc{2}$ with increasing 
    collision energy. 

    The \pt-integrated hadron yields normalized to the pion yields saturate as a function of \cme at LHC energies for the studied non-strange and single-strange hadrons. In contrast, a hint of an increase in the multi-strange hadron yields is 
    apparent as \cme increases from \rs[7] to 13~TeV, the increase being more pronounced for $\Omega$ baryons, which have the largest strange quark content among the studied hadrons. This observation is similar to our recent measurements performed 
    as a function of charged-particle multiplicity in pp collisions at \rs[7] and 13~TeV~\cite{Acharya:2018orn,Acharya:2019kyh}.

    A significant deviation from the empirical transverse mass scaling relation of the production cross sections between \pix and \kx and \kzs is observed for $\pt<\gevc{2}$. Empirical \xt scaling for \pix, \kx, \ks, and \px is well satisfied 
    (within roughly 20--40\%) in the hard scattering region of particle production.

    Next-to-leading order pQCD calculations performed at \sppt{13}, using the DSS14 FFs for \pix and the DSS FFs for \kx and \px, over-predict the measured \pt spectra both for charged pions and (anti)protons, suggesting that the fragmentation 
    functions are not well tuned in the accessible kinematic domain of the presented measurement.

    The measured hadron \pt spectra are compared with results of Monte Carlo calculations. The results of \py calculations only partially describe the measured data, while the \eplhc model describes several aspects of the data, notably strangeness 
    production. However, further tuning of the models is warranted in order to improve their descriptions of the measured trends.


\newenvironment{acknowledgement}{\relax}{\relax}
\begin{acknowledgement}
\section*{Acknowledgements}

The ALICE Collaboration would like to thank all its engineers and technicians for their invaluable contributions to the construction of the experiment and the CERN accelerator teams for the outstanding performance of the LHC complex.
The ALICE Collaboration gratefully acknowledges the resources and support provided by all Grid centres and the Worldwide LHC Computing Grid (WLCG) collaboration.
The ALICE Collaboration acknowledges the following funding agencies for their support in building and running the ALICE detector:
A. I. Alikhanyan National Science Laboratory (Yerevan Physics Institute) Foundation (ANSL), State Committee of Science and World Federation of Scientists (WFS), Armenia;
Austrian Academy of Sciences, Austrian Science Fund (FWF): [M 2467-N36] and Nationalstiftung f\"{u}r Forschung, Technologie und Entwicklung, Austria;
Ministry of Communications and High Technologies, National Nuclear Research Center, Azerbaijan;
Conselho Nacional de Desenvolvimento Cient\'{\i}fico e Tecnol\'{o}gico (CNPq), Financiadora de Estudos e Projetos (Finep), Funda\c{c}\~{a}o de Amparo \`{a} Pesquisa do Estado de S\~{a}o Paulo (FAPESP) and Universidade Federal do Rio Grande do Sul (UFRGS), Brazil;
Ministry of Education of China (MOEC) , Ministry of Science \& Technology of China (MSTC) and National Natural Science Foundation of China (NSFC), China;
Ministry of Science and Education and Croatian Science Foundation, Croatia;
Centro de Aplicaciones Tecnol\'{o}gicas y Desarrollo Nuclear (CEADEN), Cubaenerg\'{\i}a, Cuba;
Ministry of Education, Youth and Sports of the Czech Republic, Czech Republic;
The Danish Council for Independent Research | Natural Sciences, the VILLUM FONDEN and Danish National Research Foundation (DNRF), Denmark;
Helsinki Institute of Physics (HIP), Finland;
Commissariat \`{a} l'Energie Atomique (CEA) and Institut National de Physique Nucl\'{e}aire et de Physique des Particules (IN2P3) and Centre National de la Recherche Scientifique (CNRS), France;
Bundesministerium f\"{u}r Bildung und Forschung (BMBF) and GSI Helmholtzzentrum f\"{u}r Schwerionenforschung GmbH, Germany;
General Secretariat for Research and Technology, Ministry of Education, Research and Religions, Greece;
National Research, Development and Innovation Office, Hungary;
Department of Atomic Energy Government of India (DAE), Department of Science and Technology, Government of India (DST), University Grants Commission, Government of India (UGC) and Council of Scientific and Industrial Research (CSIR), India;
Indonesian Institute of Science, Indonesia;
Centro Fermi - Museo Storico della Fisica e Centro Studi e Ricerche Enrico Fermi and Istituto Nazionale di Fisica Nucleare (INFN), Italy;
Institute for Innovative Science and Technology , Nagasaki Institute of Applied Science (IIST), Japanese Ministry of Education, Culture, Sports, Science and Technology (MEXT) and Japan Society for the Promotion of Science (JSPS) KAKENHI, Japan;
Consejo Nacional de Ciencia (CONACYT) y Tecnolog\'{i}a, through Fondo de Cooperaci\'{o}n Internacional en Ciencia y Tecnolog\'{i}a (FONCICYT) and Direcci\'{o}n General de Asuntos del Personal Academico (DGAPA), Mexico;
Nederlandse Organisatie voor Wetenschappelijk Onderzoek (NWO), Netherlands;
The Research Council of Norway, Norway;
Commission on Science and Technology for Sustainable Development in the South (COMSATS), Pakistan;
Pontificia Universidad Cat\'{o}lica del Per\'{u}, Peru;
Ministry of Science and Higher Education, National Science Centre and WUT ID-UB, Poland;
Korea Institute of Science and Technology Information and National Research Foundation of Korea (NRF), Republic of Korea;
Ministry of Education and Scientific Research, Institute of Atomic Physics and Ministry of Research and Innovation and Institute of Atomic Physics, Romania;
Joint Institute for Nuclear Research (JINR), Ministry of Education and Science of the Russian Federation, National Research Centre Kurchatov Institute, Russian Science Foundation and Russian Foundation for Basic Research, Russia;
Ministry of Education, Science, Research and Sport of the Slovak Republic, Slovakia;
National Research Foundation of South Africa, South Africa;
Swedish Research Council (VR) and Knut \& Alice Wallenberg Foundation (KAW), Sweden;
European Organization for Nuclear Research, Switzerland;
Suranaree University of Technology (SUT), National Science and Technology Development Agency (NSDTA) and Office of the Higher Education Commission under NRU project of Thailand, Thailand;
Turkish Atomic Energy Agency (TAEK), Turkey;
National Academy of  Sciences of Ukraine, Ukraine;
Science and Technology Facilities Council (STFC), United Kingdom;
National Science Foundation of the United States of America (NSF) and United States Department of Energy, Office of Nuclear Physics (DOE NP), United States of America.    
\end{acknowledgement}


\bibliographystyle{utphys}   
\bibliography{biblio}


\newpage
\appendix

\section{The ALICE Collaboration}
\label{app:collab}

\begingroup
\small
\begin{flushleft}
S.~Acharya\Irefn{org141}\And 
D.~Adamov\'{a}\Irefn{org94}\And 
A.~Adler\Irefn{org73}\And 
J.~Adolfsson\Irefn{org80}\And 
M.M.~Aggarwal\Irefn{org99}\And 
G.~Aglieri Rinella\Irefn{org33}\And 
M.~Agnello\Irefn{org30}\And 
N.~Agrawal\Irefn{org10}\textsuperscript{,}\Irefn{org53}\And 
Z.~Ahammed\Irefn{org141}\And 
S.~Ahmad\Irefn{org16}
% ...
\And
\renewcommand\labelenumi{\textsuperscript{\theenumi}~}

\section*{Affiliation notes}
\renewcommand\theenumi{\roman{enumi}}
\begin{Authlist}
\item \Adef{orgI}Italian National Agency for New Technologies, Energy and Sustainable Economic Development (ENEA), Bologna, Italy
\item \Adef{orgII}Dipartimento DET del Politecnico di Torino, Turin, Italy
\item \Adef{orgIII}M.V. Lomonosov Moscow State University, D.V. Skobeltsyn Institute of Nuclear, Physics, Moscow, Russia
\item \Adef{orgIV}Department of Applied Physics, Aligarh Muslim University, Aligarh, India
\item \Adef{orgV}Institute of Theoretical Physics, University of Wroclaw, Poland
\end{Authlist}

\section*{Collaboration Institutes}
\renewcommand\theenumi{\arabic{enumi}~}
\begin{Authlist}
\item \Idef{org1}A.I. Alikhanyan National Science Laboratory (Yerevan Physics Institute) Foundation, Yerevan, Armenia
\item \Idef{org2}Bogolyubov Institute for Theoretical Physics, National Academy of Sciences of Ukraine, Kiev, Ukraine
\item \Idef{org3}Bose Institute, Department of Physics  and Centre for Astroparticle Physics and Space Science (CAPSS), Kolkata, India
\item \Idef{org4}Budker Institute for Nuclear Physics, Novosibirsk, Russia
\item \Idef{org5}California Polytechnic State University, San Luis Obispo, California, United States
\item \Idef{org6}Central China Normal University, Wuhan, China
\item \Idef{org7}Centre de Calcul de l'IN2P3, Villeurbanne, Lyon, France
\item \Idef{org8}Centro de Aplicaciones Tecnol\'{o}gicas y Desarrollo Nuclear (CEADEN), Havana, Cuba
\item \Idef{org9}Centro de Investigaci\'{o}n y de Estudios Avanzados (CINVESTAV), Mexico City and M\'{e}rida, Mexico
\item \Idef{org10}Centro Fermi - Museo Storico della Fisica e Centro Studi e Ricerche ``Enrico Fermi', Rome, Italy
\item \Idef{org11}Chicago State University, Chicago, Illinois, United States
\item \Idef{org12}China Institute of Atomic Energy, Beijing, China
\item \Idef{org13}Comenius University Bratislava, Faculty of Mathematics, Physics and Informatics, Bratislava, Slovakia
\item \Idef{org14}COMSATS University Islamabad, Islamabad, Pakistan
\item \Idef{org15}Creighton University, Omaha, Nebraska, United States
\item \Idef{org16}Department of Physics, Aligarh Muslim University, Aligarh, India
\item \Idef{org17}Department of Physics, Pusan National University, Pusan, Republic of Korea
\item \Idef{org18}Department of Physics, Sejong University, Seoul, Republic of Korea
\item \Idef{org19}Department of Physics, University of California, Berkeley, California, United States
\item \Idef{org20}Department of Physics, University of Oslo, Oslo, Norway
\item \Idef{org21}Department of Physics and Technology, University of Bergen, Bergen, Norway
\item \Idef{org22}Dipartimento di Fisica dell'Universit\`{a} 'La Sapienza' and Sezione INFN, Rome, Italy
\item \Idef{org23}Dipartimento di Fisica dell'Universit\`{a} and Sezione INFN, Cagliari, Italy
\item \Idef{org24}Dipartimento di Fisica dell'Universit\`{a} and Sezione INFN, Trieste, Italy
\item \Idef{org25}Dipartimento di Fisica dell'Universit\`{a} and Sezione INFN, Turin, Italy
\item \Idef{org26}Dipartimento di Fisica e Astronomia dell'Universit\`{a} and Sezione INFN, Bologna, Italy
\item \Idef{org27}Dipartimento di Fisica e Astronomia dell'Universit\`{a} and Sezione INFN, Catania, Italy
\item \Idef{org28}Dipartimento di Fisica e Astronomia dell'Universit\`{a} and Sezione INFN, Padova, Italy
\item \Idef{org29}Dipartimento di Fisica `E.R.~Caianiello' dell'Universit\`{a} and Gruppo Collegato INFN, Salerno, Italy
\item \Idef{org30}Dipartimento DISAT del Politecnico and Sezione INFN, Turin, Italy
\item \Idef{org31}Dipartimento di Scienze e Innovazione Tecnologica dell'Universit\`{a} del Piemonte Orientale and INFN Sezione di Torino, Alessandria, Italy
\item \Idef{org32}Dipartimento Interateneo di Fisica `M.~Merlin' and Sezione INFN, Bari, Italy
\item \Idef{org33}European Organization for Nuclear Research (CERN), Geneva, Switzerland
\item \Idef{org34}Faculty of Electrical Engineering, Mechanical Engineering and Naval Architecture, University of Split, Split, Croatia
\item \Idef{org35}Faculty of Engineering and Science, Western Norway University of Applied Sciences, Bergen, Norway
\item \Idef{org36}Faculty of Nuclear Sciences and Physical Engineering, Czech Technical University in Prague, Prague, Czech Republic
\item \Idef{org37}Faculty of Science, P.J.~\v{S}af\'{a}rik University, Ko\v{s}ice, Slovakia
\item \Idef{org38}Frankfurt Institute for Advanced Studies, Johann Wolfgang Goethe-Universit\"{a}t Frankfurt, Frankfurt, Germany
\item \Idef{org39}Fudan University, Shanghai, China
\item \Idef{org40}Gangneung-Wonju National University, Gangneung, Republic of Korea
\item \Idef{org41}Gauhati University, Department of Physics, Guwahati, India
\item \Idef{org42}Helmholtz-Institut f\"{u}r Strahlen- und Kernphysik, Rheinische Friedrich-Wilhelms-Universit\"{a}t Bonn, Bonn, Germany
\item \Idef{org43}Helsinki Institute of Physics (HIP), Helsinki, Finland
\item \Idef{org44}High Energy Physics Group,  Universidad Aut\'{o}noma de Puebla, Puebla, Mexico
\item \Idef{org45}Hiroshima University, Hiroshima, Japan
\item \Idef{org46}Hochschule Worms, Zentrum  f\"{u}r Technologietransfer und Telekommunikation (ZTT), Worms, Germany
\item \Idef{org47}Horia Hulubei National Institute of Physics and Nuclear Engineering, Bucharest, Romania
\item \Idef{org48}Indian Institute of Technology Bombay (IIT), Mumbai, India
\item \Idef{org49}Indian Institute of Technology Indore, Indore, India
\item \Idef{org50}Indonesian Institute of Sciences, Jakarta, Indonesia
\item \Idef{org51}INFN, Laboratori Nazionali di Frascati, Frascati, Italy
\item \Idef{org52}INFN, Sezione di Bari, Bari, Italy
\item \Idef{org53}INFN, Sezione di Bologna, Bologna, Italy
\item \Idef{org54}INFN, Sezione di Cagliari, Cagliari, Italy
\item \Idef{org55}INFN, Sezione di Catania, Catania, Italy
\item \Idef{org56}INFN, Sezione di Padova, Padova, Italy
\item \Idef{org57}INFN, Sezione di Roma, Rome, Italy
\item \Idef{org58}INFN, Sezione di Torino, Turin, Italy
\item \Idef{org59}INFN, Sezione di Trieste, Trieste, Italy
\item \Idef{org60}Inha University, Incheon, Republic of Korea
\item \Idef{org61}Institute for Nuclear Research, Academy of Sciences, Moscow, Russia
\item \Idef{org62}Institute for Subatomic Physics, Utrecht University/Nikhef, Utrecht, Netherlands
\item \Idef{org63}Institute of Experimental Physics, Slovak Academy of Sciences, Ko\v{s}ice, Slovakia
\item \Idef{org64}Institute of Physics, Homi Bhabha National Institute, Bhubaneswar, India
\item \Idef{org65}Institute of Physics of the Czech Academy of Sciences, Prague, Czech Republic
\item \Idef{org66}Institute of Space Science (ISS), Bucharest, Romania
\item \Idef{org67}Institut f\"{u}r Kernphysik, Johann Wolfgang Goethe-Universit\"{a}t Frankfurt, Frankfurt, Germany
\item \Idef{org68}Instituto de Ciencias Nucleares, Universidad Nacional Aut\'{o}noma de M\'{e}xico, Mexico City, Mexico
\item \Idef{org69}Instituto de F\'{i}sica, Universidade Federal do Rio Grande do Sul (UFRGS), Porto Alegre, Brazil
\item \Idef{org70}Instituto de F\'{\i}sica, Universidad Nacional Aut\'{o}noma de M\'{e}xico, Mexico City, Mexico
\item \Idef{org71}iThemba LABS, National Research Foundation, Somerset West, South Africa
\item \Idef{org72}Jeonbuk National University, Jeonju, Republic of Korea
\item \Idef{org73}Johann-Wolfgang-Goethe Universit\"{a}t Frankfurt Institut f\"{u}r Informatik, Fachbereich Informatik und Mathematik, Frankfurt, Germany
\item \Idef{org74}Joint Institute for Nuclear Research (JINR), Dubna, Russia
\item \Idef{org75}Korea Institute of Science and Technology Information, Daejeon, Republic of Korea
\item \Idef{org76}KTO Karatay University, Konya, Turkey
\item \Idef{org77}Laboratoire de Physique des 2 Infinis, Ir\`{e}ne Joliot-Curie, Orsay, France
\item \Idef{org78}Laboratoire de Physique Subatomique et de Cosmologie, Universit\'{e} Grenoble-Alpes, CNRS-IN2P3, Grenoble, France
\item \Idef{org79}Lawrence Berkeley National Laboratory, Berkeley, California, United States
\item \Idef{org80}Lund University Department of Physics, Division of Particle Physics, Lund, Sweden
\item \Idef{org81}Nagasaki Institute of Applied Science, Nagasaki, Japan
\item \Idef{org82}Nara Women{'}s University (NWU), Nara, Japan
\item \Idef{org83}National and Kapodistrian University of Athens, School of Science, Department of Physics , Athens, Greece
\item \Idef{org84}National Centre for Nuclear Research, Warsaw, Poland
\item \Idef{org85}National Institute of Science Education and Research, Homi Bhabha National Institute, Jatni, India
\item \Idef{org86}National Nuclear Research Center, Baku, Azerbaijan
\item \Idef{org87}National Research Centre Kurchatov Institute, Moscow, Russia
\item \Idef{org88}Niels Bohr Institute, University of Copenhagen, Copenhagen, Denmark
\item \Idef{org89}Nikhef, National institute for subatomic physics, Amsterdam, Netherlands
\item \Idef{org90}NRC Kurchatov Institute IHEP, Protvino, Russia
\item \Idef{org91}NRC \guillemotleft Kurchatov\guillemotright~Institute - ITEP, Moscow, Russia
\item \Idef{org92}NRNU Moscow Engineering Physics Institute, Moscow, Russia
\item \Idef{org93}Nuclear Physics Group, STFC Daresbury Laboratory, Daresbury, United Kingdom
\item \Idef{org94}Nuclear Physics Institute of the Czech Academy of Sciences, \v{R}e\v{z} u Prahy, Czech Republic
\item \Idef{org95}Oak Ridge National Laboratory, Oak Ridge, Tennessee, United States
\item \Idef{org96}Ohio State University, Columbus, Ohio, United States
\item \Idef{org97}Petersburg Nuclear Physics Institute, Gatchina, Russia
\item \Idef{org98}Physics department, Faculty of science, University of Zagreb, Zagreb, Croatia
\item \Idef{org99}Physics Department, Panjab University, Chandigarh, India
\item \Idef{org100}Physics Department, University of Jammu, Jammu, India
\item \Idef{org101}Physics Department, University of Rajasthan, Jaipur, India
\item \Idef{org102}Physikalisches Institut, Eberhard-Karls-Universit\"{a}t T\"{u}bingen, T\"{u}bingen, Germany
\item \Idef{org103}Physikalisches Institut, Ruprecht-Karls-Universit\"{a}t Heidelberg, Heidelberg, Germany
\item \Idef{org104}Physik Department, Technische Universit\"{a}t M\"{u}nchen, Munich, Germany
\item \Idef{org105}Politecnico di Bari, Bari, Italy
\item \Idef{org106}Research Division and ExtreMe Matter Institute EMMI, GSI Helmholtzzentrum f\"ur Schwerionenforschung GmbH, Darmstadt, Germany
\item \Idef{org107}Rudjer Bo\v{s}kovi\'{c} Institute, Zagreb, Croatia
\item \Idef{org108}Russian Federal Nuclear Center (VNIIEF), Sarov, Russia
\item \Idef{org109}Saha Institute of Nuclear Physics, Homi Bhabha National Institute, Kolkata, India
\item \Idef{org110}School of Physics and Astronomy, University of Birmingham, Birmingham, United Kingdom
\item \Idef{org111}Secci\'{o}n F\'{\i}sica, Departamento de Ciencias, Pontificia Universidad Cat\'{o}lica del Per\'{u}, Lima, Peru
\item \Idef{org112}St. Petersburg State University, St. Petersburg, Russia
\item \Idef{org113}Stefan Meyer Institut f\"{u}r Subatomare Physik (SMI), Vienna, Austria
\item \Idef{org114}SUBATECH, IMT Atlantique, Universit\'{e} de Nantes, CNRS-IN2P3, Nantes, France
\item \Idef{org115}Suranaree University of Technology, Nakhon Ratchasima, Thailand
\item \Idef{org116}Technical University of Ko\v{s}ice, Ko\v{s}ice, Slovakia
\item \Idef{org117}Technische Universit\"{a}t M\"{u}nchen, Excellence Cluster 'Universe', Munich, Germany
\item \Idef{org118}The Henryk Niewodniczanski Institute of Nuclear Physics, Polish Academy of Sciences, Cracow, Poland
\item \Idef{org119}The University of Texas at Austin, Austin, Texas, United States
\item \Idef{org120}Universidad Aut\'{o}noma de Sinaloa, Culiac\'{a}n, Mexico
\item \Idef{org121}Universidade de S\~{a}o Paulo (USP), S\~{a}o Paulo, Brazil
\item \Idef{org122}Universidade Estadual de Campinas (UNICAMP), Campinas, Brazil
\item \Idef{org123}Universidade Federal do ABC, Santo Andre, Brazil
\item \Idef{org124}University of Cape Town, Cape Town, South Africa
\item \Idef{org125}University of Houston, Houston, Texas, United States
\item \Idef{org126}University of Jyv\"{a}skyl\"{a}, Jyv\"{a}skyl\"{a}, Finland
\item \Idef{org127}University of Liverpool, Liverpool, United Kingdom
\item \Idef{org128}University of Science and Technology of China, Hefei, China
\item \Idef{org129}University of South-Eastern Norway, Tonsberg, Norway
\item \Idef{org130}University of Tennessee, Knoxville, Tennessee, United States
\item \Idef{org131}University of the Witwatersrand, Johannesburg, South Africa
\item \Idef{org132}University of Tokyo, Tokyo, Japan
\item \Idef{org133}University of Tsukuba, Tsukuba, Japan
\item \Idef{org134}Universit\'{e} Clermont Auvergne, CNRS/IN2P3, LPC, Clermont-Ferrand, France
\item \Idef{org135}Universit\'{e} de Lyon, Universit\'{e} Lyon 1, CNRS/IN2P3, IPN-Lyon, Villeurbanne, Lyon, France
\item \Idef{org136}Universit\'{e} de Strasbourg, CNRS, IPHC UMR 7178, F-67000 Strasbourg, France, Strasbourg, France
\item \Idef{org137}Universit\'{e} Paris-Saclay Centre d'Etudes de Saclay (CEA), IRFU, D\'{e}partment de Physique Nucl\'{e}aire (DPhN), Saclay, France
\item \Idef{org138}Universit\`{a} degli Studi di Foggia, Foggia, Italy
\item \Idef{org139}Universit\`{a} degli Studi di Pavia, Pavia, Italy
\item \Idef{org140}Universit\`{a} di Brescia, Brescia, Italy
\item \Idef{org141}Variable Energy Cyclotron Centre, Homi Bhabha National Institute, Kolkata, India
\item \Idef{org142}Warsaw University of Technology, Warsaw, Poland
\item \Idef{org143}Wayne State University, Detroit, Michigan, United States
\item \Idef{org144}Westf\"{a}lische Wilhelms-Universit\"{a}t M\"{u}nster, Institut f\"{u}r Kernphysik, M\"{u}nster, Germany
\item \Idef{org145}Wigner Research Centre for Physics, Budapest, Hungary
\item \Idef{org146}Yale University, New Haven, Connecticut, United States
\item \Idef{org147}Yonsei University, Seoul, Republic of Korea
\end{Authlist}
\endgroup

\end{document}